\documentclass[aps, a4paper, amsfonts, amssymb, amsmath, reprint, showkeys, nofootinbib, twoside, floatfix, physrev]{revtex4-2}
\usepackage[english]{babel}
\usepackage[utf8]{inputenc}
\usepackage{wrapfig}
\usepackage{float}
\usepackage{ulem}
\usepackage{bm}
\usepackage[colorinlistoftodos, color=green!40, prependcaption]{todonotes}
\usepackage{comment}
\usepackage{amsthm}
\usepackage{mathtools}
\usepackage{physics}
\usepackage{xcolor}
\usepackage{graphicx}
\usepackage[left=23mm,right=13mm,top=35mm,columnsep=15pt]{geometry} 
\usepackage{adjustbox}
\usepackage{placeins}
\usepackage[T1]{fontenc}
\usepackage{lipsum}
\usepackage{csquotes}
\usepackage[pdftex, pdftitle={Article}, pdfauthor={Author}]{hyperref} % For hyperlinks in the PDF
\begin{document}
\title{Squeezing Enhancement Through Resonant Interference in Multi-ring Resonators}

\author{M. Sloan$^1$ and J. E. Sipe $^1$}
\affiliation{$^1$Department of Physics, University of Toronto, 60 St. George Street, Toronto, ON, M5S 1A7, Canada}

\date{\today} % Leave empty to omit a date

\begin{abstract}
We develop a non-perturbative description of squeezed light generation in an arbitrary lossy structure consisting of multiple coupled microring resonators. This is applied to two ring photonic molecules where the interference of the fields in the coupled rings leads to a modification in the resonance spectrum near a shared resonance. Considering a dual-pump degenerate squeezing scheme under a five resonance approximation, we investigate two methods to suppress parasitic four-wave mixing contributions and compensate for group velocity dispersion within a primary resonator through hybridization effects with a second auxiliary resonator. In the former case, this comes from an effective splitting of the unwanted resonances supporting parasitic four-wave mixing interactions that add thermal noise to the desired degenerate squeezed state. For sufficiently strong coupling between the resonators, we demonstrate near complete suppression of such parasitic processes, resulting in near unit fidelities with the corresponding output state that would arise were the parasitic interactions neglected. In the latter case, the hybridization effectively shifts a pump resonance, realigning the desired dual-pump four-wave mixing process and leading to a significant enhancement of the signal generation and output squeezing. 
\end{abstract}

\maketitle

\section{Introduction} \label{sec:Introduction}
With the reduction in noise in one quadrature variable at the expense of increased noise in the other, quadrature squeezed light is an indispensable resource in many modern photonic technologies, seeing wide use in applications including metrology \cite{Caves1981, LIGO}, quantum computing \cite{Bourassa, Arrazola, Adcock2019, Llewellyn2020}, and non-classical state generation \cite{Hamilton2017, Larsen2025, Su2019}. Additionally, weak non-degenerate vacuum squeezed states can also provide a valuable source of heralded single photons \cite{Fasel_2004, Liu:20, Llewellyn2020}.

Strategies involving Spontaneous Parametric Down-Conversion (SPDC) or Spontaneous Four Wave Mixing (SFWM) in integrated photonic devices, utilizing $\chi_{(2)}$ and $\chi_{(3)}$ nonlinearities respectively, are 
attractive for squeezed light generation. With many components integrated on a single chip \cite{Bourassa, Bogaerts2020, Larsen2025}, there is a reduced footprint compared to bulk optics. Furthermore, integrated micro-resonators, such as photonic molecule defect cavities \cite{Banaee:08} and micro-ring resonators \cite{Vernon2019, Quesada:22, Vendromin, Vaidya}, allow for large field densities, enhancing nonlinear squeezing processes while partitioning the resulting spectrum into discrete system resonances. For degenerate squeezed light generation, dual-pump SFWM schemes \cite{Zhao2020, Zhang2021, Seifoory, Quesada:22} benefit due to the generated field being produced near the pumping frequencies, alleviating the need for complicated dispersion engineering, necessary in many SPDC implementations \cite{Fontaine, Chen, Nitiss2022}, at the expense of a reduced nonlinear strength. However, such schemes suffer from the presence of a variety of parasitic single-pump SFWM (SP-SFWM) processes, as well as hyper-parametric SFWM (HP-SFWM) and Bragg-scattering four-wave mixing (BS-FWM) processes involving additional, unwanted system resonances. The degenerate squeezed light from the desired dual-pump SFWM (DP-SFWM) interaction can be polluted with light generated by these processes, introducing thermal noise to the output state and reducing the achievable squeezing. Indeed, within a five resonance approximation and for a CW pump, these additional parasitic processes have been shown to significantly reduce the achievable output squeezing within a single ring system \cite{Seifoory}.

To avoid this one can employ photonic molecule systems constructed from multiple coupled ring resonators \cite{Zhang2021, Gentry:14, Sabattoli2021, Zatti2023, Viola:24}. They allow for the selective modification of individual system resonances through interference effects between the light propagating in the constituent rings \cite{Zhang2021, Gentry:14, Viola:24}. The resulting hybridized system can have resonance linewidths and frequencies that differ greatly from those of a single resonator, allowing for selective and tunable suppression of unwanted SFWM interactions \cite{Zhang2021, Viola:24}. Such suppression schemes involving the splitting of unwanted resonances supporting parasitic SFWM processes have been motivated by a perturbative treatment of the relative rates of each nonlinear process and demonstrated experimentally within a dual ring photonic molecule \cite{Zhang2021}. This has also been shown analytically within a single Schmidt mode approximation for a dual ring system coupled through an interferometric coupler and validated through a non-perturbative squeezing calculation \cite{Viola:24}. Consequentially, such photonic molecule systems offer robust control of squeezed light generation in integrated structures, far beyond what is possible in single ring systems.

In this paper we generalize a non-pertubative asymptotic scattering method, previously developed for non-degenerate squeezing in a single ring \cite{Sloan2024}, to describe degenerate squeezing from SFWM interactions in an arbitrary photonic molecule system. This is then applied to a dual-ring system, and we present two examples of squeezing enhancement in a dual-pump scheme, employing a five resonance model to account for parasitic SP-SFWM, HP-SFWM, and BS-FWM processes. We demonstrate suppression of undesired SFWM interactions by the splitting of unwanted idler resonances, as well as compensation of the detuning of the desired DP-SFWM by inducing an effective shift on the pump resonances. Beyond computing the maximum output squeezing, we investigate the multimode nature of the output field by constructing the reduced output state in the desired signal resonance, and analyze the effects of the thermal contributions to the achievable squeezing. Furthermore, we investigate the squeezing in the lowest order Mercer-Wolf temporal modes \cite{Wolf:82, Kopylov2025}, and observe a significant variation in the pump detuning that optimizes the achievable squeezing as a function of the coupling strength between the micro-rings.

In Section \ref{sec:SectionII}, we develop the asymptotic field expansion \cite{Breit, Liscidini} for an arbitrary system of multiple coupled resonators, including the local basis used in the nonlinear propagation. Using this, we construct the equations of motion for the asymptotic and local basis operators in Section \ref{sec:SectionIII}. In Section \ref{sec:SectionIV}, we outline the computation of the output state of the system and the decompositions to identify the squeezing and thermal contributions. The two dual-ring examples are presented in Section \ref{sec:SectionV}, after which we conclude with Section \ref{sec:Conclusion}.
     
\section{Asymptotic Fields} \label{sec:SectionII}
\begin{figure}
    \centering
    \includegraphics[width=0.98\linewidth]{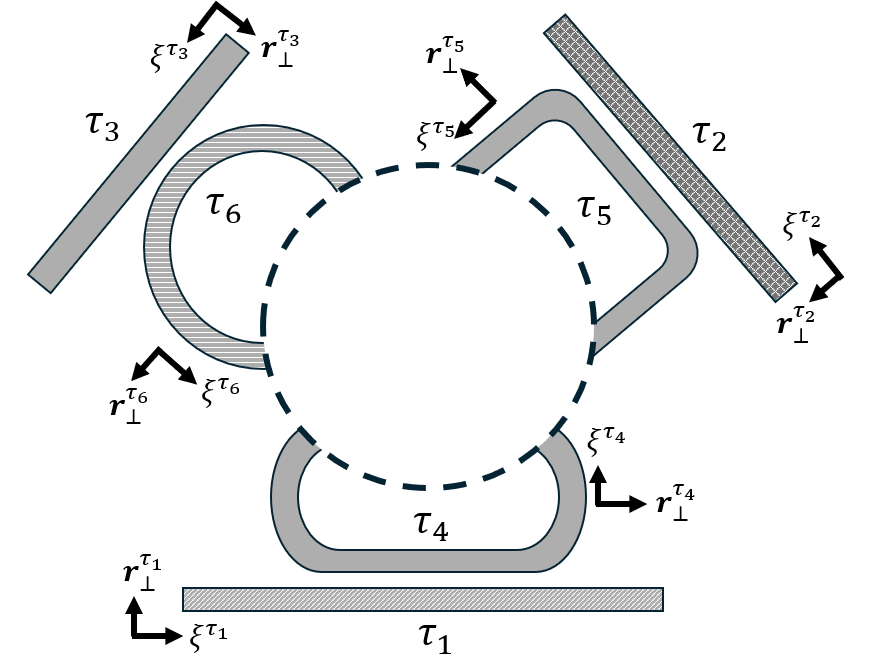}
    \caption{Diagram of a general system of coupled resonators and waveguides. The dotted central region can denote an arbitrary assortment of elements. To each ring and waveguide an index $\tau$ and a corresponding local coordinate system $\left( \textbf{r}_{\perp}^{\tau}, \xi^{\tau} \right)$ is shown.}
    \label{fig:generalSystemWithCoordinates}
\end{figure}

We begin by considering a general system consisting of multiple coupled ring resonators and waveguides, and derive the asymptotic mode expansion of the displacement field \cite{Sloan2024, Breit, Liscidini}. To each element (ring or waveguide) we assign an index $\tau$ and a local coordinate system $\textbf{r}_{\tau} = (\textbf{r}^{\tau}_{\perp}, \xi^{\tau})$, in which the coordinate $\xi^{\tau}$ measures the distance along the forward direction of propagation in the element $\tau$, and $\textbf{r}^{\tau}_{\perp}$ corresponds to the coordinates in the plane tangent to this propagation direction (see Fig. \ref{fig:generalSystemWithCoordinates}). For simplicity, each waveguide is taken to extend infinitely in the $\xi$ direction. Furthermore, for each element $\tau$ we assume that the linear permittivity in the absence of all other $\tau'$, $\epsilon_{\tau}(\textbf{r})$, depends only on $\textbf{r}^{\tau}_{\perp}$, and write $\epsilon_{\tau}(\textbf{r}) = \epsilon_{\tau}(\textbf{r}^{\tau}_{\perp})$. 

\begin{figure}
    \centering
    \includegraphics[width=\linewidth]{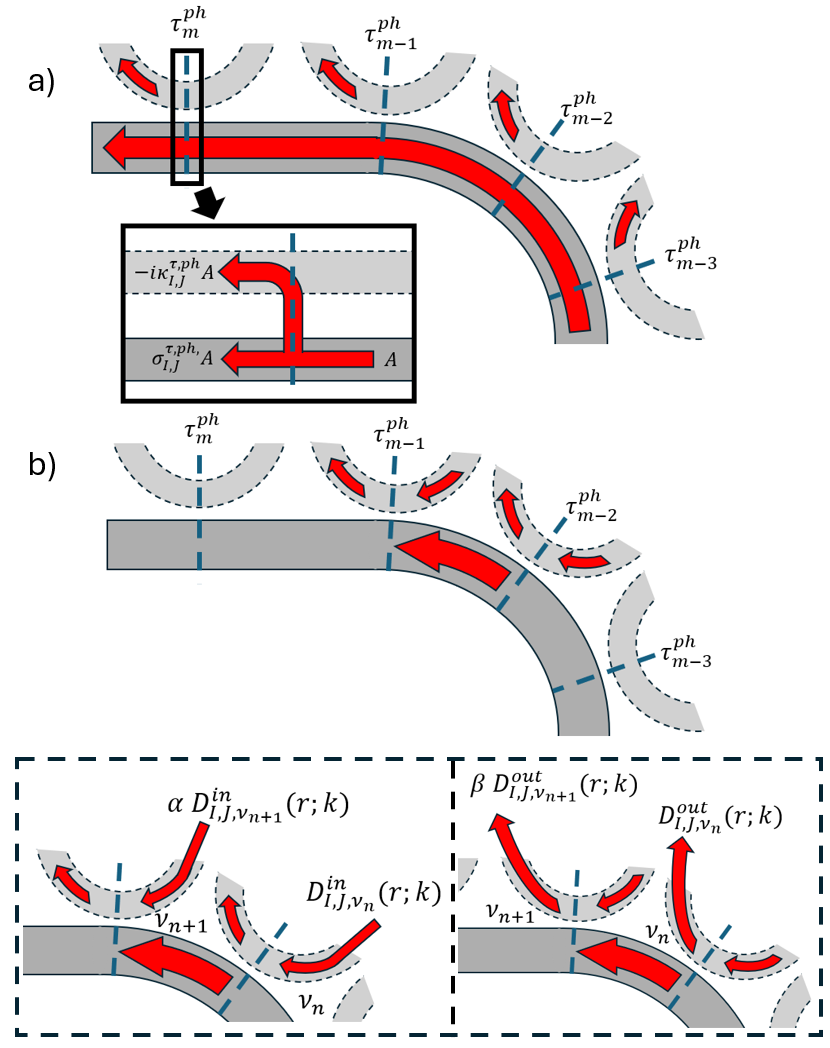}
    \caption{(a) Diagram of a section of a resonator/waveguide in which a series of phantom channels (dotted elements) are point coupled to model scattering loss. Red arrows denote the power flow corresponding to a sample asymptotic-in field distribution in which the ring energy is coupled into each of the phantom channel outputs (b) Sample local basis field distribution, constructed as a linear combination of the $\nu_{n}$ and $\nu_{n+1}$ asymptotic-in field distributions (left dotted box), or equivalently a linear combination of the $\nu_{n}$ and $\nu_{n+1}$ asymptotic-out field distributions (right dotted box), such that the non-zero field support is restricted between the adjacent phantom channel coupling points.}
    \label{fig:lossDiagrams}
\end{figure}

Scattering loss within each element is modeled through a number of fictitious phantom channels \cite{Quesada:22} point coupled along the length of each element (see Fig. \ref{fig:lossDiagrams}). In this way, the fields in the scattering channels are treated identically to the fields in the real waveguide channels; this allows us to maintain the Hermitian nature of the Hamiltonian. Indeed, the only distinction between the phantom channels and real channel will be in their coupling treatment: The real channels are allowed to couple over a finite range, whereas for simplicity the phantom channels are coupled at a point. In each case, to an element $\tau$ corresponding to a real waveguide or phantom channel we can identify an input port in which the fields are propagating towards the system, as well as an output port in which the fields are propagating away from the system. We index each of the inputs (outputs) with a label $\nu \in S = S_r \cup S_{ph}$, where $S_r$ corresponds to the real waveguide inputs (outputs) and $S_{ph}$ corresponds to the phantom channel inputs (outputs). In what is to follow, $\nu$ will be used to index the asymptotic-in and asymptotic-out modes corresponding to inputs and output in each of the system ports. Thus we understand $\nu$ to denote an \emph{input} port when indexing an asymptotic-in state, and an \emph{output} port when indexing an asymptotic-out state.

Introducing a set of disjoint frequency ranges $J \in \mathcal{J}$ such that $J \cap J' = \emptyset$ for $J \neq J'$, we write the fields of the full coupled system as \cite{Liscidini, Sloan2024}
\begin{equation} \label{eq:GeneralAsymptoticExpansion}
\begin{split}
    \textbf{D}(\textbf{r}, t) &= \sum_{I,J} \sum_{\nu \in S} \int_{R_I(J)} dk \sqrt{\frac{\hbar \omega_{J}(k)}{4 \pi}} \textbf{D}^{in}_{I, J, \nu}(\textbf{r}; k) \\
    & \qquad \qquad \qquad \qquad  \times a_{I, J, \nu}^{in}(k, t) + H. c. \\
    &= \sum_{I, J} \sum_{\nu \in S} \int_{R_I(J)} dk \sqrt{\frac{\hbar \omega_{J}(k)}{4 \pi}} \textbf{D}^{out}_{I, J, \nu}(\textbf{r}; k) \\
    & \qquad \qquad \qquad \qquad  \times a_{I, J, \nu}^{out}(k, t) + H. c. \\
\end{split}
\end{equation}
Here $I$ denotes the spatial mode of the field in the given element, with $R_I(J)$ being the k-space range for the mode $I$ corresponding to the frequencies in $J$. The pair $\textbf{D}^{in}_{I, J, \nu}(\textbf{r}; k)$ and $a_{I, J, \nu}^{in}(k, t)$ then denote the field amplitude and bosonic annihilation operator for the asymptotic-in state. A usual superposition of these, at $t = -\infty$, corresponds to a field incident in the input port $\nu$, spatial mode $I$, and wavenumber $k$. On the other hand, the pair $\textbf{D}^{out}_{I, J, \nu}(\textbf{r}; k)$ and $a_{I, J, \nu}^{out}(k, t)$ denote the field amplitude and bosonic annihilation operator for the asymptotic-out state. A usual superposition of these, at $t = +\infty$, corresponds to a field exiting in the output port $\nu$, spatial mode $I$, and wavenumber $k$ \cite{Breit}. In each case, the asymptotic-in(out) field operators satisfy the standard bosonic commutation relations \cite{Liscidini,Breit, Sloan2024},
\begin{equation}
    \begin{split}
            &\left[ a_{I,J,\nu}^{asy-in}(k, t), a_{I',J',\nu'}^{asy-in}(k', t) \right] = 0 \\
            & \qquad \qquad \qquad \qquad = \left[ a_{I,J,\nu}^{asy-out}(k, t), a_{I',J',\nu'}^{asy-out}(k', t) \right] \\
            &\left[ a_{I,J,\nu}^{asy-in \dagger}(k, t), a_{I',J',\nu'}^{asy-in}(k', t) \right] = \delta_{I, I'} \delta_{J, J'} \delta_{\nu, \nu'} \delta(k - k') \\
            & \qquad \qquad \qquad \qquad  = \left[ a_{I,J,\nu}^{asy-out \dagger}(k, t), a_{I',J',\nu'}^{asy-out}(k', t) \right] 
        \end{split}
\end{equation}

In what is to follow, for all times prior to some initial time, $t_0$, in which the energy from the input channel has yet to reach the resonator system, we choose for the inputs in each of the phantom channels to be uncorrelated, such that
\begin{equation}
\begin{split}
    &\langle a_{I, J, \nu}^{asy-in}(k, t_0) a_{I', J', \nu'}^{asy-in}(k', t) \rangle =\\
    & \qquad \langle a_{I, J, \nu}^{asy-in}(k, t_0) a_{I', J', \nu'}^{asy-in}(k', t_0) \rangle \delta_{I, I'} \delta_{J, J'} \delta_{\nu, \nu'} \delta(k - k') \\
    &\langle a_{I, J, \nu}^{asy-in \dagger}(k, t_0) a_{I', J', \nu'}^{asy-in}(k', t_0) \rangle = \\
    & \qquad  \langle a_{I, J, \nu}^{asy-in \dagger}(k, t_0) a_{I', J', \nu'}^{asy-in}(k', t_0) \rangle \delta_{I, I'} \delta_{J, J'} \delta_{\nu, \nu'} \delta(k - k')
\end{split}
\end{equation}
for all $\nu, \nu' \in S_{ph}$. Specifically, in the examples in Section \ref{sec:SectionV}, we assume that the input to each of the phantom channels is the vacuum, corresponding to
\begin{equation}
    \begin{split}
        &\langle a_{I, J, \nu}^{asy-in}(k, t_0) a_{I', J', \nu'}^{asy-in}(k', t) \rangle = 0 \\
        &\qquad \qquad \qquad = \langle a_{I, J, \nu}^{asy-in \dagger}(k, t_0) a_{I', J', \nu'}^{asy-in}(k', t) \rangle,
    \end{split}
\end{equation}
however, this can easily be generalized to a more exotic environment if needed.

\begin{figure*}
    \centering
    \includegraphics[width=0.98\linewidth]{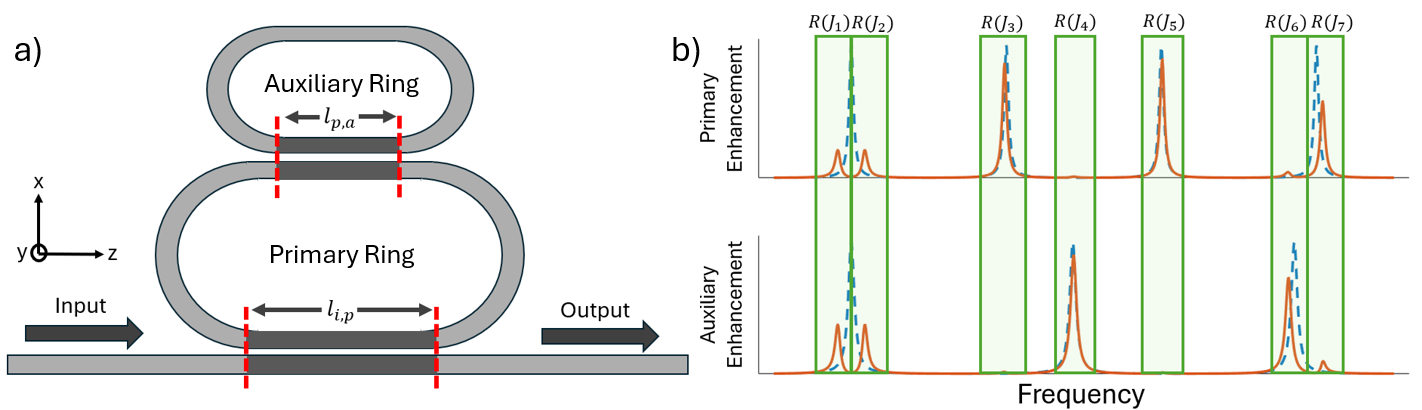}
    \caption{(a) Diagram of a two ring photonic molecule. Here, an input/output waveguide and the auxiliary ring are side coupled to the primary ring along a length $l_{i,p}$ and $l_{p,a}$ respectively. (b) Sample field enhancement in the primary ring (top) and the auxiliary ring (bottom) when the rings are uncoupled (dotted blue) and coupled (solid red). A set of frequency ranges $J_i$ are superimposed.}
    \label{fig:ringDiagrams}
\end{figure*}

In terms of the frequency ranges, we choose each $J$ to be centered at a single resonance of the coupled system of central frequency $\omega_J$, with a span large enough to fully contain the excitations of the resonance, but small enough to remain disjoint from each other $J'$. This is shown diagrammatically in Fig. \ref{fig:ringDiagrams}b for a dual-ring photonic molecule (Fig. \ref{fig:ringDiagrams}a) that will be studied in Section \ref{sec:SectionV}, in which each range $J$ corresponds to an isolated resonance of the primary resonator, an isolated resonance of the auxiliary resonator, or a hybridized resonance of the coupled system.

Note that for a general lossless system with many coupled resonators, there may also exist modes which are confined to a single resonator element or a combination of resonator elements. Such bound modes would be distinct from the asymptotic modes included in equation (\ref{eq:GeneralAsymptoticExpansion}), as the asymptotic character of the fields in these modes could not be phrased in terms of a field incident into or exiting out of the resonator system. However, when including scattering and radiation loss through the addition of phantom channels, as considered here, any finite field intensity within a resonator element $\tau$ must necessarily decay to zero at $t=+\infty$ regardless of the coupling to the real waveguide channels, due to energy being carried away from the system through the phantom channels. Consequentially, one could fully describe the state of the system in terms of the asymptotic-out modes (and by extension the asymptotic-in modes) corresponding to the real waveguides and phantom channels, with equation (\ref{eq:GeneralAsymptoticExpansion}) representing an expansion of the displacement field in terms of a complete set of mode fields. 

For our current treatment, we consider systems in which the coupling between resonators and input/output channels is weak, with sufficiently large bends that radiation loss within the curved sections can be safely neglected. We show in Appendix \ref{sec:asyFields} that the mode distributions $\textbf{D}^{in(out)}_{I, J, \nu}(\textbf{r}; k)$ around the element $\tau$ can then be decomposed as
\begin{equation} \label{eq:AsymptoticFieldDistExpansion}
    \textbf{D}^{in(out)}_{I, J, \nu}(\textbf{r}; k) = \textbf{d}_{I, J}^{\tau}(\textbf{r}_{\perp}; \xi) h_{I, J, k}^{in(out),\nu, \tau}(\xi) e^{i \beta_{I, J}^{\tau} \xi^{\tau}}.
\end{equation}
In what is to follow, to avoid clutter we omit the superscript on the local coordinates $\xi$ and $\textbf{r}_{\perp}$ when they are evaluated within functions in which $\tau$ is present.

Here and below the quantity $\beta_{I, J}^{\tau}$ is the effective wave number of the field propagating in the element $\tau$, spatial mode $I$, and central frequency of the $J^{th}$ frequency bin, $\omega_J$. Additionally, $\textbf{d}_{I, J}^{\tau}(\textbf{r}_{\perp}; \xi)$ gives the polarization direction and field amplitude in the plane tangent to the propagation direction, and from the above assumption can be normalized as (see Appendix \ref{sec:asyFields}) 
\begin{equation} \label{eq:modeNormalization}
    \int \frac{\textbf{d}^{\tau *}_{I, J}(\textbf{r}_{\perp}; \xi) \cdot \textbf{d}^{\tau}_{I, J}(\textbf{r}_{\perp}; \xi)}{\epsilon_0 \epsilon_{\tau}(\textbf{r}_{\perp})} d\textbf{r}_{\perp}^{\tau} = 1.
\end{equation}
Finally, $h_{I, J, k}^{\nu, \tau}(\xi)$ gives the $\xi$ dependent field amplitude in the element $\tau$. Note that from the point coupling assumption of the phantom channels, $h_{I, J, k}^{\nu, \tau}(\xi)$ is discontinuous about the coupling points with each $\tau$ corresponding to a phantom channel.

For a general system containing many resonators, waveguides, and phantom channels, it can be tedious to write down the particular form of $h_{I, J, k}^{\nu, \tau}(\xi)$ at all points $\textbf{r}$. Indeed, in most cases this function would need to be found numerically. However, in some special cases a simple, closed form expression can be identified within a given region of the system. A particularly insightful example is when we consider shrinking the coupling regions between the input waveguide and the primary ring, as well as shrinking the coupling region between the primary and auxiliary resonators in the dual ring photonic molecule in Fig \ref{fig:ringDiagrams}a, to single points; then the coupling can be described by effective self-coupling coefficients  $\varsigma^{wg}_{I, J}$ and $\varsigma^{aux}_{I, J}$ respectively. Taking $\tau = wg, pr, ar$ to denote the input/output waveguide, primary ring, and auxiliary ring respectively, with  $\nu_{wg}$ to represent the input port of the real waveguide, and introducing a self-coupling coefficient $\sigma^{\nu, ph}_{I,J}$ for each $\nu \in S_{ph}$, we show in Appendix \ref{sec:SimpilifiedTransmission} that within the output port of the waveguide we can write
\begin{equation} \label{eq:SimplifiedTransmissionField}
    h_{I, J, k}^{in,\nu_{wg}, wg}(\xi) = \frac{\varsigma^{wg}_{I, J} - T_{I,J}^{ar}(k) \Gamma_{I,J}^{pr} e^{i \phi_{I,J}^{pr}(k)}}{1 - \varsigma^{wg}_{I, J} T_{I,J}^{ar}(k) \Gamma_{I,J}^{pr} e^{i \phi_{I,J}^{pr}(k)}} e^{i(k - \beta_{I,J}^{wg}) \xi},
\end{equation}
where 
\begin{equation} \label{eq:AuxiliaryRingTransmission}
    T_{I,J}^{ar}(k) = \frac{\varsigma^{aux}_{I, J} - \Gamma_{I,J}^{ar} e^{i \phi_{I,J}^{ar}(k)}}{1 - \varsigma^{aux}_{I, J} \Gamma_{I,J}^{ar} e^{i \phi_{I,J}^{ar}(k)}},
\end{equation}
and the factors $\Gamma_{I,J}^{pr}$ and $\Gamma_{I,J}^{ar}$ are defined as
\begin{equation}
    \begin{split}
        \Gamma_{I,J}^{pr} &= \prod_{\nu \in S_{ph} \cap pr} \sigma_{I,J}^{\nu, ph}, \\
        \Gamma_{I,J}^{ar} &= \prod_{\nu \in S_{ph} \cap ar} \sigma_{I,J}^{\nu, ph}, \\
    \end{split}
\end{equation}
where we use $S_{ph} \cap pr$ and $S_{ph} \cap ar$ as 
short hand to denote the set of input (output) ports corresponding respectively to phantom channels coupled to the primary and auxiliary ring. Finally, the factors $\phi_{I,J}^{pr}(k)$ and $\phi_{I,J}^{ar}(k)$ represent the round trip phase for a single pass of the primary and auxiliary ring respectively.

\begin{figure}
    \centering
    \includegraphics[width=0.95\linewidth]{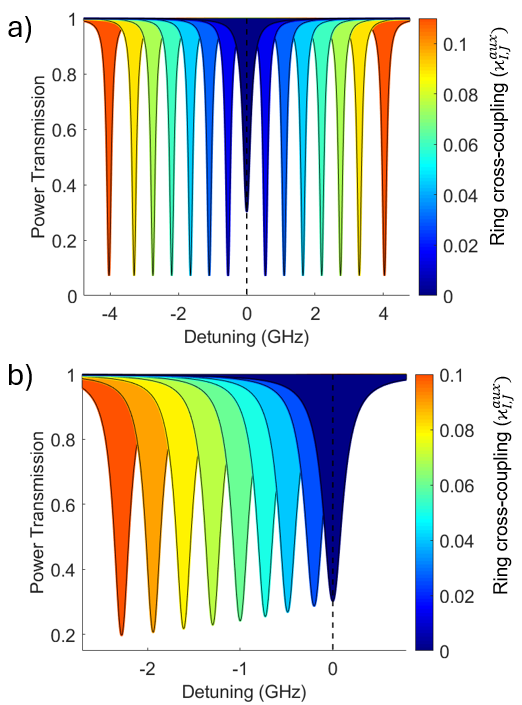}
    \caption{(a) Hybridized system resonance leading to resonance splitting with $\delta_a = 0$ GHz, $\varsigma_{I,J}^{wg} = 0.997$, $\Gamma_p = 0.9991$, and $\Gamma_a = 0.99935$. Primary and auxiliary ring lengths, $L_p$ and $L_a$, are $L_p = 2\pi \times 120$ $\mu$m $ = \frac{4}{3} L_a$ (b) Effective shift in primary resonance due to resonance hybridization with $\delta_a = 4.77$ GHz, $L_p = 2\pi \times 120$ $\mu$m, $ L_a = 2 \pi \times 75$ $\mu$m, $\varsigma_{I,J}^{wg} = 0.997$, $\Gamma_p = 0.9991$, and $\Gamma_a = 0.99945$.}
    \label{fig:transmissionDiagrams}
\end{figure}

We note that $\left| h_{I, J, k}^{in,wg, wg}(\xi) \right|^2$ identifies the linear power transmission of the structure subject to an input in the spatial mode $I$ and wavenumber $k$ in the waveguide channel. For a low loss, weakly coupled auxiliary ring ($\varsigma^{aux}_{I, J}$, $\Gamma_{I,J}^{ar} \approx 1$) with an input field far from an auxiliary ring resonance, we have $T_{I,J}^{ar}(k) \approx 1$, leading to a transmission spectrum resembling that of an equivalent system with the auxiliary ring decoupled from the primary ring. However, when a resonance of the primary resonator is sufficiently close to a resonance of the auxiliary ring, the two resonances can hybridize, producing system resonances distinct from those of each isolated ring. This is shown in Fig. \ref{fig:transmissionDiagrams}, where for a fixed $\varsigma^{wg}_{I, J}$ we vary the magnitude of the ring cross-coupling $\varkappa_{I,J}^{aux} = \sqrt{1 - \left(\varsigma^{aux}_{I, J} \right)^2}$. Indeed, for a shared resonance between the primary and auxiliary ring, the resonances of the coupled system split into a pair of resonances equally spaced from the resonance of the uncoupled system. When the resonances of the uncoupled primary and auxiliary ring are detuned by a small amount $\delta_a$, the transmission profile of the two hybridized system resonances are no longer symmetric. And when $\delta_a$ is sufficiently large, the hybridization of the resonances resembles an effective shift of the primary ring resonance. Importantly, the resonances of the full coupled photonic molecule may differ in location compared to that of each uncoupled resonator, allowing for selective detuning of nonlinear processes involving fields within these resonances.

The above simplified examples also demonstrate the benefit of framing the system in terms of asymptotic fields: all the linear dynamics, including the interference effects from the coupling between the ring resonators, is folded into the definitions of the modes themselves through, for example, the functions $h_{I, J, k}^{in(out), \nu, \tau}(\xi)$. As such, the mode expansion of the relevant fields can be readily interpreted in terms of modes describing the full, linear system response, rather than expanding in terms of mode of the constituent elements. In this way, we need only choose the $J$ and associated ranges $R(J)$ which correspond to significant field enhancement in the resonator structure, which, as show in Fig. \ref{fig:transmissionDiagrams} and as will be the case in the examples of Section \ref{sec:SectionV}, may or may not correspond to the resonances of any individual, isolated ring resonance.

To conclude this section, we introduce a local basis in which the field amplitude within the resonators are non-zero only between a pair of adjacent phantom channel coupling points \cite{Sloan2024}. This is shown diagrammatically in Fig. \ref{fig:lossDiagrams}b and can be understood simply as taking a linear combination of the asymptotic-in (or equivalently asymptotic-out) field distributions tuned to evaluate to zero following a given phantom channel coupling point. Consequently for $\nu \in S$ we can denote by $\textbf{D}_{I, J, \nu}^{loc}(\textbf{r}; k)$ the local basis field amplitude constructed as a superposition of $\textbf{D}_{I, J, \nu}^{in}(\textbf{r}; k)$ and $\textbf{D}_{I, J, \nu'}^{in}(\textbf{r}; k)$, for for neighboring ports $\nu'$, such that the field amplitude cancel at the $\nu'$ coupling point. Thus the index $\nu$ serves to identify the location where $\textbf{D}_{I, J, \nu}^{loc}(\textbf{r}; k)$ is nonzero (see Fig.\ref{fig:lossDiagrams}). As discussed earlier \cite{Sloan2024}, within the region in which two real elements, $\tau$ and $\tau'$, have non-zero coupling, a local basis construction as outlined above can be formed only by aligning the coupling points of the phantom channels coupled to $\tau$ and $\tau'$ within this region. In what is to follow, we assume such an alignment has been adopted.

With the local basis field amplitudes set, we can introduce transformations $\mathcal{L}_{I, J, k}^{in}$ and $\mathcal{L}_{I, J, k}^{out}$ such that
\begin{equation} \label{eq:LocalBasisDefinition}
\begin{split}
    \textbf{D}_{I, J, \nu}^{loc}(\textbf{r}; k) &= \sum_{\nu' \in S} \left[ \mathcal{L}_{I, J, k}^{in} \right]_{\nu, \nu'} \textbf{D}_{I, J, \nu'}^{in}(\textbf{r}; k) \\
    &= \sum_{\nu' \in S} \left[ \mathcal{L}_{I, J, k}^{out} \right]_{\nu, \nu'} \textbf{D}_{I, J, \nu'}^{out}(\textbf{r}; k),
\end{split}
\end{equation}
and expand the local basis field amplitudes in the element $\tau$ as
\begin{equation} \label{eq:LocalBasisExpansion}
    \textbf{D}_{I, J, \nu}^{loc}(\textbf{r}; k) = \textbf{d}_{I, J}^{\tau}(\textbf{r}_{\perp}; \xi) h_{I, J, k}^{loc,\nu, \tau}(\xi) e^{i \beta_{I, J}^{\tau} \xi^{\tau}}
\end{equation}
where,
\begin{equation}
    \begin{split}
        h_{I, J, k}^{loc,\nu, \tau}(\xi) &= \sum_{\nu' \in S} \left[ \mathcal{L}_{I, J, k}^{in} \right]_{\nu, \nu'} h_{I, J, k}^{in,\nu', \tau}(\xi) \\
        & = \sum_{\nu' \in S} \left[ \mathcal{L}_{I, J, k}^{out} \right]_{\nu, \nu'} h_{I, J, k}^{out,\nu', \tau}(\xi).
    \end{split}
\end{equation}
Then the operator associated with the field amplitude $\textbf{D}_{I, J, \nu}^{loc}(\textbf{r}; k)$, $a^{loc}_{I, J, \nu}(k, t)$, can be related to the bosonic asymptotic-in/out operators through
\begin{equation} \label{eq:localBasisOperatorTransformation}
    \begin{split}
    a^{loc}_{I, J, \nu}(k, t) &= \sum_{\nu' \in S} \left[ \left(\mathcal{L}_{I, J, k}^{in} \right)^{-1} \right]_{\nu', \nu} a^{in}_{I, J, \nu'}(k, t), \\
    &= \sum_{\nu' \in S} \left[ \left(\mathcal{L}_{I, J, k}^{out} \right)^{-1} \right]_{\nu', \nu} a^{out}_{I, J, \nu'}(k, t).
    \end{split}
\end{equation}
However, the $a^{loc}_{I, J, \nu}(k, t)$ do not satisfy the standard bosonic commutation relations, but rather 
\begin{equation}
\begin{split}
    &\left[ a^{loc}_{I, J, \nu}(k, t), a^{loc}_{I', J', \nu'}(k', t) \right] = 0 \\
    & \qquad \qquad\qquad \qquad = \left[ a^{loc \dagger}_{I, J, \nu}(k, t), a^{loc \dagger}_{I', J', \nu'}(k', t) \right], \\
    &\left[ a^{loc}_{I, J, \nu}(k, t), a^{loc \dagger}_{I', J', \nu'}(k', t) \right] \\
    &\qquad \qquad \qquad \qquad  =  \delta_{J,J'} \delta(k - k') \left[\mathcal{C}_{J}^{I, I'}(k) \right]_{\nu, \nu'}, \\
\end{split}
\end{equation}
where 
\begin{equation}
\begin{split}
    \left[\mathcal{C}_{J}^{I, I'}(k) \right]_{\nu, \nu'} &= \sum_{\nu'' \in S} \left[ \left(\mathcal{L}_{I, J, k}^{in} \right)^{-1} \right]_{\nu'', \nu} \left[ \left(\mathcal{L}_{I', J', k'}^{in} \right)^{-1} \right]_{\nu'', \nu'} \\
    &= \sum_{\nu'' \in S} \left[ \left(\mathcal{L}_{I, J, k}^{out} \right)^{-1} \right]_{\nu'', \nu} \left[ \left(\mathcal{L}_{I', J', k'}^{out} \right)^{-1} \right]_{\nu'', \nu'}
\end{split}
\end{equation}
Despite this, the local basis can provide a more convenient expansion for calculating the nonlinear evolution \cite{Sloan2024}. Indeed, as will be seen in the following section, at the expense of phrasing the nonlinear evolution in terms of a non-orthogonal basis of modes, one can greatly simplify the nonlinear overlaps as a result of the non-zero field support of the local basis modes corresponding to an input $\nu$ being mostly disjoint from all other local basis modes corresponding to an input $\nu'$ with $\nu \neq \nu'$. Furthermore, within the local basis, unlike the asymptotic-in(out) basis, the field enhancement resulting from the resonant structure of the system is mainly contained within the commutators of the field operators rather than $|h_{I, J, k}^{\nu, \tau, loc}(\xi)|$ which, in the case when the frequency dependence of the coupling to the phantom channel is negligible within the range $R(J)$, is often independent of $k$.
\section{Nonlinear Hamiltonian and Field Evolution} \label{sec:SectionIII}
With the field expansion in each of the bases of interest in hand, we now expand the Hamiltonian for the system and derive the equations of motion for each of the operators. For simplicity, we assume that the fields driving the system are restricted to a single spatial mode $I$, with the coupling between elements sufficiently weak, and bend radii sufficiently large, that the coupling of the mode $I$ to the higher order modes $I'$ can be neglected. As such, we drop the label $I$ in each subscript or superscript, but emphasize that the following treatment can be generalized to include additional spatial modes in a straightforward manner if needed.

To start, we write the full Hamiltonian of the system as
\begin{equation}
    H = H_{L} + H_{NL},
\end{equation}
where $H_L$ contains the terms leading to linear evolution of the field operators, with $H_{NL}$ providing the additional terms to describe the nonlinear evolution. From the definition of the asymptotic-in(out) and local bases, the linear Hamiltonian can be written as \cite{Liscidini, Sloan2024}
\begin{equation}
\begin{split}
    H_{L} &= \sum_{J, \nu} \int_{R(J)} dk \hbar \omega_J(k) a^{in(out)\dagger}_{J, \nu}(k, t) a^{in(out)}_{J, \nu}(k, t) \\
    &= \sum_{J, \nu, \nu'} \int_{R(J)} dk \hbar \omega_J(k) \left[\mathcal{C}^{-1}_J(k) \right]_{\nu, \nu'} \\
    & \qquad \qquad \qquad \qquad \qquad \times a^{loc \dagger}_{J,\nu}(k, t) a^{loc}_{J,\nu'}(k, t),
\end{split}
\end{equation}
such that, in the absence of the nonlinearity, 
the equations of motion would be given by
\begin{equation}
\label{eq:GeneralDynamics}
\begin{split}
    \frac{\partial}{\partial t} a_{J, \nu}(k,t) &= \frac{1}{i\hbar} \left[ a_{J, \nu}(k,t), H_{L} \right] \\
    &= -i\omega_J(k) a_{J, \nu}(k,t).
\end{split}
\end{equation}
where we have omitted the superscript to emphasize that the above equation holds in any of the three bases. Indeed, from equation (\ref{eq:localBasisOperatorTransformation}), the operator $a_{J,\nu}^{loc}(k)$ is constructed from a linear combination of the asymptotic-in(out) operators with the same $k$ and $J$, and so oscillates at a frequency $\omega_J(k)$.

For the nonlinear interaction, we take the waveguides and rings to be made from a centro-symmetric material with the second order contribution to the permittivity along the surface of each element sufficiently small that we can take $\chi_{(2)}^{ijk}(\textbf{r}) \cong 0$ for all $\textbf{r}$. We then define $\vec{J} = (J_1, J_2, J_3, J_4)$ and $\vec{\nu} = (\nu_1, \nu_2, \nu_3, \nu_4)$, and expand the nonlinear Hamiltonian to third order, resulting in \cite{Quesada:22, Sloan2024}
\begin{equation} \label{eq:GeneralHamiltonian}
\begin{split}
    H_{NL} &= \frac{-1}{4\epsilon_0} \int d\textbf{r} \Gamma_{(3)}^{ijkl}(\textbf{r}) D^i(\textbf{r}, t) D^j(\textbf{r}, t) D^k(\textbf{r}, t) D^l(\textbf{r}, t) \\
    &= -\hbar \sum_{\vec{J}, \vec{\nu}, \tau} \int d\textbf{k} \int d\xi^{\tau} \Lambda_{\vec{J}}^{\tau}(\xi) \mathcal{J}^{\tau}_{\vec{J}, \vec{k}, \vec{\nu}}(\xi) e^{-i \Delta k_{\vec{J}}^{\tau} \xi^{\tau}} \\
    & \qquad \qquad \qquad \qquad \qquad \times a_{J_1, \nu_1}^{\dagger}(k_1, t) a_{J_2, \nu_2}^{\dagger}(k_2, t) \\ 
    & \qquad \qquad \qquad \qquad \qquad \times a_{J_3, \nu_3}(k_3, t) a_{J_4, \nu_4}(k_4, t),
\end{split}
\end{equation}
which can be expressed in any of the three bases by expanding the displacement fields in terms of Eq. (\ref{eq:AsymptoticFieldDistExpansion}) or Eq. (\ref{eq:LocalBasisExpansion}). The sum over $\tau$ indicates that the integral $\int d\xi^{\tau}$ should be performed over each element along the nonlinear region, which we take to be each of the resonators and the coupling regions with the real waveguide channels, where the field enhancement can also be large. The nonlinear strength, $\Lambda_{\vec{J}}^{\tau}(\xi)$, is then related to the integral of the product of the field components and the nonlinear tensor $\Gamma_{(3)}^{ijkl}(\textbf{r})$, which itself can be related to the standard permittivity tensor $\chi_{(3)}^{ijkl}(\textbf{r})$, and is written as
\begin{equation}
    \Lambda_{\vec{J}}^{\tau}(\xi) = \frac{1}{8\pi^2} \hbar \omega_{\vec{J}} (v^{\tau}_{\vec{J}})^2 \gamma_{NL}^{\vec{J}, \tau}(\xi),
\end{equation}
where,
\begin{equation}
\begin{split}
    \gamma_{NL}^{\vec{J}, \tau}(\xi) &= \frac{3 \omega_{\vec{J}}}{4 \epsilon_0 (v^{\tau}_{\vec{J}})^2} \int d\textbf{r}_{\perp} \Gamma_{(3)}^{ijkl}(\textbf{r}_{\perp}) \textbf{d}^{\tau, i *}_{J_1}(\textbf{r}_{\perp}; \xi) \\
    & \qquad \qquad \qquad \times \textbf{d}^{\tau, j *}_{J_2}(\textbf{r}_{\perp}; \xi) \textbf{d}^{\tau, k}_{J_3}(\textbf{r}_{\perp}; \xi) \textbf{d}^{\tau, l}_{J_4}(\textbf{r}_{\perp}; \xi),
\end{split}
\end{equation}
and $\omega_{\vec{J}}$ and $v^{\tau}_{\vec{J}}$ are given by
\begin{equation}
\begin{split}
    \omega_{\vec{J}} &= \left( \omega_{J_1} \omega_{J_2} \omega_{J_3} \omega_{J_4} \right)^{1/4}, \\
    v^{\tau}_{\vec{J}} &= \left( v^{\tau}_{J_1} v^{\tau}_{J_2} v^{\tau}_{J_3} v^{\tau}_{J_4} \right)^{1/4},
\end{split}
\end{equation}
for $v_J^{\tau}$ the group velocity of the fields in the element $\tau$ and frequency $\omega_J$.

Note that in expanding the displacement fields in Eq. (\ref{eq:GeneralHamiltonian}), we have made the assumption that the factors of $\omega$ in the square root terms of Eq. (\ref{eq:GeneralAsymptoticExpansion}) can be approximated as $\omega_J(k) \cong \omega_J$ when evaluated within the nonlinear Hamiltonian. This allows us to write the nonlinear strength, $\Lambda_{\vec{J}}^{\tau}(\xi)$, as independent of the detunings of the fields from the resonance centers of the resonances contain in $\vec{J}$; however, we maintain the $k$-dependence of $\omega_J(k)$ within the linear Hamiltonian.

The exponential term in Eq. (\ref{eq:GeneralHamiltonian}) is related to the phase mismatch between each of the fields in the integral, with 
\begin{equation}
    \Delta k_{\vec{J}}^{\tau} = k^{\tau}_{J_1} + k^{\tau}_{J_2} - k^{\tau}_{J_3} - k^{\tau}_{J_4},
\end{equation}
and the final relevant term in Eq. (\ref{eq:GeneralHamiltonian}) is the function $\mathcal{J}^{\tau}_{\vec{J}, \vec{k}, \vec{\nu}}(\xi)$, which contains the slowly varying field amplitudes, 
\begin{equation}
\begin{split}
    \mathcal{J}^{\tau}_{\vec{J}, \vec{k}, \vec{\nu}}(\xi) &= h^{\nu_1, \tau *}_{J_1, k_1} (\xi) h^{\nu_2, \tau *}_{J_2, k_2} (\xi) h^{\nu_3, \tau}_{J_3, k_3} (\xi) h^{\nu_4, \tau}_{J_4, k_4} (\xi).
\end{split}
\end{equation}
As with the operators $a_{J,\nu}(k,t)$ and $a^{\dagger}_{J,\nu}(k,t)$ in equations (\ref{eq:GeneralDynamics}) and  (\ref{eq:GeneralHamiltonian}), we have omitted the basis in which we have expanded the displacement field as we can do this in any of the three bases we have introduced. However, in what is to follow it will be convenient to perform the field evolution within the local basis, since $\mathcal{J}^{\tau}_{\vec{J}, \vec{k}, \vec{\nu}}(\xi)$ vanishes for almost all instances of $\vec{\nu}$ with $\nu_i \neq \nu_j$ for some pair of $i, j \in \{1, 2, 3, 4\}$. Furthermore, in the local basis each $\mathcal{J}^{\tau}_{\vec{J}, \vec{k}, \vec{\nu}}(\xi)$ is non-zero only along a small length of the nonlinear region, in most cases between adjacent phantom channel coupling points. So in many practical cases, such as when there are large, high finesse resonators as we consider here, the $\mathcal{J}^{\tau}_{\vec{J}, \vec{k}, \vec{\nu}}(\xi)$ tend not to vary significantly as $\vec{k}$ ranges within the resonances ranges $\vec{J}$, even when only a modest number of phantom channels are coupled to each resonator (5-7 per ring for the parameters considered in Section \ref{sec:SectionV}). Consequently, we can make the approximation \cite{Sloan2024}
\begin{equation} \label{eq:phaseMatchingAssumption}
\begin{split}
    &\int d\xi \Lambda_{\vec{J}}^{\tau}(\xi) \mathcal{J}^{\tau, loc}_{\vec{J}, \vec{k}, \vec{\nu}}(\xi) e^{-i \Delta k_{\vec{J}}^{\tau} \xi} \\
    & \qquad \cong \int d\xi \Lambda_{\vec{J}}^{\tau}(\xi) \mathcal{J}^{\tau, loc}_{\vec{J}, \vec{k}^{\tau}_{\vec{J}}, \vec{\nu}}(\xi) e^{-i \Delta k_{\vec{J}}^{\tau} \xi} = \tilde{\Lambda}_{\vec{J}, \vec{\nu}}^{\tau, loc},
\end{split}
\end{equation}
where $\vec{k}^{\tau}_{\vec{J}} = (k^{\tau}_{J_1}, k^{\tau}_{J_2}, k^{\tau}_{J_3}, k^{\tau}_{J_4})$ correspond to the central wavevectors for the frequency ranges in $\vec{J}$.

With the above approximation in hand, we look to use the asymptotic-in(out) bases to analyze the field inputs(outputs), but utilize the local basis to perform the time evolution. As such, we write the nonlinear Hamiltonian as
\begin{equation} \label{eq:SimplifiedNonlinearHamiltonian}
\begin{split}
    H_{NL} &= -\hbar \sum_{\vec{J}, \vec{\nu}, \tau} \int d\textbf{k} \tilde{\Lambda}_{\vec{J}, \vec{\nu}}^{\tau, loc} a_{J_1, \nu_1}^{loc \dagger}(k_1, t) a_{J_2, \nu_2}^{loc \dagger}(k_2, t) \\
    & \qquad \qquad \qquad \qquad  \times a_{J_3, \nu_3}^{loc}(k_3, t) a_{J_4, \nu_4}^{loc}(k_4, t),
\end{split}
\end{equation}
and upon defining the $\tilde{a}^{loc}_{J,\nu}(k,t) = a^{loc}_{J,\nu}(k,t) e^{i \omega_J t}$, compute the equations of motion for the slowly varying local basis operators as
\begin{widetext}
    \begin{equation} \label{eq:generalEOM}
    \begin{split}
        \frac{\partial}{\partial t} \tilde{a}^{loc}_{J,\nu}(k,t) = -i \Delta \omega_J(k) \tilde{a}^{loc}_{J,\nu}(k,t) + i \sum_{\vec{J}, \vec{\nu}} \delta_{J, J_1} [\mathcal{C}_J(k)]_{\nu,\nu_1} \bar{\Lambda}_{\vec{J}, \vec{\nu}}^{loc} \int d\textbf{k} \tilde{a}_{J_2, \nu_2}^{loc \dagger}(k_2, t) \tilde{a}_{J_3, \nu_3}^{loc}(k_3, t) \tilde{a}_{J_4, \nu_4}^{loc}(k_4, t) e^{i \Delta \Omega_{\vec{J}} t}.
    \end{split}
    \end{equation}
\end{widetext}
with $\Delta \omega_J(k) = \omega_J(k) - \omega_J$ the detuning of the field from the central frequency of the resonance $J$, as well as $\Delta \Omega_{\vec{J}}$ defined by
\begin{equation}
    \Delta \Omega_{\vec{J}} = -\omega_{J_1} - \omega_{J_2} + \omega_{J_3} + \omega_{J_4}
\end{equation}
and $d\textbf{k} = dk_2 dk_3 dk_4$. Additionally, we have written 
\begin{equation}
    \bar{\Lambda}^{loc}_{\vec{J}, \vec{\nu}} = \left(1 + \delta_{J_1, J_2}\right) \sum_{\tau} \tilde{\Lambda}^{\tau, loc}_{\vec{J}, \vec{\nu}}
\end{equation}
with the $\delta_{J_1, J_2}$ term included to take into account the combinatorial factor when evaluating the commutator using equation (\ref{eq:SimplifiedNonlinearHamiltonian}). 

\begin{figure}
    \centering
    \includegraphics[width=0.95\linewidth]{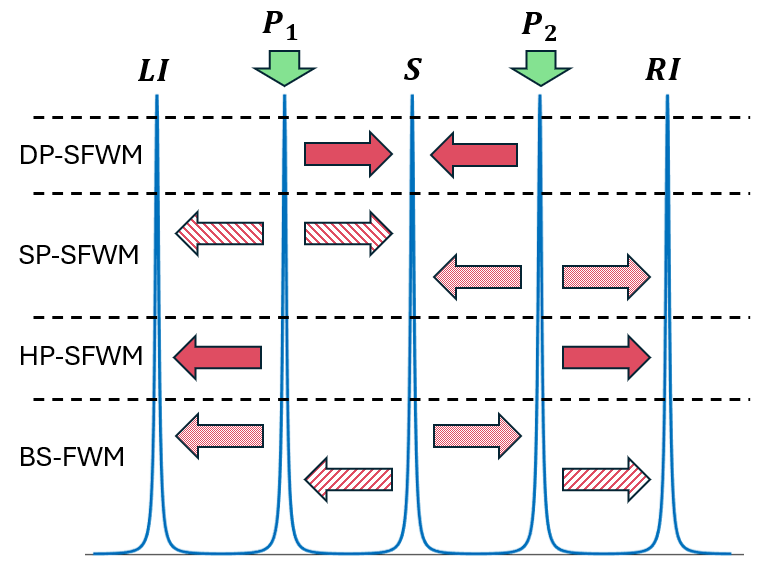}
    \caption{Diagram of an effective five resonance model in which the resonances $P_1$ and $P_2$ are driven with a classical field (denoted by downward green arrows), with photons generated in the $LI$, $S$, and $RI$ resonances. Horizontal arrows describe the energy preserving third order iterations which are appreciable in the weak generated field limit, with photon pair generation following the arrow direction.}
    \label{fig:FiveResonanceModel}
\end{figure}

For our current treatment, we will be interested in generating degenerate vacuum squeezed light in a single resonance $J = S$ corresponding to a 'signal' field. To do this, we will adopt a five resonance model as show in Fig. \ref{fig:FiveResonanceModel}. In particular, the center resonance will correspond to the signal field in which the desired squeezing is generated, with the two adjacent resonances, labeled $J= P_1, P_2$, being driven by classical pumps. The remaining two 'idler' resonances, labeled $J=LI, RI$, allow for unwanted parasitic SFWM interactions that can degrade the squeezing in the signal resonance. Our aim is to suppress these by adopting the dual ring structure discussed previously. We emphasize that each of these resonances will correspond to a system resonance, and as such can correspond to a resonance of the isolated primary ring, isolated auxiliary ring, or a hybrid resonance of the coupled system. We will identify the specific nature of each resonance in each of the examples to come.

In all cases, we consider strong classical fields driving the two pump resonances ($J=P_1, P_2$), but still sufficiently weak that pump depletion, as well as cross-phase modulation (XPM) from the generated fields in signal and idler resonances (initially in the vacuum), is negligible. This allows us to treat the evolution of the pump fields separately from the signal and idler fields. In particular, by replacing the pump operators with their expectation values as 
\begin{equation}
\begin{split}
    \tilde{a}^{loc}_{P_1,\nu}(k,t) &\rightarrow \tilde{\alpha}^{loc}_{P_1,\nu}(k,t) \cong \langle \tilde{a}^{loc}_{P_1,\nu}(k,t) \rangle, \\
    \tilde{a}^{loc}_{P_2,\nu}(k,t) &\rightarrow \tilde{\alpha}^{loc}_{P_2,\nu}(k,t) \cong \langle \tilde{a}^{loc}_{P_2,\nu}(k,t) \rangle,
\end{split}
\end{equation}
we can restrict the sum over $\vec{J}$ in equation (\ref{eq:generalEOM}) to only those contributions that can lead to an energy conserving interaction, with $J_i \in \{P_1, P_2 \}$ for each $i$, and write 
\begin{widetext}
    \begin{equation}
    \begin{split}
        \frac{\partial}{\partial t} \tilde{\alpha}_{P_1, \nu}^{loc}(k,t) = -i\Delta \omega_J(k) \tilde{\alpha}_{P_1, \nu}^{loc}(k,t) + i \sum_{\vec{\nu}} [\mathcal{C}_{P_1}(k)]_{\nu,\nu_1} & \left\{ \bar{\Lambda}_{P_1, \vec{\nu}}^{SPM} \int d\textbf{k} \tilde{\alpha}_{P_1, \nu_2}^{loc *}(k_2,t) \tilde{\alpha}_{P_1, \nu_3}^{loc}(k_3,t) \tilde{\alpha}_{P_1, \nu_4}^{loc}(k_4,t) \right. \\
        &\left. + \bar{\Lambda}_{P_1, \vec{\nu}}^{XPM} \int d\textbf{k} \tilde{\alpha}_{P_2, \nu_2}^{loc *}(k_2,t) \tilde{\alpha}_{P_1, \nu_3}^{loc}(k_3,t) \tilde{\alpha}_{P_2, \nu_4}^{loc}(k_4,t) \right\}, \\
        \frac{\partial}{\partial t} \tilde{\alpha}_{P_2, \nu}^{loc}(k,t) = -i\Delta \omega_J(k) \tilde{\alpha}_{P_2, \nu}^{loc}(k,t) + i \sum_{\vec{\nu}} [\mathcal{C}_{P_2}(k)]_{\nu,\nu_1} & \left\{ \bar{\Lambda}_{P_2, \vec{\nu}}^{SPM} \int d\textbf{k} \tilde{\alpha}_{P_2, \nu_2}^{loc *}(k_2,t) \tilde{\alpha}_{P_2, \nu_3}^{loc}(k_3,t) \tilde{\alpha}_{P_2, \nu_4}^{loc}(k_4,t) \right. \\
        &\left. + \bar{\Lambda}_{P_2, \vec{\nu}}^{XPM} \int d\textbf{k} \tilde{\alpha}_{P_1, \nu_2}^{loc *}(k_2,t) \tilde{\alpha}_{P_2, \nu_3}^{loc}(k_3,t) \tilde{\alpha}_{P_1, \nu_4}^{loc}(k_4,t) \right\}. \\
    \end{split}
    \end{equation}
\end{widetext}
Here the first term in the brackets describes SPM of each pump, and the second term corresponds to XPM induced by the other pump resonance. The nonlinear coefficients are given by
\begin{equation} \label{eq:ClassicalPumpEOM}
    \begin{split}
        \bar{\Lambda}_{P_1, \vec{\nu}}^{SPM} &= \bar{\Lambda}_{(P_1, P_1, P_1, P_1), \vec{\nu}}^{loc} \\
        \bar{\Lambda}_{P_1, \vec{\nu}}^{XPM} &= \bar{\Lambda}_{(P_1, P_2, P_1, P_2), \vec{\nu}}^{loc} + \bar{\Lambda}_{(P_1, P_2, P_2, P_1), \vec{\nu}}^{loc} \\
        &= 2\bar{\Lambda}_{(P_1, P_2, P_1, P_2), \vec{\nu}}^{loc} \\
        \bar{\Lambda}_{P_2, \vec{\nu}}^{SPM} &= \bar{\Lambda}_{(P_2, P_2, P_2, P_2), \vec{\nu}}^{loc} \\
        \bar{\Lambda}_{P_2, \vec{\nu}}^{XPM} &= \bar{\Lambda}_{(P_2, P_1, P_2, P_1), \vec{\nu}}^{loc} + \bar{\Lambda}_{(P_2, P_1, P_1, P_2), \vec{\nu}}^{loc} \\
        &= 2\bar{\Lambda}_{(P_2, P_1, P_2, P_1), \vec{\nu}}^{loc}, \\
    \end{split}
\end{equation}
where we have used the fact that $\bar{\Lambda}^{loc}_{\vec{J}, \vec{\nu}}$ is unchanged when swapping $J_1 \leftrightarrow J_2$ or $J_3 \leftrightarrow J_4$.

The coupled equations in (\ref{eq:ClassicalPumpEOM}) are now in terms of classical variables, and as such, can be solved from some initial time $t_0$ to some final time $t_f$, using a linear multi-step method \cite{linearMultiStep}. The initial conditions for each of the $\tilde{\alpha}^{loc}_{P_i, \nu}(k_i, t_0)$ will, in general, be chosen to correspond to pulses contained within the real waveguide input channels. For each $\nu \in S_r$ we define the initial pulse spectrum in $\nu$ stimulating the $P_i$ resonance by $f_{P_i}^{\nu}(k)$, leading to the field operators in the asymptotic-in basis at $t=t_0$ being given by
\begin{equation} \label{eq:ClassicalInitialConditionAsyIn}
    \tilde{\alpha}^{in}_{P_i, \nu}(k, t_0) = \begin{cases}
        f_{P_i}^{\nu}(k) & \textit{for } \nu \in S_r ,\\
        0 & \textit{otherwise}.
    \end{cases}
\end{equation}
From equation (\ref{eq:localBasisOperatorTransformation}) it then follows that
\begin{equation} \label{eq:LocalBasisPulseDef}
    \tilde{\alpha}_{P_i, \nu}^{loc} (k, t_0) = \sum_{\nu' \in S_r} \left[ \left( \mathcal{L}^{in}_{P_i, k} \right)^{-1} \right]_{\nu', \nu} f_{P_i}^{\nu'}(k).
\end{equation}

With $\{\tilde{\alpha}^{loc}_{P_1,\nu}(k,t)\}_{\nu,k}$ and $\{\tilde{\alpha}^{loc}_{P_2,\nu}(k,t)\}_{\nu,k}$ in hand for all times between $t_0$ and $t_f$, the calculation of the signal and idler fields then follows. In particular, when the generation rate of the signal and idler photons are sufficiently weak that stimulated effects from the generated fields are negligible, we can neglect terms in equation (\ref{eq:generalEOM}) that are quadratic in the non-pump operators. The equations of motion for the signal and idler fields, now linearized, are
\begin{widetext}
    \begin{equation} \label{eq:NonClassicalEOM}
    \begin{split}
        \frac{\partial}{\partial t} \tilde{a}^{loc}_{LI,\nu}(k,t) &= i \sum_{\nu'} \int dk' \left\{ \left(A^{XPM_1}_{LI, \nu, \nu'}(k, t) + A^{XPM_2}_{LI, \nu, \nu'}(k, t) \right) \tilde{a}^{loc}_{LI,\nu'}(k',t) + A^{SP_1}_{LI, \nu, \nu'}(k, t) \tilde{a}^{loc \dagger}_{S,\nu'}(k',t) \right. \\
        & \qquad \qquad \qquad \quad \left. + A^{HP}_{LI, \nu, \nu'}(k, t) \tilde{a}^{loc \dagger}_{RI,\nu'}(k',t) + A^{BS_1}_{LI, \nu, \nu'}(k, t) \tilde{a}^{loc}_{S,\nu'}(k',t)\right\} - i\Delta \omega_J(k) \tilde{a}^{loc}_{LI,\nu}(k,t) \\
        \frac{\partial}{\partial t} \tilde{a}^{loc}_{S,\nu}(k,t) &= i \sum_{\nu'} \int dk' \left\{ \left(A^{XPM_1}_{S, \nu, \nu'}(k, t) + A^{XPM_2}_{S, \nu, \nu'}(k, t) \right) \tilde{a}^{loc}_{S,\nu'}(k',t) + A^{SP_1}_{S, \nu, \nu'}(k, t) \tilde{a}^{loc \dagger}_{LI,\nu'}(k',t) \right. \\
        & \qquad \qquad \qquad \quad \left. + A^{SP_2}_{S, \nu, \nu'}(k, t) \tilde{a}^{loc \dagger}_{RI,\nu'}(k',t) + A^{BS_1*}_{S, \nu, \nu'}(k, t) \tilde{a}^{loc}_{LI,\nu'}(k',t) + A^{BS_2*}_{S, \nu, \nu'}(k, t) \tilde{a}^{loc}_{RI,\nu'}(k',t) \right\} \\
        & \qquad - i\Delta \omega_J(k) \tilde{a}^{loc}_{LI,\nu}(k,t) \\
        \frac{\partial}{\partial t} \tilde{a}^{loc}_{RI,\nu}(k,t) &= i \sum_{\nu'} \int dk' \left\{ \left(A^{XPM_1}_{RI, \nu, \nu'}(k, t) + A^{XPM_2}_{RI, \nu, \nu'}(k, t) \right) \tilde{a}^{loc}_{RI,\nu'}(k',t) + A^{SP_2}_{RI, \nu, \nu'}(k, t) \tilde{a}^{loc \dagger}_{S,\nu'}(k',t) \right. \\
        & \qquad \qquad \qquad \quad \left. + A^{HP}_{RI, \nu, \nu'}(k, t) \tilde{a}^{loc \dagger}_{LI,\nu'}(k',t) + A^{BS_2}_{RI, \nu, \nu'}(k, t) \tilde{a}^{loc}_{S,\nu'}(k',t)\right\} - i\Delta \omega_J(k) \tilde{a}^{loc}_{RI,\nu}(k,t)
    \end{split}
    \end{equation}
\end{widetext}
where the factors $A^{\chi}_{J, \nu, \nu'}(k,t)$ contain the contributions from the pump fields, the commutators, and the nonlinear coefficients. They are each given in Appendix \ref{sec:EOMappendix}.

We then discretize the signal and idler operators such that
\begin{equation}
    \int_{R(J)} dk \tilde{a}^{loc}_{J, \nu}(k, t) \rightarrow \sum_{i=1}^{N_k} \tilde{a}^{loc}_{J, \nu, k_i}(t),
\end{equation}
where we have partitioned the range $R(J)$ into $N_k$ equally spaced bins of width $\delta k_J$, and associate the operator $\tilde{a}^{loc}_{J, \nu, k_i}(t)$ with the $i^{th}$ wavenumber in the range $R(J)$. Following a similar procedure introduced earlier for a single-pump non-degenerate squeezing interaction \cite{Sloan2024}, we can write the equations of motion compactly by first gathering each of the $\tilde{a}^{loc}_{J, \nu, k_i}(t)$ together for each channel as
\begin{equation}
    \textbf{a}_{J,k_i}^{loc}(t) = \begin{pmatrix}
        \tilde{a}^{loc}_{J, \nu_1, k_i}(t) \\ \vdots \\ \tilde{a}^{loc}_{J, \nu_N, k_i}(t)
    \end{pmatrix},
\end{equation}
then gathering each of the resonances as
\begin{equation}
    \textbf{b}^{loc}_{k_i}(t) = \begin{pmatrix}
        \textbf{a}_{LI,k_i}^{loc}(t) \\
        \textbf{a}_{S,k_i}^{loc}(t) \\
        \textbf{a}_{RI,k_i}^{loc}(t) \\
        \textbf{a}_{LI,k_i}^{loc \dagger}(t) \\
        \textbf{a}_{S,k_i}^{loc \dagger}(t) \\
        \textbf{a}_{RI,k_i}^{loc \dagger}(t)
    \end{pmatrix},
\end{equation}
allowing us to write
\begin{equation}
    \frac{\partial}{\partial t} \textbf{b}^{loc}_{k_i}(t) = i\mathcal{A}^L_i \textbf{b}^{loc}_{k_i}(t) + i\mathcal{A}^{NL}_i(t) \sum_j \textbf{b}^{loc}_{k_j}(t),
\end{equation}
where the matrices $\mathcal{A}_i^L$ are diagonal and contain the time independent linear contributions to the equations of motion, while $\mathcal{A}_i^{NL}(t)$ is constructed from the time dependent nonlinear terms.

Utilizing a split-step approach to separate the linear and nonlinear evolution, and considering a time step $\Delta t$ small enough that $\mathcal{A}_i^{NL}(t)$ is approximately constant for $t \in [t, t+\Delta t]$, we can write the short-time evolution of the field operators in the local basis as \cite{Sloan2024}
\begin{equation}
    \textbf{b}^{loc}_{k_i}(t+\Delta t) = \sum_j K_{i,j}^{loc}(t, t+\Delta t) \textbf{b}^{loc}_{k_j}(t),
\end{equation}
where 
\begin{equation} \label{eq:ApproximateShortTimeStep}
    K^{loc}_{i, j}(t, t+\Delta t) = e^{i \mathcal{A}_i t/2} \left( \mathbb{I}\delta_{i,j} + K^{NL}_{i}(t, t+\Delta t) \right) e^{i \mathcal{A}_j t/2},
\end{equation}
and the nonlinear evolution operator is given by
\begin{equation}
    K^{NL}_i(t, t+\Delta t) = i\mathcal{A}_i^{NL}(t) \sum_n \frac{\Delta t^{n+1}}{(n+1)!} \left( i\sum_j \mathcal{A}^{NL}_j(t) \right)^n.
\end{equation}
In many practical cases -- in particular when $\tilde{\mathcal{A}}^{NL}(t) = \sum_j \mathcal{A}^{NL}_j(t)$ is invertible -- the term involving the sum over $n$ can be expressed in terms of a matrix exponential of $\tilde{\mathcal{A}}^{NL}(t)$ \cite{Sloan2024}.

Gathering all of the operators together, we can write
\begin{equation} \label{eq:FullShortTimeStep}
    \begin{pmatrix}
        \textbf{b}^{loc}_{k_1}(t+\Delta t) \\ \vdots \\ \textbf{b}^{loc}_{k_{N_k}}(t+\Delta t)
    \end{pmatrix} = \mathcal{K}^{loc}(t, t+\Delta t) \begin{pmatrix}
        \textbf{b}^{loc}_{k_1}(t) \\ \vdots \\ \textbf{b}^{loc}_{k_{N_k}}(t)
    \end{pmatrix} 
\end{equation}
with the blocks of $\mathcal{K}^{loc}(t, t+\Delta t)$ being constructed from the $K^{loc}_{i,j}(t, t+\Delta t)$. We then build up the full time evolution matrix from some initial time $t_0$ to a final time $t_f$ as
\begin{equation} \label{eq:FullTimeEvolutionOperator}
    \mathcal{K}^{loc}(t_0, t_f) = \mathcal{K}^{loc}(t_f-\Delta t, t_f) \dots \mathcal{K}^{loc}(t_0, t_0+\Delta t)
\end{equation}
where we have discretized the interval $[t_0, t_f]$ into time steps of length $\Delta t$ short enough to approximate each step using equation (\ref{eq:ApproximateShortTimeStep}). Hence, the operators at $t_f$ can be written, in the local basis, as
\begin{equation}
    \begin{pmatrix}
        \textbf{b}^{loc}_{k_1}(t_f) \\ \vdots \\ \textbf{b}^{loc}_{k_{N_k}}(t_f)
    \end{pmatrix} = \mathcal{K}^{loc}(t_0, t_f) \begin{pmatrix}
        \textbf{b}^{loc}_{k_1}(t_0) \\ \vdots \\ \textbf{b}^{loc}_{k_{N_k}}(t_0)
    \end{pmatrix}.
\end{equation}
We can then utilize equation (\ref{eq:localBasisOperatorTransformation}) to convert to the asymptotic-out basis, leading to
\begin{equation} \label{eq:FullEvoAsyOutBasis}
    \begin{pmatrix}
        \textbf{b}^{out}_{k_1}(t_f) \\ \vdots \\ \textbf{b}^{out}_{k_{N_k}}(t_f)
    \end{pmatrix} = \mathcal{K}^{out}(t_0, t_f) \begin{pmatrix}
        \textbf{b}^{out}_{k_1}(t_0) \\ \vdots \\ \textbf{b}^{out}_{k_{N_k}}(t_0)
    \end{pmatrix}.
\end{equation}
This gives us a description of the fields at the time $t_f$ in a mode basis corresponding to fields exiting each of the real and phantom channels.

Indeed, the simpler single ring application of the more general construction illustrated above has been shown to match well with previous SFWM treatments, including a pertubative asymptotic field approach \cite{Liscidini, Banic-PhysRevA.106.043707} as well as a commonly used non-perturbative coupled mode formulation \cite{Quesada:22, Seifoory, Vernon2015}, in the case of a single pump drive generating non-degenerate squeezed light \cite{Sloan2024}.
\section{Obtaining the Output Field} \label{sec:SectionIV}
With the full time evolution operator $\mathcal{K}^{out}(t_0, t_f)$ in hand, we can now construct the state describing the field exiting the waveguide output channel. To do this, we first note that from equation (\ref{eq:FullEvoAsyOutBasis}), we can write 
\begin{equation} \label{eq:ExpandedEvolution}
\begin{split}
    a^{out}_{J, \nu, k}(t_f) = \sum_{J'}' \sum_{\nu', k'} & \left[ V_{\nu, \nu'}^{J, J'}(k, k') a^{out}_{J', \nu', k'}(t_0) \right. \\
    & \left.  + W_{\nu, \nu'}^{J, J'}(k, k') a^{out \dagger}_{J', \nu', k'}(t_0) \right] 
\end{split}
\end{equation}
where the $V_{\nu, \nu'}^{J, J'}(k, k')$ and $W_{\nu, \nu'}^{J, J'}(k, k')$ are the appropriate elements of the matrix $\mathcal{K}^{out}(t_0, t_f)$, and the prime on the sum over $J$ indicates that the sum is over the non-pump resonances, $J \in \{LI, S, RI\}$. If the signal and idler resonances are initially vacuum at $t_0$, it follows that the first order moments of the operators at $t_f$ are
\begin{equation} \label{eq:FirstOrderMoments}
    \langle a^{out}_{J, \nu, k}(t_f) \rangle =  0 = \langle a^{out \dagger}_{J, \nu, k}(t_f) \rangle,
\end{equation}
with the second order moments written as
\begin{equation}
\begin{split} \label{eq:SecondOrderMoments}
    N_{\nu,\nu'}^{J,J'}(k,k') & \equiv \langle a^{out \dagger}_{J, \nu, k}(t_f) a^{out}_{J', \nu', k'}(t_f) \rangle \\
    & = \sum_{J''} \sum_{\nu'', k''} W^{J', J''}_{\nu', \nu''}(k', k'') W^{J, J'' *}_{\nu, \nu''}(k, k''), \\
    M_{\nu,\nu'}^{J,J'}(k,k') & \equiv \langle a^{out}_{J, \nu, k}(t_f) a^{out}_{J', \nu', k'}(t_f) \rangle \\
    & = \sum_{J''} \sum_{\nu'', k''} W^{J', J''}_{\nu', \nu''}(k', k'') V^{J, J''}_{\nu, \nu''}(k, k''). \\
\end{split}
\end{equation}
In Appendix \ref{sec:appendix1} we show that the state of the signal and idler fields at the time $t_f$ is a Gaussian state, and as such can be completely described by the first and second order operator moments from equations (\ref{eq:FirstOrderMoments}) and (\ref{eq:SecondOrderMoments}). Indeed, for the coordinates $\alpha_{J,\nu}^k$ corresponding to the operators $a^{out}_{J, \nu, k}(t_f)$, we can write the Wigner characteristic function as
\begin{widetext}
\begin{equation} \label{eq:CharacteristicFunction_NMdef}
\begin{split}
    \chi_W(\{\alpha_{J,\nu}^k\}) &= \exp \left\{ -\frac{1}{2} \sum_{J, J'}' \sum_{\nu,\nu',k,k'} \alpha_{J,\nu}^k \alpha_{J',\nu'}^{k' *} \left[   \delta_{J,J'} \delta_{\nu,\nu'} \delta_{k,k'} + N_{\nu,\nu'}^{J,J'}(k,k') + N_{\nu',\nu}^{J',J *}(k',k) \right] \right. \\
    & \qquad \qquad  \left. + \frac{1}{2} \sum_{J, J'}' \sum_{\nu,\nu',k,k'} \alpha_{J,\nu}^k \alpha_{J',\nu'}^{k'} M_{\nu,\nu'}^{J,J' *}(k,k') + \frac{1}{2} \sum_{J, J'}' \sum_{\nu,\nu',k,k'} \alpha_{J,\nu}^{k*} \alpha_{J',\nu'}^{k' *} M_{\nu,\nu'}^{J,J'}(k,k') \right\} \\
    &= \exp \left\{ -\frac{1}{2} \begin{pmatrix}
        \pmb{\alpha}^T & \pmb{\alpha}^{\dagger}
    \end{pmatrix} \begin{pmatrix}
        \frac{1}{2} \mathbb{I} + \textbf{N}^* & -\textbf{M} \\ -\textbf{M}^* & \frac{1}{2} \mathbb{I} + \textbf{N}
    \end{pmatrix} \begin{pmatrix}
        \pmb{\alpha} \\ \pmb{\alpha}^*
    \end{pmatrix}  
    \right\} \equiv \exp \left\{ -\frac{1}{2} \begin{pmatrix}
        \pmb{\alpha}^T & \pmb{\alpha}^{\dagger}
    \end{pmatrix} \pmb{\Phi} \begin{pmatrix}
        \pmb{\alpha} \\ \pmb{\alpha}^*
    \end{pmatrix}  
    \right\}
\end{split}
\end{equation}
\end{widetext}
where $\pmb{\alpha}$ is some ordering of the variable $\{\alpha_{J,n}^k\}$ and the matrices $\textbf{N}$ and $\textbf{M}$ are constructed from $N_{\nu,\nu'}^{J,J'}(k,k')$ and $M_{\nu,\nu'}^{J,J'}(k,k')$ respectively, with the rows and columns ordered the same way as $\pmb{\alpha}$.

Furthermore, we can construct the quadrature operators $(\hat{X}_{J,\nu}^k(t), \hat{Y}_{J,\nu}^k(t))$ as
\begin{equation}
\begin{split}
    \hat{X}_{J,\nu}^k(t) &= \frac{1}{\sqrt{2}} \left( a^{out}_{J,\nu,k}(t) + a^{out \dagger}_{J,\nu,k}(t) \right), \\
    \hat{Y}_{J,\nu}^k(t) &= \frac{-i}{\sqrt{2}} \left( a^{out}_{J,\nu,k}(t) - a^{out \dagger}_{J,\nu,k}(t) \right),
\end{split}
\end{equation}
with the corresponding coordinate transformation, $\mathcal{L}$, such that
\begin{equation}
    \begin{pmatrix}
        \textbf{X} \\ \textbf{Y}
    \end{pmatrix} = \mathcal{L} \begin{pmatrix}
        \pmb{\alpha} \\
        \pmb{\alpha}^*
    \end{pmatrix}  = \frac{1}{\sqrt{2}} \begin{pmatrix}
        \mathbb{I} & \mathbb{I} \\ -i\mathbb{I} & i\mathbb{I}
    \end{pmatrix} \begin{pmatrix}
        \pmb{\alpha} \\
        \pmb{\alpha}^*
    \end{pmatrix},
\end{equation}
to write the characteristic function in terms of the quadrature variables as
\begin{equation} \label{eq:QuadratureCharacteristicFunction}
\begin{split}
    &\chi_W(\{\alpha_{J,n}^k \}) = \chi_W(\textbf{X}, \textbf{Y}) \\
    & \qquad = \exp \left\{ -\frac{1}{2} \begin{pmatrix}
        \textbf{X}^T & \textbf{Y}^T
    \end{pmatrix} \pmb{\Sigma} \begin{pmatrix}
        \textbf{X} \\ \textbf{Y}
    \end{pmatrix} \right\} \\
    & \qquad = \exp \left\{ -\frac{1}{2} \begin{pmatrix}
        \textbf{X}^T & \textbf{Y}^T
    \end{pmatrix} \begin{pmatrix}
        \Sigma_{XX} & \Sigma_{XY} \\ \Sigma_{YX} & \Sigma_{YY}
    \end{pmatrix} \begin{pmatrix}
        \textbf{X} \\ \textbf{Y}
    \end{pmatrix} \right\}.
\end{split}
\end{equation}
Within the quadrature basis, we have $\pmb{\Sigma} = \mathcal{L} \pmb{\Phi} \mathcal{L}^{\dagger}$, with components given by
\begin{equation} \label{eq:SigmaBlocks}
    \begin{split}
        \Sigma_{XX} &= \frac{1}{2} \mathbb{I} + \Re \{\textbf{N}\} + \Re \{\textbf{M}\} \\
        \Sigma_{XY} &= \Im \{\textbf{N}\} + \Im \{\textbf{M}\} \\
        \Sigma_{YX} &= -\Im \{\textbf{N}\} + \Im \{\textbf{M}\} \\
        \Sigma_{YY} &= \frac{1}{2} \mathbb{I} + \Re \{\textbf{N}\} - \Re \{\textbf{M}\}
    \end{split}
\end{equation} 

We note from the definition of $N_{\nu,\nu'}^{J,J'}(k,k')$ and $M_{\nu,\nu'}^{J,J'}(k,k')$ in equation (\ref{eq:SecondOrderMoments}) that $\textbf{N}$ is Hermitian and $\textbf{M}$ is symmetric, from which it is clear that $\pmb{\Sigma}$ is real and symmetric. Additionally, we show in Appendix \ref{sec:appendix2} that $\pmb{\Sigma}$ can be decomposed as
\begin{equation} \label{eq:FullSqueezedState}
    \pmb{\Sigma} = \mathcal{O}_{\Sigma} \begin{pmatrix}
        \oplus_i \frac{1}{2} e^{2r_i} & 0 \\ 0 & \oplus_i \frac{1}{2} e^{-2r_i}
    \end{pmatrix} \mathcal{O}^T_{\Sigma}
\end{equation}
where $\mathcal{O}_{\Sigma}$ is real, orthogonal, and symplectic, and the parameters $r_i$ are real and nonnegative. Thus the matrix $\mathcal{O}_{\Sigma}$ defines a basis of temporal modes that preserves the commutation relations between the quadrature operators. The maximum reduction in quadrature variance is then associated with the temporal mode with eigenvalue $\frac{1}{2} e^{-2 r_{max}}$ for $r_{max} = \max \{r_i\}$. However, such a temporal mode will, in general, contain contributions from the $LI$ and $RI$ resonances (which we envision filtering out), as well as from the modes describing the fields exiting the phantom channels (which are not accessible). Rather, it is more useful to consider the reduced states corresponding to fields exiting from a real waveguide channel.

For this reason, we consider partitioning the Hilbert space, $\mathcal{H}$, as $\mathcal{H} = \mathcal{H}_A \otimes \mathcal{H}_B$. In Appendix \ref{sec:appendix1}, we derive the Wigner characteristic function for the reduced density operator $\rho_{R} = \Tr_B [\rho]$, where $\rho$ is the full density operator, describing the signal (here $S$) and idler (here $LI$ and $RI$) fields at time $t_f$, and $\mathcal{H}_A$ and $\mathcal{H}_B$ represent an arbitrary partitioning of $\mathcal{H}$ into orthogonal subspaces. Of interest here is the choice of $\mathcal{H}_A$ to contain only those modes $a^{out}_{J, \nu, k}(t)$ corresponding to the signal resonance exiting a real output channel. We then find the reduced characteristic function to be given by 
\begin{equation} \label{eq:ReducedCharacteristicFunction}
\begin{split}
    &\chi_W^R(\{\alpha_{S,\nu_{out}}^k\}) = \exp \left\{ -\frac{1}{2} \begin{pmatrix}
        \tilde{\pmb{\alpha}}^T & \tilde{\pmb{\alpha}}^{\dagger}
    \end{pmatrix} \tilde{\pmb{\Phi}} \begin{pmatrix}
        \tilde{\pmb{\alpha}} \\ \tilde{\pmb{\alpha}}^*
    \end{pmatrix}  
    \right\} \\ 
    & \qquad = \exp \left\{ -\frac{1}{2} \begin{pmatrix}
        \tilde{\pmb{\alpha}}^T & \tilde{\pmb{\alpha}}^{\dagger}
    \end{pmatrix} \begin{pmatrix}
        \frac{1}{2} \mathbb{I} + \tilde{\textbf{N}}^* & - \tilde{\textbf{M}} \\ -\tilde{\textbf{M}}^* & \frac{1}{2} \mathbb{I} + \tilde{\textbf{N}}
    \end{pmatrix} \begin{pmatrix}
        \tilde{\pmb{\alpha}} \\ \tilde{\pmb{\alpha}}^*
    \end{pmatrix}  
    \right\}
\end{split}
\end{equation}
where $\tilde{\pmb{\alpha}}$ is the restriction of $\pmb{\alpha}$ to those variables $\alpha_{J,\nu}^k$ with $\nu=\nu_{out}$ and $J=S$, and likewise $\tilde{\textbf{N}}$ and $\tilde{\textbf{M}}$ being the restrictions of $\textbf{N}$ and $\textbf{M}$ to only those terms $N_{\nu,\nu'}^{J,J'}(k,k')$ and $M_{\nu,\nu'}^{J,J'}(k,k')$ with $\nu=\nu_{out}=\nu'$ and $J=S=J'$ respectively.

In the same manner denoting $(\tilde{\textbf{X}}, \tilde{\textbf{Y}})$ to be the restriction of $(\textbf{X}, \textbf{Y})$ to the output signal operators, we can also express the reduced characteristic function in the quadrature basis as
\begin{equation}
    \chi^R_W(\tilde{\textbf{X}}, \tilde{\textbf{Y}}) = \exp \left\{ -\frac{1}{2} \begin{pmatrix}
        \tilde{\textbf{X}}^T & \tilde{\textbf{Y}}^T
    \end{pmatrix} \tilde{\pmb{\Sigma}} \begin{pmatrix}
        \tilde{\textbf{X}} \\ \tilde{\textbf{Y}}
    \end{pmatrix} \right\},
\end{equation}
with $\tilde{\pmb{\Sigma}}$ defined identically to equation (\ref{eq:SigmaBlocks}) upon making the replacement $\textbf{N} \rightarrow \tilde{\textbf{N}}$ and $\textbf{M} \rightarrow \tilde{\textbf{M}}$.

With a structure that follows that of $\pmb{\Sigma}$, the reduced matrix $\tilde{\pmb{\Sigma}}$ is real, symmetric, and positive definite, meaning we can write
\begin{equation} \label{eq:DiagonalizationOfReducedState}
    \tilde{\pmb{\Sigma}} = \tilde{\mathcal{O}}_{\Sigma} \tilde{D}_{\Sigma} \tilde{\mathcal{O}}_{\Sigma}^T,
\end{equation}
with $\tilde{D}_{\Sigma} > 0$ diagonal and $\tilde{\mathcal{O}}_{\Sigma}$ orthogonal. However, it is no longer true that $\tilde{\mathcal{O}}_{\Sigma}$ is in general symplectic, and it therefore does not define a basis of temporal modes that preserves the commutation relations of the quadrature variables. Despite this, writing $\tilde{D}_{\Sigma} = diag \{\tilde{d}_1, \tilde{d}_2, \dots, \tilde{d}_{2N_k}\}$ with $\tilde{d}_1 \geq \tilde{d}_2 \geq \dots \geq \tilde{d}_{2N_k}$, one could construct a basis of temporal modes in which a quadrature variance of $\tilde{d}_{2N_k}$ is achievable; however, this would mark the maximum squeezing available in the output state \cite{Kopylov2025}. Furthermore, denoting the eigenvalues of $\pmb{\Sigma}$ as $d_1 \geq d_2 \geq \dots \geq d_{2N}$ for $N$ the total number of modes in $\mathcal{H}$, it follows that $\tilde{d}_{2N_k} \geq d_{2N}$ (Appendix \ref{sec:appendix2}). Hence the maximum squeezing in the reduced state is bounded by the maximum squeezing in the full state.

Indeed, from equation (\ref{eq:DiagonalizationOfReducedState}) it can be seen that the reduced output signal state is no longer a pure squeezed state, but rather a multimode squeezed thermal state \cite{Paraoanu}. To gain insight into the form of the squeezing and thermal contributions, we can first utilize a Williamson decomposition \cite{Williamson, Houde2024} to write
\begin{equation}
    \tilde{\pmb{\Sigma}} = \tilde{S} (\tilde{D}_S \oplus \tilde{D}_S) \tilde{S}^T,
\end{equation}
where $\tilde{S}$ is a real and symplectic matrix and $\tilde{D}_S$ is a diagonal matrix with $\tilde{D}_S \geq \frac{1}{2} \mathbb{I}$ (see Appendix \ref{sec:appendix2}).
Utilizing a Bloch-Messiah decomposition \cite{Serafini, Houde2024}, we can further decompose $\tilde{S}$ as
\begin{equation}
    \tilde{S} = \tilde{\mathcal{O}}_S (\tilde{R}_S \oplus \tilde{R}_S^{-1}) \tilde{\mathcal{O}}_S',
\end{equation}
with both $\tilde{\mathcal{O}}_S$ and $\tilde{\mathcal{O}}_S'$ being real, symplectic, and orthogonal -- and so defining a quadrature space transformation -- and the matrix $\tilde{R}_S$ being diagonal with $\tilde{R}_S > 0$. Thus we can write 
\begin{equation} \label{eq:ReducedCharacteristicDecomp}
    \tilde{\pmb{\Sigma}} = \tilde{\mathcal{O}}_S (\tilde{R}_S \oplus \tilde{R}_S^{-1}) \tilde{\mathcal{O}}_S' (\tilde{D}_S \oplus \tilde{D}_S) \tilde{\mathcal{O}}_S'^T (\tilde{R}_S \oplus \tilde{R}_S^{-1}) \tilde{\mathcal{O}}_S^T.
\end{equation}

This provides us with a convenient way of analyzing the field output as, in the case when $\tilde{D}_S = \frac{1}{2} \mathbb{I}$, the matrix $\tilde{\pmb{\Sigma}}$ reduces to
\begin{equation} \label{eq:PureSqueezedState}
    \tilde{\pmb{\Sigma}} \rightarrow \tilde{\pmb{\Sigma}}_{sq} = \frac{1}{2} \tilde{\mathcal{O}}_S (\tilde{R}_S^2 \oplus \tilde{R}_S^{-2}) \tilde{\mathcal{O}}_S^T,
\end{equation}
which describes a multimode squeezed state, squeezed about the vacuum variance, with squeezing parameters related to $\tilde{R}_S$. The matrices $\tilde{\mathcal{O}}_S^T$ then define the temporal mode basis in which the squeezing is applied. Indeed, by writing $\tilde{R}_S = \oplus_i e^{\tilde{r}_i}$, we recover the same form of the solution as in equation (\ref{eq:FullSqueezedState}).

On the other hand, when $\tilde{R}_S = \mathbb{I}$, we have
\begin{equation} \label{eq:PureThermalState}
    \tilde{\pmb{\Sigma}} \rightarrow \tilde{\pmb{\Sigma}}_{therm} = (\tilde{\mathcal{O}}_S \tilde{\mathcal{O}}_S') (\tilde{D}_S \oplus \tilde{D}_S) (\tilde{\mathcal{O}}_S \tilde{\mathcal{O}}_S')^T,
\end{equation}
which corresponds to a thermal state with a temporal mode basis given by $(\tilde{\mathcal{O}}_S \tilde{\mathcal{O}}_S')^T$ and thermal parameters related to $\tilde{D}_S$. 

Consequently, the matrix $\tilde{\mathcal{O}}'^T_S$ can be interpreted as the quadrature transformation from the squeezing basis to the thermal basis. In the case when $\tilde{\mathcal{O}}'^T_S = \mathbb{I}$, one can see from equation (\ref{eq:ReducedCharacteristicDecomp}) that the reduced output signal state can be written as a product of independent single mode squeezed thermal states, with $\tilde{\mathcal{O}}^T_S$ providing the analogue of a transformation into a basis of Schmidt modes \cite{Christ_2011, Quesada:22}. In such a case, one could assign a `thermal photon number', $n_i^{th}$, and a `squeezing photon number', $n_i^{sq}$, to each temporal mode $i$ related to the separate squeezing and thermal contributions \cite{Seifoory:17}. Writing $\tilde{D}_S = \frac{1}{2}\otimes_i \hat{d}_i$, the minimum variance of the independent squeezed thermal mode $i$, $V_{min}^i$, would then be given by \cite{Seifoory:17}
\begin{equation}
    V_{min}^i = \frac{1}{2} \hat{d}_i e^{-2 \tilde{r}_i} = (2 n_i^{th} + 1) e^{-2 \tilde{r}_i},
\end{equation}
which would result in squeezing below the vacuum whenever $\hat{d}_ie^{-2 \tilde{r}_i} < 1$.

In the case of a general output signal state $\tilde{\mathcal{O}}'^T_S \neq \mathbb{I}$, the field cannot be written as a product of independent squeezed thermal states, but the number of photons associated with the thermal state $\tilde{\pmb{\Sigma}}_{therm}$, $n^{th}_{tot}$, and the number of photons of the squeezed state $\tilde{\pmb{\Sigma}}_{sq}$, $n^{sq}_{tot}$, can still be computed as
\begin{equation} \label{eq:ThermAndSqueezedPhotonDef}
\begin{split}
    n^{th}_{tot} &= \Tr [\tilde{D}_S - \frac{1}{2} \mathbb{I}] \\
    n^{sq}_{tot} &= \Tr [\sinh^2 \left(\log(\tilde{R}_S) \right)]
\end{split}
\end{equation}
From this, we define a unit-less value $F_s \in [0, 1]$ as
\begin{equation} \label{eq:FsDefinition}
    F_s = \frac{n_{tot}^{sq}}{n_{tot}^{sq} + n_{tot}^{th}},
\end{equation}
which corresponds to a multimode squeezed state ($\tilde{D}_S = \frac{1}{2} \mathbb{I}$) when $F_s = 1$ and to a multimode thermal state ($\tilde{R}_S = \mathbb{I}$) when $F_s = 0$.

We emphasize that the definition of $n^{th}_{tot}$ and $n^{sq}_{tot}$ should not be thought of as a partitioning of the photons in the signal output. Indeed, the total number of signal photons exiting the waveguide output is given by $n_{tot} = \Tr [\tilde{\textbf{N}}]$, which is not equal to $n^{th}_{tot} + n^{sq}_{tot}$ in general. Rather, these quantities provide an estimate of the relative strength of the process that provides squeezing of the signal resonance, and of the processes that add thermal noise to the output state. In particular, for the case of a lossless ring in which one neglects the contributions from all SFWM processes except DP-SFWM, the resulting state after tracing out $\mathcal{H}_B$ necessarily would have $n^{th}_{tot} = 0$ (see Appendix \ref{sec:appendix2}). Indeed, this would correspond to the subspaces $\mathcal{H}_A$ and $\mathcal{H}_B$ being disjoint, with the trace over $\mathcal{H}_B$ leaving the reduced state in $\mathcal{H}_A$ unchanged. It is only when there is some coupling between the operators in $\mathcal{H}_A$ and $\mathcal{H}_B$, be it through the output operators coupling to the loss channels, or through the signal fields coupling with the idler fields that are subsequently filtered out, that the thermal photon number has $n^{th}_{tot} > 0$. 
\section{Example 
Calculations} \label{sec:SectionV}
In this section we consider two schemes to enhance the achievable degenerate squeezing in a signal resonance through the coupling of an additional resonator to the primary ring. We work within the five resonance approximation detailed above (see Fig. \ref{fig:FiveResonanceModel}) as this allows for a conceptually simpler setup while highlighting the salient features of each suppression schemes. The limitations of the five resonance approximations and generalizations to schemes involving higher order SFWM interactions are discussed in Appendix \ref{sec:appendix_nineMode}.

In each example we consider a dual ring photonic molecule as show in Fig. \ref{fig:ringDiagrams}(a), where a primary resonator is side coupled to an input/output waveguide as well as to a single auxiliary resonator. Fixing the properties of the primary ring, we tune the auxiliary resonator to suppress unwanted parasitic SFWM processes, or to compensate for detunings of the desired DP-SFWM interaction.

For simplicity, we consider the input waveguide and the primary and auxiliary resonators to be constructed from the same material, with the same cross-sectional dimensions. As such, for a given $J$, we can approximate the dispersion properties to be the same within each of the elements and suppress the $\tau$ superscript on the group velocity. Furthermore, taking the input pump field and the generated field to be polarized in the direction perpendicular to the plane of the resonator system, we can approximate the nonlinear factor $\gamma_{NL}^{\vec{J}, \tau}(\xi)$ to be independent of the coordinate $\xi$ and the element $\tau$. Additionally, when the resonances are sufficiently close in frequency, as will be the case here, we expect $\textbf{d}_J^{\tau}(\textbf{r}_{\perp}; \xi)$ to vary little with the choice of $J$; we can then write $\gamma_{NL}^{\vec{J}, \tau}(\xi) \cong \gamma_{NL}$ for some value $\gamma_{NL}$. In particular, choosing $\lambda_S = 1550$ nm we will take $\gamma_{NL} = 1.0$ $(Wm)^{-1}$, which is consistent with previous squeezing calculations in silicon nitride rings \cite{Vernon2015, Sloan2024}.

As in the simplified point coupling scenario, to the waveguide/primary ring coupler we can introduce an effective self-coupling coefficient, $\varsigma_{J}^{wg}$, and cross-coupling coefficient, $\varkappa_{J}^{wg} = \sqrt{1 - (\varsigma_{J}^{wg})^2}$, corresponding to the amplitude factor induced by a single pass of the coupling region. Similarly, to the coupling region between the primary and auxiliary ring, we denote by $\varsigma_J^{aux}$ and $\varkappa_J^{aux} = \sqrt{1 - (\varsigma_{J}^{aux})^2}$ the self- and cross-coupling coefficients. We refer to the latter of these quantities as the ``ring cross-coupling,'' which will control the resonance splitting and shifting that will allow for the squeezing enhancement (see Fig. \ref{fig:transmissionDiagrams}). 

In both squeezing schemes, the five resonance spectrum will resemble that shown in Fig. \ref{fig:FiveResonanceModel} in which pumps drive the $J= P_1$ and $J=P_2$ resonances surrounding a central $J=S$ signal resonance, with two additional idler resonances $J= LI, RI$ on either side of the pumps that allow parasitic SFWM interactions. Because of the coupling with the auxiliary resonator, the shape and width of each resonance and their associated ranges $R(J)$ may not be identical. However, each will be chosen such that, in the absence of group velocity dispersion and coupling to the auxiliary ring, the five resonance set would correspond to equally space peaks of the primary ring. We discuss the effect of the auxiliary resonator and its modification of the resonance spectrum in each example. 

In all cases, we set the discretized bin width, $\delta k_J$, of each resonance and the length of the discretized time step, $\Delta t$, sufficiently small so as to ensure convergence of the resulting output field. A sample sweep of these parameters for a representative setup using the scheme detailed in Example 1 is shown in Appendix \ref{section:appendix_convergence}. To each ring resonator, we couple 5 phantom loss channels equally space along the length of each ring with equal self-coupling coefficients, $\sigma_{J}^{\nu, ph}$, such that the total round trip attenuation in the primary and auxiliary rings, $\Gamma_J^{pr}$ and $\Gamma_J^{ar}$ respectively, correspond to the values detailed in each example.

Finally, we take the pulses stimulating each of the pump resonances to have a Gaussian power distribution, leading to the asymptotic-in distributions corresponding to the real waveguide input channel, $\nu_{in}$, at $t=t_0$ being given by
\begin{equation}
    \alpha_{J, \nu_{in}}^{in}(k, t_0) = \left( \frac{2}{\pi} \right)^{\frac{1}{4}} \sqrt{\frac{E_J \delta t_J v_J}{\hbar \omega_J}} e^{-v_J^2 \delta t_J^2 (k-k_J^p)^2} e^{-i(k-k_J^p)\mu_J},
\end{equation}
for $J \in \{P_1, P_2 \}$. Here, the values $E_J$, $\delta t_J$, $k_J^p$, and $\mu_J$ correspond respectively to the total energy of the pulse, standard deviation of the gaussian power distribution, central wavenumber of the pulse spectrum, and $\xi$ position of the pulse center at $t=t_0$ for the pump stimulating the resonance $J$. The initial pump fields in the local basis, $\alpha_{J, \nu}^{loc}(k, t_0)$, can then be computed from equation (\ref{eq:LocalBasisPulseDef}).

\begin{figure*}
    \centering
    \includegraphics[width=0.95\linewidth]{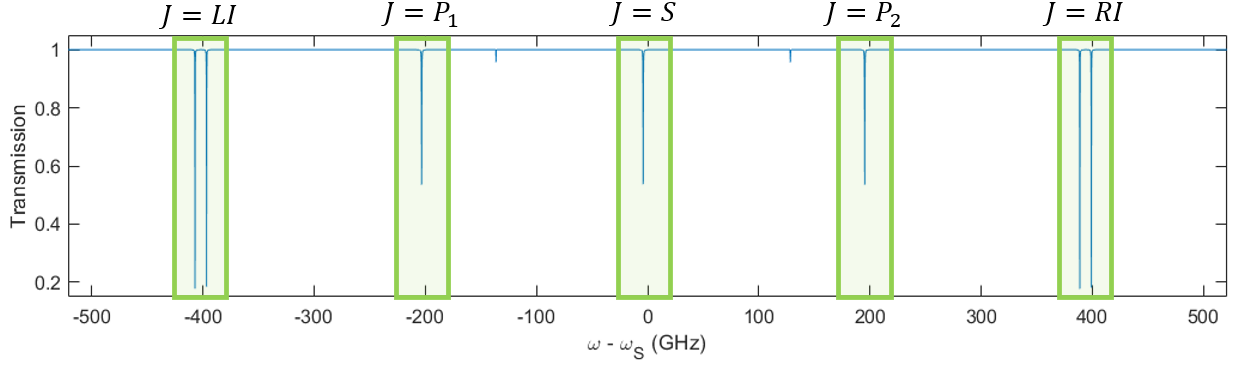}
    \caption{Diagram of the system resonance spectrum corresponding to the first example. Here, the length of the auxiliary resonator is tuned such that the LI and RI resonances are symmetrically split about the uncoupled primary ring resonance positions.}
    \label{fig:splittingSpectrum_withLabels}
\end{figure*}

\subsection{Example 1: Resonance Splitting}
In this first example, we assume the group velocity dispersion in each element can be taken to be negligible within a frequency range containing the five resonances of interest, and adopt a parasitic suppression scheme investigated earlier \cite{Zhang2021}. Were the auxiliary resonator completely decoupled from the primary resonator, this would result in a set of equally spaced resonances with the FWHM determined by the coupling strength between the waveguide and the primary resonator, as well as the scattering loss within the primary resonator. 

We consider degenerate squeezing, where the goal is to generate squeezed light in one resonance ($J=S$), which we label the ``signal'' resonance, with two pump fields. For simplicity, we take the pumps ($J=P_1, P_2$) to correspond to the higher and lower frequency nearest-neighbor resonances of the signal resonance (see Fig. \ref{fig:splittingSpectrum_withLabels}). Such a scheme can be plagued by the fact that pump $P_1$ can generate a pair of photons, only one in $J=S$ and the other in $J=LI$, leading to unpaired photons in $J=S$; similarly, the pump $P_2$ can generate a pair of photons, only one in $J=S$ and the other in $J=RI$; we take the ``idler'' frequencies here ($J = LI, RI$) to be the next-nearest-neighbor resonances. These parasitic single pump interactions are what we refer to as SP-SFWM (see Fig. \ref{fig:FiveResonanceModel}). Additionally, nonlinear processes such as BS-FWM (shown schematically in Fig. \ref{fig:FiveResonanceModel}) can further mix the fields in the signal resonance with those in the idler resonances. The strategy we adopt is to adjust the coupling between the primary and auxiliary rings so that the resonances $J=RI, LI$ and the nearby auxiliary ring resonances are split, destroying energy conservation for the unwanted, parasitic processes producing photons at ($S,LI$) from $P_1$ and photons at ($S,RI$) from $P_2$, as well as the BS-FWM interactions. 

For sufficiently high pump powers, additional higher order SFWM involving resonances beyond the five resonance approximation, originating from sufficiently strong classical fields being generated in other system resonances due to stimulated single pump SFWM, can be appreciable. These strong classical fields can then act as a secondary, generated pump, resulting in additional parasitic interactions with the signal fields. In Appendix \ref{sec:appendix_nineMode} we discuss the effects of these higher order processes and show that by generalizing the dual-ring structure considered here to a three ring photonic molecule, we can suppress this classical field generation, reducing the system to one which is well approximated by the five resonance scheme considered here.

Letting $L_p$ and $L_a$ correspond to the effective lengths of the primary and auxiliary resonators respectively, we then set $L_p = 2 \pi \times 120$ $\mu$m, $L_a = \frac{3}{4} L_p$, and tune the auxiliary ring such that it shares a resonance with the primary ring at both $\omega_{LI}$ and $\omega_{RI}$. Reintroducing the coupling between the primary and auxiliary resonator then results in a set of system resonances resembling Fig. \ref{fig:splittingSpectrum_withLabels}, in which the $LI$ and $RI$ system resonances are symmetrically split about the position of the isolated primary resonances.

\begin{figure*}
    \centering
    \includegraphics[width=0.98\linewidth]{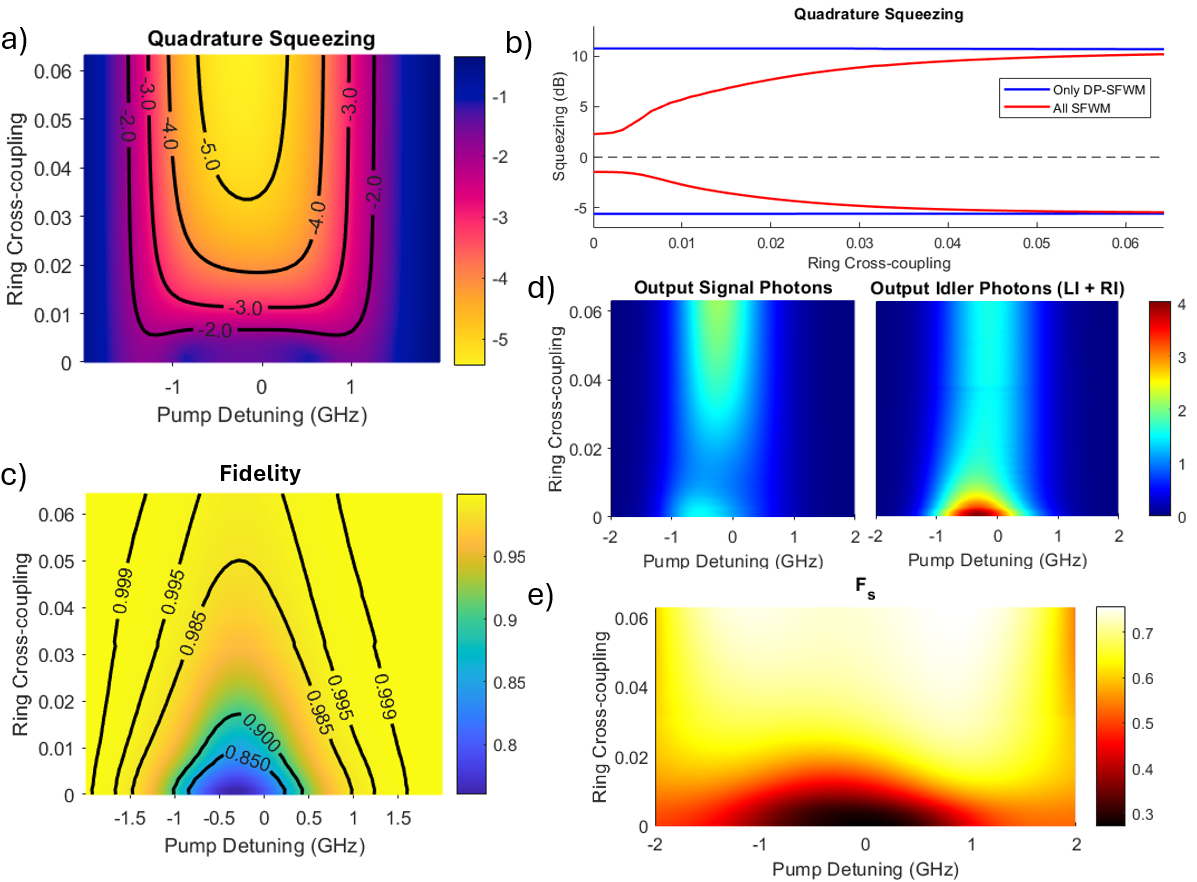}
    \caption{(a) Quadrature Squeezing of the maximally squeezed temporal mode of the reduced output signal state. (b) Quadrature squeezing and anti-squeezing of the maximally squeezed temporal mode corresponding to the reduced output signal state for a fixed pump detuning of -284 MHz when including all SFWM interactions (red), and when only including the desired DP-SFWM interaction (blue). (c) Fidelity of the reduced output signal state when including all SFWM interactions with the ideal (DP-SFWM only) reduced output signal state. (d) Number of signal (left) and idler (right) photons exiting the real waveguide channel. (e) Relative fractions of squeezing photons in the reduced output state (see Eq. (\ref{eq:FsDefinition})).}
    \label{fig:splittingPlotsMerged}
\end{figure*}

By varying the coupling strength between the primary and auxiliary rings, we can modify the splitting of the LI and RI resonances, allowing for selective suppression of the SFWM that mixes the fields in the these split resonances with the fields in the unsplit signal resonance. For the ring parameters quoted above, the magnitude of the splitting as a function of the ring cross-coupling is shown in Fig. \ref{fig:transmissionDiagrams}(a). Note that when the splitting is sufficiently large the $LI$ and $RI$ resonances can be effectively treated as two separate system resonances. However, to maintain consistency between the resonance description, regardless of the splitting magnitude, we continue to refer to each of the full two-peak split resonances with a single $J$. In practice, the distance between the two peaks of the split resonance will be much less than the free spectral range of the primary resonator, and so this identification arises naturally.

In Fig. \ref{fig:splittingPlotsMerged}a we show the maximum quadrature squeezing achievable for the given dual pump scheme with $E_{P_1} = E_{P_2} = 100$ pJ, $\delta t_{P_1} = \delta t_{P_2} = 70$ ps, and equal pump delays $\mu_{P_1} = \mu_{P_2}$, as a function of the pump detuning from the resonance center (taken to be equal for both pumps), as well as the cross-coupling strength between the primary and auxiliary rings. Here we have chosen the coupling strength between the waveguide and primary ring such that $|\varsigma_J^{wg}| = 0.997$, with the round trip attenuation in each ring set to $\Gamma_J^{pr} = 0.9991$ and $\Gamma_J^{ar} = 0.99935$, leading to an over-coupled primary ring. Were the auxiliary ring completely decoupled from the primary ring, this would result in an intrinsic and loaded quality factors of $Q_{int} = 3.4 \times 10^6$ and $Q_{load} = 7.83 \times 10^5$ for the primary ring, which is currently achievable in silicon nitride systems \cite{Stern:17}.

Note that when the splitting of the idler resonances become large relative to the FWHM of the unsplit pump and signal resonances, the number of idler photons exiting through the output channel is reduced, but does not vanish (see Fig. \ref{fig:splittingPlotsMerged}d). Indeed, the splitting of the idler resonances detune the SP-SFWM and BS-FWM interactions, yet the HP-SFWM interaction can still preserve energy in each of the split lobes regardless of the magnitude of the splitting. Importantly, this interaction does not mix the idler fields with the desired signal field. Thus, despite the presence of the remaining idler fields, the squeezing in the reduced signal output state does approach the ideal case in which all SFWM are neglected except DP-SFWM, as shown in Fig. \ref{fig:splittingPlotsMerged}b. The reduction in the coupling of the signal field to the idler fields can be further verified by calculating the fidelity, $\mathcal{F}$, of the reduced signal output state \cite{Paraoanu} including all SFWM interactions and that of the ideal DP-SFWM only state. For the parameters detailed above, we find $\mathcal{F} > 0.985$ when $\varkappa_J^{aux} > 0.05$ with pump detunings around $\Delta \nu = -284$ MHz which optimize the squeezing in the signal output. This can be compared with $\mathcal{F} < 0.77$ for the same detuning with a decoupled auxiliary ring ($\varkappa_J^{aux} = 0$). We note, however, that while fidelities approaching $\mathcal{F} = 1$ can be seen for sufficiently large pump detunings, this would also correspond to a regime of parameter space in which the output signal field approaches the vacuum with negligible squeezing and thus little interest for a practical squeezing device. Regardless, we see that the resulting output state closely resembles the ideal DP-SFWM only case near the optimal pump detuning when the splitting is sufficiently large, resulting in a dramatic improvement in the measurable squeezing.

\begin{figure*}
    \centering
    \includegraphics[width=0.98\linewidth]{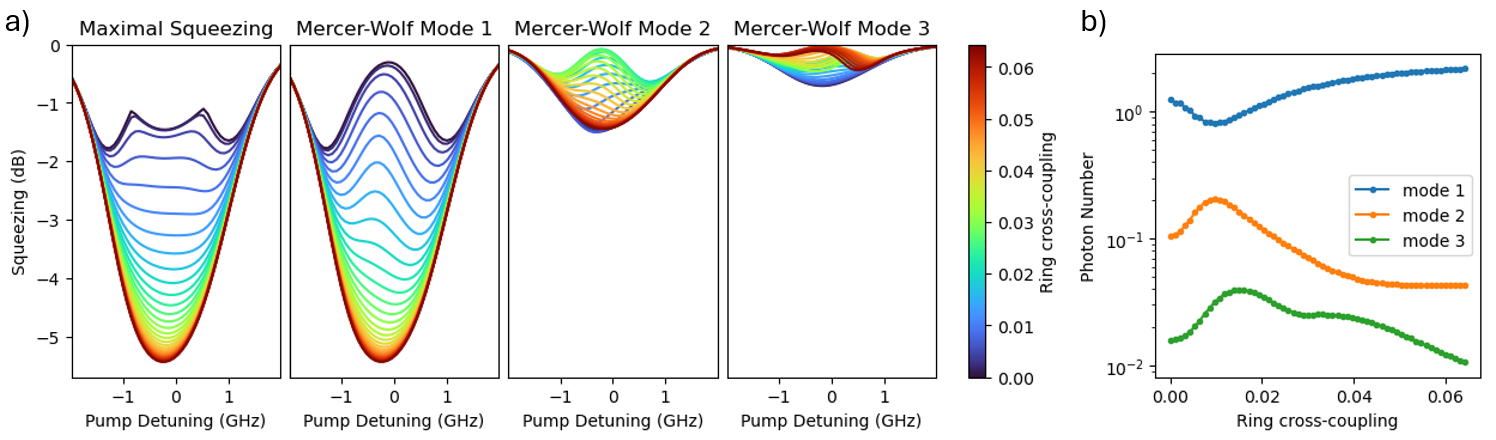}
    \caption{(a) Squeezing of the maximally squeezed temporal mode and the first three Mercer-Wolf temporal modes as a function of the ring cross-coupling and pump detuning. (b) Number of output signal photons in the first three Mercer-Wolf temporal modes at the pump detuning which optimizes the maximal squeezing in the limit of large ring cross-coupling.}
    \label{fig:temporalModeSqueezing}
\end{figure*}

Interestingly, the effect on the squeezing of the parasitic SFWM contributions can vary significantly for different temporal modes of the reduced signal state. In Fig. \ref{fig:temporalModeSqueezing}a we show the squeezing in the maximally squeezed temporal mode, as well as the first three Mercer-Wolf modes \cite{Wolf:82, Kopylov2025} of the reduced signal output state, the latter corresponding to the temporal mode basis which diagonalizes $\tilde{N}$ (see Sec. \ref{sec:SectionIV}), for different values of the ring cross-coupling. In the simple case in which a given field is a direct product of a number of independent vacuum squeezed temporal modes, the squeezing in each temporal mode increases monotonically with the expected number of photons in the mode \cite{Gerry_Knight_2004}. However, due to the presence of the parasitic interactions mixing the signal field with the idler fields, for sufficiently weak splitting we observe more squeezing in the second and third order Mercer-Wolf modes around the pump detuning which optimizes the total number of generated signal photons. This is despite more than an order of magnitude difference in the photon number in the secondary and tertiary temporal modes compared to the primary temporal mode for those values of the ring cross-coupling (see Fig. \ref{fig:temporalModeSqueezing}b).

\begin{figure*}
    \centering
    \includegraphics[width=0.9\linewidth]{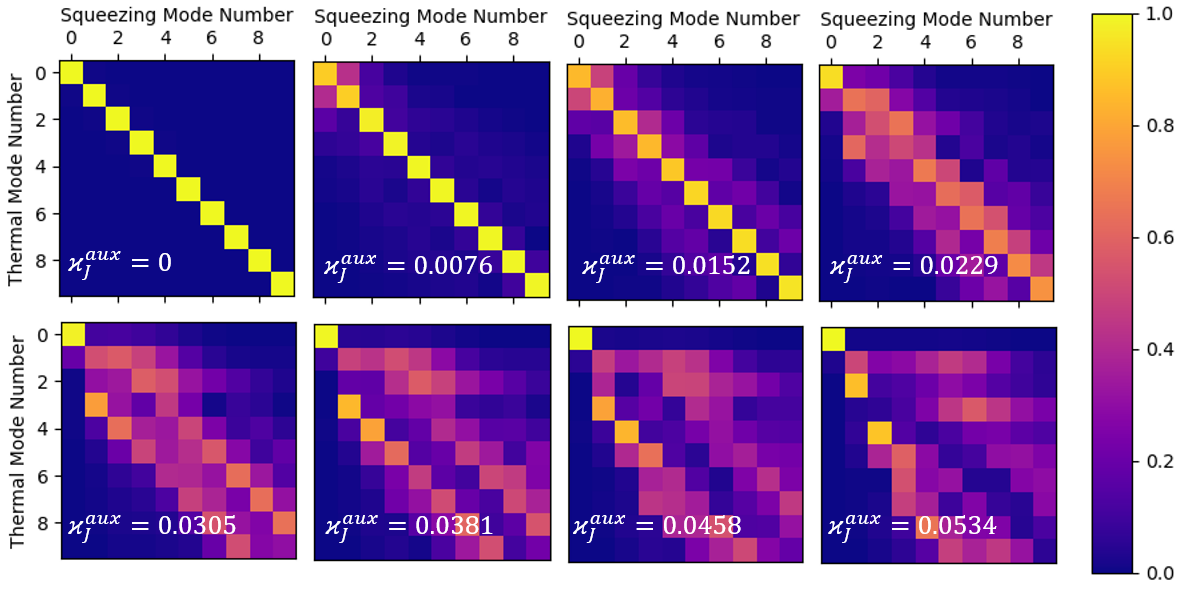}
    \caption{Overlap of the basis of squeezing modes and the basis of thermal modes constructed from the covariance matrix decomposition (see eq. (\ref{eq:ReducedCharacteristicDecomp})) for different ring cross-couplings.}
    \label{fig:TemporalModeDecomposition}
\end{figure*}

Furthermore, the impact of the parasitic processes on the squeezing can differ among the temporal modes for a given value of the ring cross-coupling. Indeed, for very weak coupling, the squeezing of the primary Mercer-Wolf mode is heavily suppressed near zero detuning of the pumps, but is enhanced as the coupling strength is increased. On the other hand, the secondary temporal mode initially sees a reduction in the peak squeezing up to a cross coupling near $\varkappa^{aux}_J = 0.03$, after which it increases back to the unsplit squeezing value. This can be understood by appealing to the squeezing and thermal basis constructed from the $\tilde{\bm{\Sigma}}_{sq}$ and $\tilde{\bm{\Sigma}}_{therm}$ respectively. In Fig. \ref{fig:TemporalModeDecomposition} we show the overlap between the first ten temporal modes of the squeezing and thermal basis for different values of the ring cross-coupling. When the auxiliary ring is decoupled from the primary, corresponding to a cross-coupling of $\varkappa^{aux}_{J} = 0$ and unsplit idler resonances, the squeezing and thermal bases are seen to be identical. For this particular case, this can be understood as a result of the high amount of symmetry in the linearized coupling coefficient in Eq. (\ref{eq:NonClassicalEOM}), $A_{J, \nu, \nu'}^{\chi}(k, t)$, stemming from the resonances being equally spaced with identical lineshapes, as well as the pumps having identical pulse shapes and detunings. Consequentially, for an uncoupled auxiliary ring in this scheme, each $A_{J, \nu, \nu'}^{\chi}(k, t)$ would be identical up to combinatorial factors and conjugations of the classical pump fields. This coincidence of the squeezing and thermal basis would then correspond to the output signal state being described by a product of independent squeezed thermal states, with the primary mode being dominated by the thermal contributions arising from the parasitic SFWM interactions when the pumps are near resonance. As the cross-coupling is increased, the detuning of the SP-SFWM and BS-FWM interactions results in signal field generated further from the resonance center. The thermal noise then prioritizes the higher order squeezing modes of the signal field with support along a broader range of resonance frequencies. For all values of ring cross-coupling, the overlap of the primary squeezing mode and the primary thermal mode remains near unity due to the scattering loss, which represents the dominant thermal contribution at large idler splittings.

\subsection{Example 2: Resonance shifting}
\begin{figure*}
    \centering
    \includegraphics[width=0.98\linewidth]{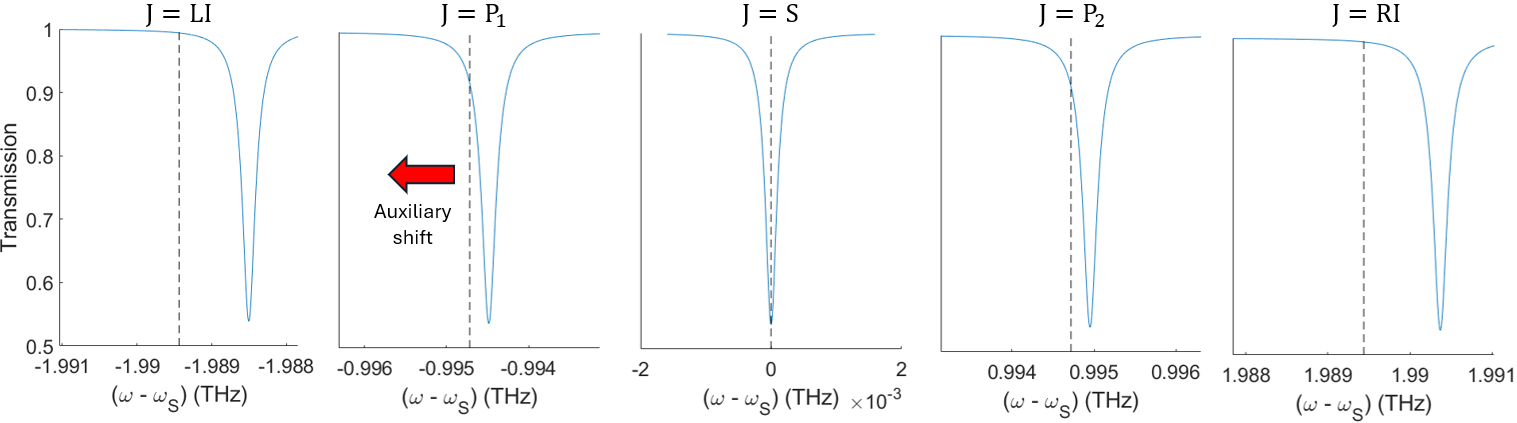}
    \caption{Spectrum of the five resonances system of an uncoupled auxiliary resonator ($\varkappa_{J}^{aux} = 0$) with $\beta = 0.5$ ps$^2$ /m. Black dashed lines denote equally spaced resonance positions from the central $J=S$ resonance corresponding to the resonance centers when neglecting GVD. Red arrow shows the direction of the $J=P_1$ resonance shift induced by the coupling to the auxiliary resonator.}
    \label{fig:pushingSpectrum}
\end{figure*}

As a second example, we consider a system which has significant group velocity dispersion within the frequency range containing the five resonances of interest. In particular, for $\lambda_0 = 1550$ nm and the corresponding frequency $\bar{\omega}$, we expand the effective wavenumber in each element as
\begin{equation}
    k(\omega) = k_0 + \frac{1}{\bar{v}} (\omega - \bar{\omega}) + \frac{\bar{\beta}}{2} (\omega - \bar{\omega})^2
\end{equation}
for $k_0$ and $\bar{v}$ the corresponding wavenumber and group velocity at $\lambda_0 = 1550$ nm. Taking $\bar{\beta} = 0.5$ $ps^2/m$ (which is in line with measurements of GVD in silicon nitride waveguides near 1550 nm \cite{Tan2010}), the signal field to be centered near $\lambda_S = 1550$ nm, and the pump and idler resonances to be located 5 and 10 primary resonance peaks on either side of the signal respectively, results in a spectrum shown in Fig. \ref{fig:pushingSpectrum}. Due to the presence of GVD, the five primary ring resonances are no longer equally spaced in frequency, resulting in some of the SFWM interactions being suppressed even without the inclusion of the auxiliary resonator. Indeed, for such a system, even the desired DP-SFWM interaction has a detuning of $\Delta \Omega_{(S, S, P_1, P_2)} = -453$ MHz, which would reduce the number of generated signal photons and, as a result, the squeezing.

\begin{figure}
    \centering
    \includegraphics[width=0.98\linewidth]{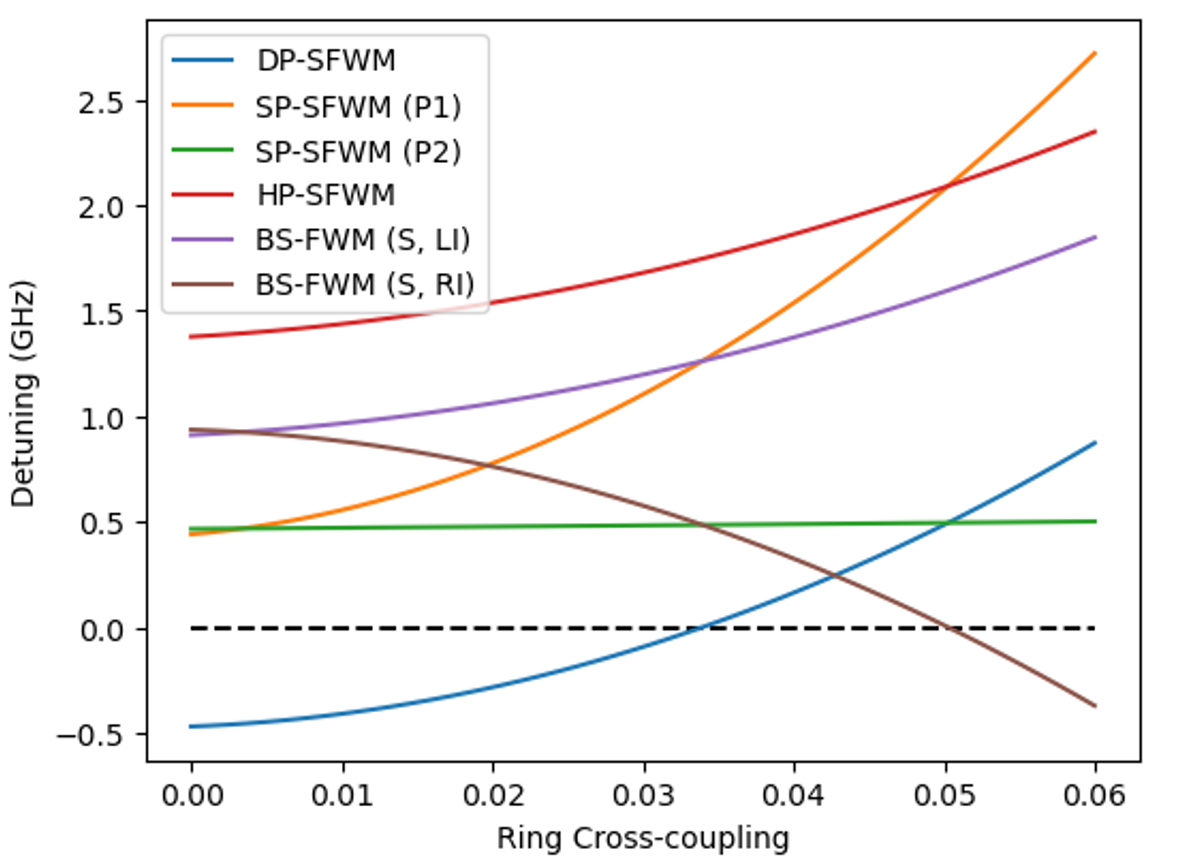}
    \caption{Detuning of the SFWM processes as a function of the ring cross-coupling.}
    \label{fig:SFWMdetunings}
\end{figure}

Here the strategy is to tune the auxiliary ring such that a single auxiliary resonance lies near the $J=P_1$ resonance of the primary ring with some detuning $\delta_a$, but far from any of the other 4 resonances of interest. By choosing $\delta_a$ sufficiently large, we can induce an effective shift of the $P_1$ resonance, tuning the system such that the $P_1$, $P_2$, and $S$ resonances of the full coupled system become equally spaced. As such, we could selectively correct the desired dual pump process while leaving the other interactions suppressed. This can be seen in Fig. \ref{fig:SFWMdetunings}, where we show the detunings of each of the SFWM interactions, $\Delta \Omega_{\vec{J}}$, as a function of the ring cross-coupling.

This scheme shares similarities with previous GVD compensation schemes from Gentry \textit{et al.} \cite{Gentry:14} applied to non-degenerate seeded four-wave mixing. However, our approach differs in that we consider hybridizing one of the pump resonances rather than a resonance in which the desired field will be generated. With the $P_2$ pump (and $S$ resonance) remaining far from an auxiliary resonance, the overlap between the $P_1$ and $P_2$ pumps, and consequentially the generated signal field, is constrained mainly in the primary resonator. As such, the escape efficiency of the generated squeezed light remains near that of the isolated primary resonator, reducing loss from the signal field coupling into, and subsequently scattering out of, the auxiliary ring.

As in the previous example, we take the length of the primary resonator to be given by $L_p = 2 \pi \times 120$ $\mu$m with $\varsigma_{J}^{wg} = 0.997$ and $\Gamma_{p} = 0.9991$. We then set the length of the auxiliary resonator to be $L_a = 2 \pi \times 75$ $\mu$m, with $\Gamma_a = 0.99935$ and fix $\delta_a = 4.77$ GHz. For these values, the shifting of the $J = P_1$ resonance due to the resonance hybridization as a function of the ring cross-coupling is shown in Fig. \ref{fig:transmissionDiagrams}b.

\begin{figure*}
    \centering
    \includegraphics[width=0.98\linewidth]{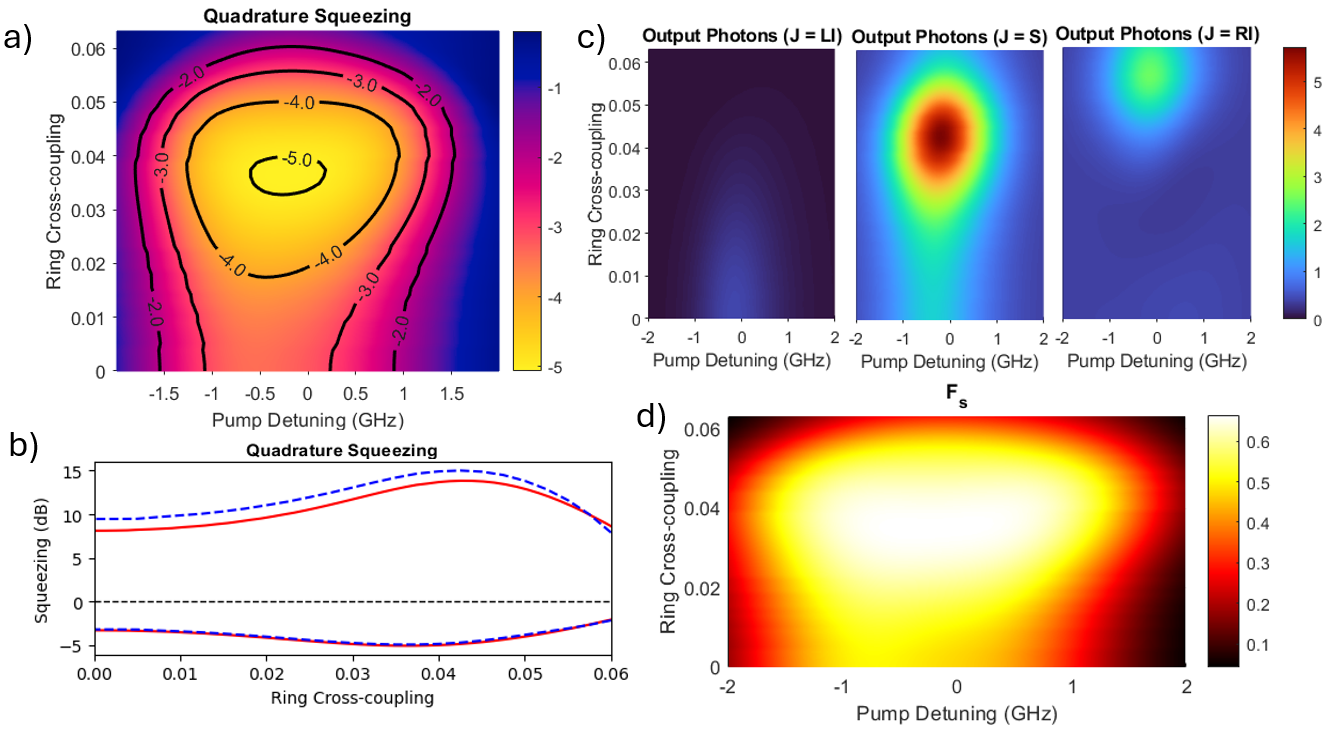}
    \caption{(a) Quadrature squeezing of the most squeezed temporal mode of the reduced signal output state as a function of the ring cross-coupling and $P_1$ pump detuning for a fixed $P_2$ pump detuning of -170 MHz. (b) Quadrature squeezing and anti-squeezing of the most squeezed temporal mode of the reduced signal output state for a fixed symmetric detuning of the $P_1$ and $P_2$ pumps of -170 MHz as a function of the ring cross-coupling (red solid lines). Blue dashed lines correspond to the squeezing and anti-squeezing when including higher order parasitic SFWM interaction in a nine resonance approximation (see Appendix \ref{sec:appendix_nineMode}). (c) Number of photons exiting through the output channel for the $J= LI$ (left), $J=S$ (center), and $J=RI$ resonances. (d) Relative fractions of squeezing photons in the reduced output state (see Eq. (\ref{eq:FsDefinition})).}
    \label{fig:pushingPlotsMerged}
\end{figure*}

In Fig. \ref{fig:pushingPlotsMerged} we show the maximum squeezing in the reduced signal output state as a function of pump detuning and ring cross-coupling for $E_{P_1} = E_{P_2} = 100$ pJ, $\delta t_{P_1} = \delta t_{P_2} = 70$ ps, $\mu_{P_1} = \mu_{P_2}$, and the detuning of the $P_2$ resonance fixed at $\Delta \nu = -170$ MHz. Indeed we see that as the $J=P_1$ resonance is shifted due to the coupling with the auxiliary ring, the DP-SFWM process is enhanced leading to a larger number of photons generated in the signal resonance due to the DP-SFWM interaction (Fig. \ref{fig:pushingPlotsMerged}c), and correspondingly to a larger achievable squeezing. However, unlike the previous example in which increasing the coupling of the auxiliary resonator suppressed the parasitic SFWM contributions but did not detune the DP-SFWM, here the squeezing decreases past an optimal cross-coupling due to the desired DP-SFWM becoming detuned once more (see Fig. \ref{fig:SFWMdetunings}). Furthermore, since interactions such as BS-FWM and SP-SFWM of the $P_2$ pump are not completely suppressed near the cross-coupling that optimizes the DP-SFWM detuning, we find that the anti-squeezing of the most squeezed temporal mode continues to increase beyond the optimal cross-coupling value of $\varkappa^{aux}_J = 0.0343$, as the BS-SFWM involving the $J=S$ and $J=RI$ resonances become realigned.

Unlike the first example, here the higher order contributions from SFWM processes beyond the five resonance model are much less significant. This is due to the stimulated SFWM interactions leading to a strong classical field generation in resonance beyond the frequency range considered here (see Appendix \ref{sec:appendix_nineMode}) being further detuned by GVD than those interaction contained in the five resonance approximation. This is shown in Fig. \ref{fig:pushingPlotsMerged}b in which generalizing the treatment to a nine mode approximation to include such higher order effects only results in a slight decrease of the squeezing in the maximally squeezed temporal mode of the reduced signal output.

\begin{figure*}
    \centering
    \includegraphics[width=0.98\linewidth]{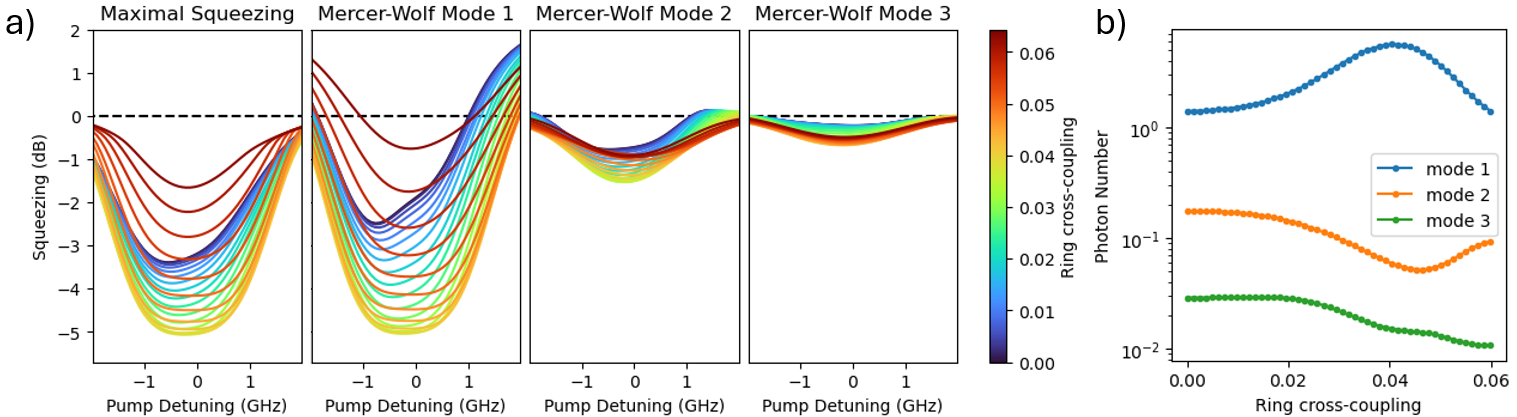}
    \caption{(a) Squeezing of the maximally squeezed temporal mode, as well as the first three Mercer-Wolf temporal modes as a function of the ring cross-coupling and pump detuning for example 2. (b) Number of output signal photons in the first three Mercer-Wolf temporal modes at the $P_1$ detuning which optimizes the maximal squeezing.}
    \label{fig:temporalModeSqueezing_push}
\end{figure*}

Decomposing the reduced output signal state into the Mercer-Wolf mode basis, we show the quadrature squeezing and expected photon number of the primary, secondary, and tertiary temporal modes in Fig. \ref{fig:temporalModeSqueezing_push}. For certain pump detunings and auxiliary shifts some SP-SFWM and BS-FWM interactions can be enhanced relative to the desired DP-SFWM interaction, leading to a deterioration in the signal mode squeezing in the primary Mercer-Wolf temporal mode and resulting in thermal contributions dominating. Despite this, properly tuning the auxiliary ring coupling and pump detuning shows an increase of more than 2.5 dB of squeezing in the primary Mercer-Wolf mode, from -2.48 dB for a decoupled auxiliary ring to -5.04 dB for a ring cross-coupling of $\varkappa^{aux}_J = 0.0355$, with an increase of more than 1.7 dB for the maximally squeezed temporal mode from -3.30 dB to -5.06 dB. Interestingly, this ring cross-coupling, which optimizes the squeezing in the primary Mercer-Wolf modes, is slightly larger that that which realigns the DP-SFWM interaction ($\varkappa^{aux}_J = 0.0340$) and leads to the largest available squeezing discussed previously.

\section{Conclusion} \label{sec:Conclusion}
We have generalized a non-perturbative asymptotic scattering method \cite{Sloan2024} to describe degenerate dual-pump squeezing in an arbitrary multi-ring photonic molecule system, including parasitic non-degenerate SFWM interactions. By adopting a five resonances model, we investigated two examples in which the achievable squeezing through the desired DP-SFWM can be enhanced within a two ring system by utilizing the auxiliary resonator to modify the resonance spectrum of the primary resonator in which the squeezing interaction occurs. In particular, this was done by either splitting the unwanted sideband resonances, which allow for the parasitic non-degenerate SFWM interactions, or by shifting one of the pump resonances to compensate for GVD effects that detune the desired DP-SFWM interaction.

In our first example, we employed the parasitic SFWM suppression scheme considered earlier \cite{Zhang2021}. Here the additional auxiliary resonator is tuned to share resonances with two undesired sideband idler resonances of the primary resonator, resulting in hybridized resonances split symmetrically about the uncoupled resonance positions. We have shown that for a sufficiently strong coupling between the primary and auxiliary resonators, leading to splitting much larger then the primary resonator linewidth, SP-SFWM and BS-FWM interactions involving the idler sidebands and the desired signal resonance become detuned and sufficiently suppressed such that the reduced signal output state resembles that of an equivalent three resonance system in which all SFWM, except the desired DP-SFWM, is absent. This is validated by computing the fidelity of the output signal state, including parasitic suppression, with that of the ideal DP-SFWM only case, leading to values greater than 0.991 for 4.77 GHz of splitting when detuning the pump to optimize the output squeezing. Consequentially, the achievable squeezing is significantly enhanced compared to the unsplit idler system, asymptotically approaching that of the DP-SFWM-only case. This is despite the HP-SFWM interaction, leading to photon generation in both idler resonances, being preserved for any symmetric splitting, since the interaction does not result in mixing of the fields between the signal and idlers.

As a second example, we considered five resonances of interest sufficiently separated that GVD in the resonator system would lead to an appreciable detuning of the desired DP-SFWM interaction. By tuning the auxiliary resonator such that a single resonance lies near one of the pump resonance, but sufficiently far so as to not induce significant splitting, the hybridization of the resonances would result in an effective shift of the pump, allowing for realignment of the degenerate DP-SFWM interaction. By optimizing the pump detuning and the coupling between the primary and auxiliary rings, we demonstrate a significant improvement in squeezing of the output signal state, up to 2.5 dB in the primary Mercer-Wolf mode (1.7 dB in the maximally squeezed mode) for the resonator parameters considered here. This is limited, in part, to some SP-SFWM and BS-FWM interactions not being fully suppressed for the ring couplings that lead to pump shifts optimizing the DP-SFWM interaction. 

By utilizing both resonance hybridization schemes in tandem, one could envision multi-ring resonator systems employing three or more micro-rings, allowing for parasitic suppression and squeezing enhancement beyond the five resonance model considered here. Such systems would allow tunable control over individual resonances, leading to further optimization of nonlinear processes and squeezing in integrated micro-systems beyond current capabilities.

\section*{Acknowledgements} \label{sec:acknowledgements}
    MS acknowledges support from the Mitacs Accelerate program and the Ontario Graduate Scholarship program. JES acknowledges support from the Natural Sciences and Engineering Research Council of Canada.

\bibliographystyle{unsrt}
\bibliography{References}

\appendix
\section{Asymptotic Fields} \label{sec:asyFields}
In this appendix, we motivate the decomposition of the asymptotic field amplitudes as shown in equation (\ref{eq:AsymptoticFieldDistExpansion}). We do this by partitioning the system into a set of segments, within each of which we can develop the quantization of the fields. Matching the boundaries of adjacent segments, we then extend the description to the full system.

\begin{figure*}
    \centering
    \includegraphics[width=0.85\linewidth]{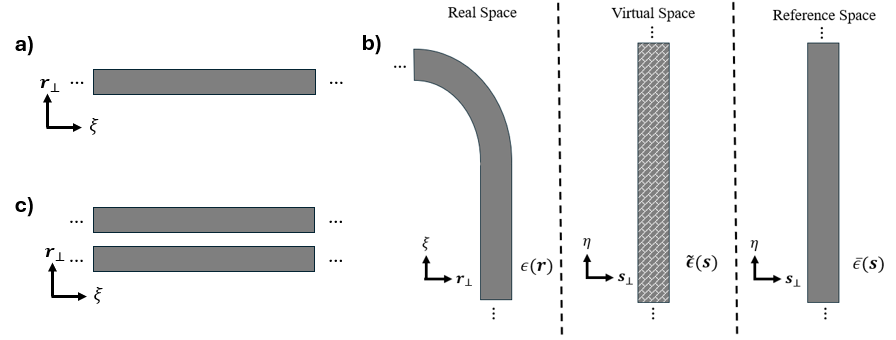}
    \caption{Diagram of (a) a straight segment of an isolated waveguide, (b) a segment of an isolated waveguide with some arbitrary bend profile, the corresponding straight waveguide in virtual space, as well as in the reference space, and (c) a segment of a uniform parallel directional coupler.}
    \label{fig:segmentDiagrams}
\end{figure*}

Consider first a straight and uniform segment of a single element $\tau$, labeled $V_{\tau}$, sufficiently far from any other element $\tau'$ such that there is negligible coupling between $\tau$ and $\tau'$ (Fig. \ref{fig:segmentDiagrams}(a)). This could, as an example, represent the region of the waveguide or phantom channels containing the input/output fields. Then we can expand the displacement field locally around the segment of $V_{\tau}$ as \cite{Quesada:22}
\begin{equation} \label{eq:straightWaveguideModeExpansion}
\begin{split}
    \textbf{D}(\textbf{r}, t) &\cong \sum_{I, J} \int_{\mathcal{R}(J)} d\omega \sqrt{\frac{\hbar \omega_J}{4 \pi v_{I,J}^{V_{\tau}}}} \tilde{\textbf{d}}_{I,J}^{V_{\tau}}(\textbf{r}_{\perp}) \\
    & \qquad \qquad \qquad \times a_{I, J}^{V_{\tau}}(\xi ,t; \omega) + H.c.,
\end{split}
\end{equation}
changing from the variable $k$ to a surrogate variable $\omega$ related to $k$ by the dispersion relation. Now $\mathcal{R}(J)$ is the range of $\omega$ corresponding to $J$ and $v_{I, J}^{V_{\tau}}$ is the group velocity of the spatial mode $I$ at the frequency $\omega_J$, which we have assumed does not vary significantly within the range $J$. The normalization of each $\tilde{\textbf{d}}_{I,J}^{V_{\tau}}(\textbf{r}_{\perp})$ can then be chosen to satisfy equation (\ref{eq:modeNormalization}). In this simple case, the $\xi$ dependence of the operator $a_{I, J}^{V_{\tau}}(\xi ,t; \omega)$ is given by
\begin{equation}
    a_{I, J}^{V_{\tau}}(\xi ,t; \omega) = a_{I, J}^{V_{\tau}}(\xi_0 ,t; \omega) e^{i k_{I,J}^{V_{\tau}}(\omega) (\xi - \xi_0)},
\end{equation}
with $k_{I,J}^{V_{\tau}}(\omega)$ being the effective wavenumber for the mode $I$, at frequency $\omega$, within the segment $V_{\tau}$. Indeed, when the segment $V_{\tau}$ contains the incoming (outgoing) fields, we can interpret $a_{I, J}^{V_{\tau}}(\xi_0 ,t; \omega)$ as describing the $I^{th}$ mode input (output) to (from) the system. 

More generally, an arbitrary resonator structure can exhibit a $\xi$ dependent variation in the curvature of the guiding structure. This would lead to more complicated dynamics, as the field profiles along the plane normal to the propagation direction would vary with the bend curvature, leading to coupling between each pair $(I, I')$ of spatial modes. Because of this, let us consider another segment, labeled $V_{\tau}'$, which may exhibit an arbitrary bend profile, but is still sufficiently far from any other element $\tau'$ that the coupling of the fields between $\tau$ and $\tau'$ is negligible. Utilizing a transformation optics scheme \cite{Pendry, Kundtz, Guerra}, we can introduce a transformation on the real space coordinates $\textbf{r} = (\textbf{r}_{\perp}, \xi)$, to define a virtual space $\textbf{s} = (\textbf{s}_{\perp}, \eta)$ as $\textbf{s} = f(\textbf{r})$, in which the geometry of the guiding structure within the virtual space is straight and independent of $\eta$ (see Fig. \ref{fig:segmentDiagrams}b). Maxwell's equations in the virtual space can be phrased in terms of Cartesian differential operators, but this necessitates the replacement of the isotropic and $\xi$ independent permittivity and permeability considered here, $\epsilon(\textbf{r}_{\perp})$ and $\mu(\textbf{r}_{\perp})$, with a transformed permittivity and permeability, $\tilde{\bm{\epsilon}}(\textbf{s})$ and $\tilde{\bm{\mu}}(\textbf{s})$, that in general is anisotropic and $\eta$ dependent.

Utilizing the completeness of the modes of a uniform and isotropic waveguide, we can then introduce a reference system which matches the geometry of the virtual system $V_{\tau}'$, but with an isotropic and $\eta$ independent permittivity and permeability, $\bar{\epsilon}(\textbf{s}_{\perp})$ and $\bar{\mu}(\textbf{s}_{\perp})$ (see Fig. \ref{fig:segmentDiagrams}b). Expanding the fields in terms of the modes of the reference system, $\left\{\bar{\textbf{d}}_{I, J}^{V_{\tau}'}(\textbf{s}_{\perp}) \right\}_{I,J}$, we can write the displacement field locally in the virtual space, in a manner similar to equation (\ref{eq:straightWaveguideModeExpansion}), as
\begin{equation} \label{eq:bendWaveguideExpansion}
\begin{split}
    \textbf{D}(\textbf{s}, t) &\cong \sum_{I,J} \int_{R_I(J)} d \omega \sqrt{\frac{\hbar \omega_J}{4 \pi v_{I,J}^{V_{\tau}'}}} \bar{\textbf{d}}_{I, J}^{V_{\tau}'}(\textbf{s}_{\perp}) \\
    & \qquad \qquad \qquad \times a_{I,J}^{V_{\tau}'}(\eta, \omega, t) + H.c.,
\end{split}
\end{equation}
where again we choose the mode normalization in the reference system to be such that $\bar{\textbf{d}}_{I, J}^{V_{\tau}'}(\textbf{s}_{\perp})$ satisfies equation (\ref{eq:modeNormalization}). Importantly, the field distributions $\left\{\bar{\textbf{d}}_{I, J}^{V_{\tau}'}(\textbf{s}_{\perp}) \right\}_{I,J}$ do not constitute a basis of eigenmodes in the virtual space, but rather in the reference space. As such, propagation along the $\eta$ direction leads to coupling of the operators $\{a_{I,J}^{V_{\tau}'}(\eta, \omega, t) \}_{I}$, which can be described through an effective coupled mode treatment
\cite{Guerra} as
\begin{equation}
    \partial_{\eta}\textbf{a}^{V_{\tau}'}_J(\eta, \omega, t) = -i\textbf{M}^{V_{\tau}'}_J(\eta; \omega) \textbf{a}^{V_{\tau}'}_J(\eta, \omega, t),
\end{equation}
where $\textbf{a}^{V_{\tau}'}_J(\eta, \omega, t)$ is a vector of the operators $a_{I,J}^{V_{\tau}'}(\eta, \omega, t)$ for a fixed $J$ and a given ordering of the spatial modes $I$, and $\textbf{M}^{V_{\tau}'}_J(\eta; \omega)$ contains both terms related to the propagation phase and coupling between the modes. Consequently, we can write
\begin{equation}
    a_{I,J}^{V_{\tau}'}(\eta, \omega, t) = \sum_{I'} g^{V_{\tau}'}_{I, I', J}(\eta, \eta_0; \omega) a_{I,J}^{V_{\tau}'}(\eta_0, \omega, t),
\end{equation}
where $g_{I,I',J}^{V_{\tau}'}(\eta_0, \eta_0; \omega) = \delta_{I, I'}$.

Note that when transforming back into the real space the field distribution $\tilde{\textbf{d}}_{I, J}^{V_{\tau}'}(\textbf{r}_{\perp}; \xi)$, corresponding to $\bar{\textbf{d}}_{I, J}^{V_{\tau}'}(\textbf{s}_{\perp}; \eta)$ in the virtual space, would have the some modification of both the magnitude and direction of the field vector relative to $\bar{\textbf{d}}_{I, J}^{V_{\tau}'}(\textbf{s}_{\perp}; \eta)$. The latter of these, while adding an extra $\xi$ dependence to the mode fields, would not affect the normalization of the modes. On the other hand, letting $\textbf{s}_{\perp} = (s_x, s_y)$, where $s_y$ is oriented normal to the plane of the resonator system, it can be shown that the transformation of the field resulting from the straightening of a freeform waveguide can be expressed in terms of the quantity $\gamma(s_x, \eta) = 1 - s_x/R(\eta)$, where $R(\eta)$ is the instantaneous radius of curvature of the center line of the waveguide at the point given by $\eta$ \cite{Nesic:22}. As such, for resonators with radii of curvature much larger than the cross-sectional width of the waveguides, as considered here, $\gamma(s_x, \eta) \cong 1$ within the region of the plane in which the field is bound. Thus, by properly choosing the reference space permittivity and permeability, $\bar{\epsilon}(\textbf{s}_{\perp})$ and $\bar{\mu}(\textbf{s}_{\perp})$, we can make the approximation that
\begin{equation}
    \int \frac{\tilde{\textbf{d}}^{V_{\tau}' *}_{I,J}(\textbf{r}_{\perp}; \xi) \cdot \tilde{\textbf{d}}^{V_{\tau}'}_{I,J}(\textbf{r}_{\perp}; \xi)}{\epsilon_0 \epsilon_{\tau}(\textbf{r}_{\perp})} d \textbf{r}_{\perp} \cong 1
\end{equation}

Finally, consider a region of the resonator/waveguide system, labeled $V_{\tau, \tau'}$, in which two elements $\tau$ and $\tau'$ are sufficiently close that energy can be transferred between the two elements through their evanescent coupling. For simplicity, we will approximate such a region as a parallel and uniform directional coupler between the two elements (see Fig. \ref{fig:segmentDiagrams}c); however, the following treatment can be generalized to a wider variety of structures \cite{Huang:94, microRingBook}. Letting the coordinates $(\textbf{r}_{\perp}, \xi)$ correspond to joint local coordinates ranging over the cross section of both $\tau$ and $\tau'$, we could expand the displacement field in terms of the normal modes of a dual waveguide system, as done for a single isolated waveguide. But for the squeezed light generation strategies considered here, we are interested in weakly coupled systems in which the input waveguides and resonators are constructed from the same materials with similar cross-sectional dimensions. Thus, we can instead approximate the field in terms of the modes of each isolated element, and adopt a coupled mode approach, writing \cite{Huang:94, microRingBook}
\begin{equation}
\begin{split}
    \textbf{D}(\textbf{r}, t) &\cong \sum_J \left\{ \sum_{I \in \mathcal{I}_{\tau}} \int_{\mathcal{R}(J)} d \omega \sqrt{\frac{\hbar \omega_J}{4 \pi v_{I,J}^{V_{\tau, \tau'}}}} \tilde{\textbf{d}}_{I,J, \tau}^{V_{\tau, \tau'}}(\textbf{r}_{\perp}) \right. \\
    & \qquad \qquad \qquad \qquad \times a_{I,J, \tau}^{V_{\tau, \tau'}}(\xi, \omega, t) \\
    & \qquad \qquad  + \sum_{I \in \mathcal{I}_{\tau'}} \int_{\mathcal{R}(J)} d \omega \sqrt{\frac{\hbar \omega_J}{4 \pi v_{I,J}^{V_{\tau, \tau'}}}} \tilde{\textbf{d}}_{I,J, \tau'}^{V_{\tau, \tau'}}(\textbf{r}_{\perp}) \\
    & \left. \qquad \qquad \qquad \qquad  \times a_{I,J, \tau'}^{V_{\tau, \tau'}}(\xi, \omega, t) \right\} + H.c.
\end{split}
\end{equation} 
where
\begin{equation}
\begin{split}
    \begin{pmatrix}
        \textbf{a}^{V_{\tau, \tau'}}_{J,\tau}(\xi, t; \omega) \\ \textbf{a}^{V_{\tau, \tau'}}_{J,\tau'}(\xi, t; \omega)
    \end{pmatrix} = \textbf{G}_J^{V_{\tau, \tau'}}(\xi, \xi_0; \omega) \begin{pmatrix}
        \textbf{a}^{V_{\tau, \tau'}}_{J,\tau}(\xi_0, t; \omega) \\ \textbf{a}^{V_{\tau, \tau'}}_{J,\tau'}(\xi_0, t; \omega)
    \end{pmatrix}.
\end{split}
\end{equation}

With this, we can approximate our general resonator system as being constructed from a number of disjoint segments of the three kinds discussed above. By matching the boundary conditions at the interfaces of each adjacent segment we can express each of the operators, $a^{V_{\tau}}_{I,J}(\xi, t; \omega)$ and $a^{V_{\tau, \tau'}}_{I,J, \tau}(\xi, t; \omega)$, in terms of the annihilation operators corresponding to the input segments of the real waveguide and phantom channels. Indeed, letting $\mathcal{T}_{in}$ denote the set of elements corresponding to a real waveguide or phantom channel, and defining $\nu(\tau)$  and $V_{\tau}^{in}$ for $\tau \in \mathcal{T}_{in}$ as the input port and segment corresponding to the input in element $\tau$, we can write $a^{in}_{I,J,\nu(\tau)}(\omega,t) = a^{V_{\tau}^{in}}_{I,J}(\xi_0, t; \omega)$. This allows us to expand the field locally in the element $\tau$ as
\begin{equation} \label{eq:mergedDescription}
\begin{split}
    \textbf{D}(\textbf{r}, t) &= \sum_{I, J} \sum_{I'} \int_{\mathcal{R}(J)} \sqrt{\frac{\hbar \omega_J}{4 \pi v_{I,J}^{\nu}}} \bar{\textbf{d}}_{I,J}(\textbf{r}_{\perp}, \xi) \\
    & \qquad \qquad \qquad \times \bar{g}_{I,I',J}^{\nu, \tau}(\xi; \omega) a_{I',J, \nu}^{in}(\omega, t) + H.c. \\
    &= \sum_{I, J} \int_{\mathcal{R}(J)} \sqrt{\frac{\hbar \omega_J}{4 \pi v_{I,J}^{\nu}}} \textbf{d}_{I,J}^{\tau}(\textbf{r}_{\perp}, \xi) h_{I, J, \omega}^{in, \nu, \tau}(\xi) \\
    & \qquad \qquad \qquad \times a_{I,J, \nu}^{in}(\omega, t) e^{i \beta_{I, J}^{\tau} \xi^{\tau}} + H.c. \\
\end{split}
\end{equation}
for some $\bar{g}_{I,I',J}^{\nu, \tau}(\xi; \omega)$, where we have chosen $h_{I, J, \omega}^{in, \nu, \tau}(\xi)$, $\textbf{d}_{I,J}^{\tau}(\textbf{r}_{\perp}, \xi)$, and $\beta_{I,J}^{\tau}$ such that
\begin{equation}
\begin{split}
    &\sqrt{\frac{1}{v_{I,J}^{\nu}}} \textbf{d}_{I,J}^{\tau}(\textbf{r}_{\perp}, \xi) h_{I, J, \omega}^{in, \nu, \tau}(\xi) e^{i \beta_{I, J}^{\tau} \xi^{\tau}}  \\
    & \qquad = \sum_{I'} \sqrt{\frac{1}{v_{I',J}^{\nu}}} \bar{\textbf{d}}_{I,J}(\textbf{r}_{\perp}, \xi) \bar{g}_{I,I',J}^{\nu, \tau}(\xi; \omega),
\end{split}
\end{equation}
with $h_{I, J, \omega}^{in, \nu, \tau}(\xi)$ having a slowly varying phase and $\textbf{d}_{I,J}^{\tau}(\textbf{r}_{\perp}, \xi)$ satisfying the normalization in equation (\ref{eq:modeNormalization}).

Finally, by setting $a_{I,J, \nu}^{in}(k, t) = \sqrt{v_{I,J}^{\nu}} a_{I,J, \nu}^{in}(\omega_{I, J}(k), t)$, we can express the integration in equation (\ref{eq:mergedDescription}) in terms of $k$, from which we recover equation (\ref{eq:GeneralAsymptoticExpansion}) with the mode field expansion in equation (\ref{eq:AsymptoticFieldDistExpansion}).

\section{Phantom Loss Channels} \label{sec:LossChannelEnvironment}
In this section, we provide more details about the use of the phantom channels to model the environment to which the fields in the resonator system can couple. To begin, we can consider a section of a resonator system, supporting a single mode, around the coupling point to a phantom channel. For clarity, we follow the notation of appendix \ref{sec:asyFields}, but focus only on a single resonance range, $J$, and write the ring field operator as $a(\xi, t; \omega)$ and the phantom channel operator as $b(\xi, t; \omega)$ rather than introducing the segment index $V_{\tau}$, and drop the mode index $I$ for the moment. Letting $\xi = 0$ denote the coupling point between the ring segment and the phantom channel, we relate the fields at $\xi = 0^{\pm}$ through
\begin{equation} \label{eq:singlaModePhantomChannelCoupling}
    \begin{pmatrix}
        a(0^+, t; \omega) \\ b(0^+, t; \omega)
    \end{pmatrix} = \begin{pmatrix}
        \sigma & -i\kappa \\ -i\kappa^* & \sigma^*
    \end{pmatrix} \begin{pmatrix}
        a(0^-, t; \omega) \\ b(0^-, t; \omega)
    \end{pmatrix}.
\end{equation}
Since each phantom channel represents a distinct pathway for the fields in the ring to scatter, and from the assumptions that the input fields from each phantom channel are uncorrelated, the noise from each loss channel can be injected into the ring through the $b(z < 0, t; \omega)$ operators, but the field from the ring would only be contained within the $b(z>0, t; \omega)$ operators which are subsequently carried off to infinity without being recoupled into the system. As such, the environment introduced through the inclusion of the discrete set of phantom channels would be memory-less (Markovian). Furthermore, by taking the spacing between adjacent phantom channels to be sufficiently small, one could describe such an environment through a Langevin equation \cite{Kopylov2025}, however we do not further develop this here.

Note that by expanding the relevant fields in terms of the asymptotic-in states, we simply decompose the system in terms of the operators corresponding to the physical and phantom channel inputs, which does not affect the treatment of the environment. On the other hand, the asymptotic-out and local basis can be simply understood as linear combinations of the asymptotic-in states.

Of course a realistic description of the environment of the resonator systems considered here would necessitate the inclusion of a continuum of modes that the ring fields could couple into. Despite the relative simplicity of the phantom channel treatment, such a model can indeed represent such an environment. To see this, we can consider a generalized description of the ring to phantom channel coupling in which the single ring mode $a(\xi, t; \omega)$ can couple into a set of modes $\{b_{\lambda}(\xi, t; \omega)\}_{\lambda}$ where $\lambda$ is a continuous index denoting a variety of scattering modes. In such a case, the ring field immediately after the coupling point to the phantom channel would be written as
\begin{equation}
    a(0^+, t; \omega) = \sigma a(0^-, t; \omega) + \int d\lambda g(\lambda) b_{\lambda}(0^-, t; \omega),
\end{equation}
for some coupling strength $g(\lambda)$. Instead, such a coupling can be written as an effective single mode interaction by introducing the environment super-mode operator, $B(\xi, t; \omega)$, as
\begin{equation}
    B(\xi, t; \omega) = \frac{1}{G} \int d\lambda g(\lambda) b_{\lambda}(\xi, t;\omega), 
\end{equation}
for a normalization constant $G$. After this, the interaction at the phantom channel coupling point would be written as
\begin{equation}
    a(0^+, t; \omega) = \sigma a(0^-, t; \omega) + G B(0^-, t; \omega).
\end{equation}
From the Hermitian nature of the coupling description, it follows that one could write the coupling interaction as in Eq. (\ref{eq:singlaModePhantomChannelCoupling}) by relating $G$ to the cross-coupling coefficient $\kappa$.

In the case when one considers a set of ring modes indexed by $I$, the ring field immediately after the coupling point could be written generally as
\begin{equation}
\begin{split}
    a_I(0^+, t; \omega) &= \sum_{I'} \sigma_{I, I'} a_{I'}(0^-, t; \omega) \\
    & \qquad + \int d\lambda g_I(\lambda) b_{\lambda}(0^-, t; \omega),
\end{split}
\end{equation}
which, in a similar way to the single mode case, would lead to a set of environment super-modes $\{B_I(\xi, t; \omega)\}_I$ corresponding to the different coupling strengths $g_I(\lambda)$. However, the set of $\{B_I(\xi, t; \omega)\}_I$ may not, in general, correspond to an orthogonal basis of modes. Rather, one could determine the span of the set $\{B_I(\xi, t; \omega)\}_I$, then define an orthogonal basis of corresponding to this subspace, $\{\tilde{B}_{\tilde{I}}(\xi, t; \omega)\}_{\tilde{I}}$, after which the coupling interaction could be reduced to an interaction between the set of ring modes $\{a_I(\xi, t;\omega)\}_I$ and the set of environment super-modes $\{\tilde{B}_{\tilde{I}}(\xi, t; \omega)\}_{\tilde{I}}$.
\section{Dual ring Transmission} \label{sec:SimpilifiedTransmission}
\begin{figure*}
    \centering
    \includegraphics[width=0.95\linewidth]{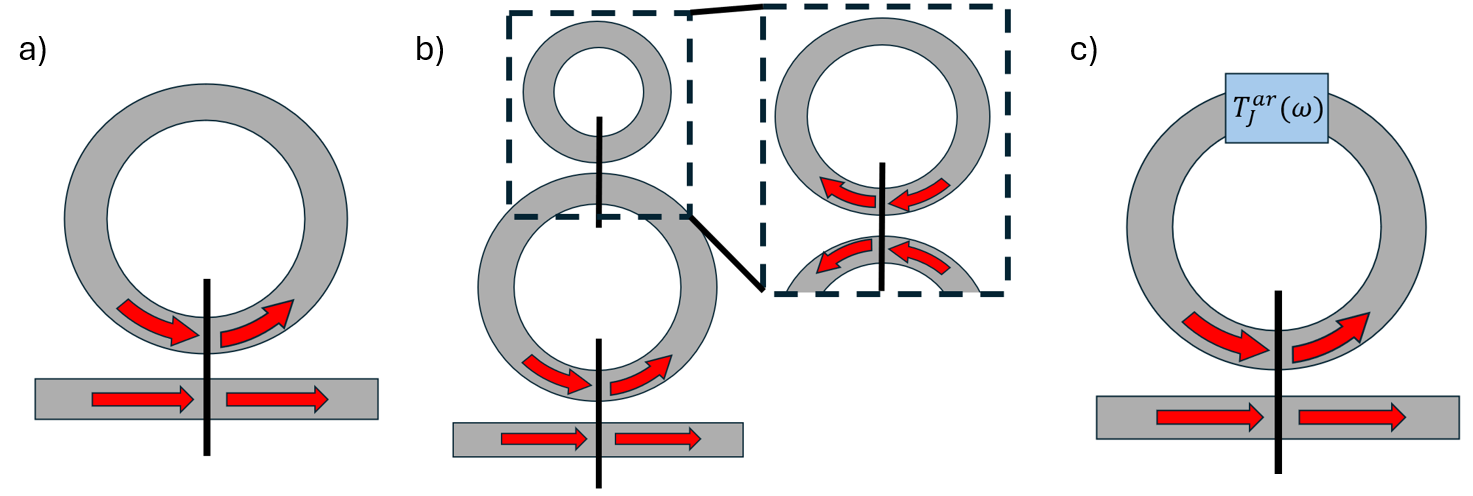}
    \caption{(a) Diagram of a single ring resonator point coupled to an input/output waveguide. (b) Simplified dual ring structure in which the coupling between each element is taken to occur at a single point. (c) Effective single ring structure used to derive the transmission of the dual ring system, with the effect of the auxiliary ring captured by the $T_J^{ar}(\omega)$ factor.}
    \label{fig:simplifiedAppendixRings}
\end{figure*}

Here we derive the transmission of the simplified dual ring structure in which we take the coupling between the input/output waveguide and the primary ring, as well as that between the primary ring and the auxiliary ring, to occur at a point (see Fig. \ref{fig:simplifiedAppendixRings}(b)). Furthermore, we take the coupling between different waveguide modes along the resonator bends and coupler to be negligible, and only consider a single $I$.

To begin, we consider the simpler example of a single lossless ring point-coupled to the input/output waveguide as shown in Fig. \ref{fig:simplifiedAppendixRings}(a); this corresponds to an auxiliary ring completely decoupled from the photonic molecule system. In such a case, we can relate the fields on either side of the coupling point with the waveguide through
\begin{equation} \label{eq:PointCoupler}
    \begin{pmatrix}
        a_{J}^{V_{wg,out}}(0^+, t; \omega) \\ a_{J}^{V_{ring}}(0^+, t; \omega)
    \end{pmatrix} = \begin{pmatrix}
        \varsigma & i\varkappa \\ i\varkappa & \varsigma
    \end{pmatrix} \begin{pmatrix}
        a_{J}^{V_{wg,in}}(0^-, t; \omega) \\ a_{J}^{V_{ring}}(L_r^-, t; \omega)
    \end{pmatrix},
\end{equation}
where, using the notation of Appendix \ref{sec:asyFields}, we have $V_{wg,in(out)}$ corresponding to the input (output) sections of the waveguide and $V_{ring}$ the segment corresponding to the full length of the ring. Additionally, we take $L_r$ as the length of the ring, and $\varsigma$ and $\varkappa$ as the effective self- and cross-coupling coefficients associated with the coupler. Having assumed the system is lossless and single mode, we can then relate $a_{J}^{V_{ring}}(L_r^-, t; \omega)$ to $a_{J}^{V_{ring}}(0^+, t; \omega)$ as
\begin{equation} \label{eq:propagationPhase}
    a_{J}^{V_{ring}}(L_r^-, t; \omega) = a_{J}^{V_{ring}}(0^+, t; \omega) e^{i \phi_J(\omega)},
\end{equation}
where $\phi_J(\omega)$ corresponds to the propagation phase around the length of the ring. Using equation (\ref{eq:propagationPhase}) in equation (\ref{eq:PointCoupler}), we can solve for $a_{J}^{V_{wg,out}}(0^+, t; \omega)$ in terms of $a_{J}^{V_{wg,in}}(0^-, t; \omega)$, resulting in
\begin{equation} \label{eq:LosslessSingleRingTransmission}
    \begin{split}
        a_{J}^{V_{wg,out}}(0^+, t; \omega) &= \frac{\varsigma - e^{i \phi_J(\omega)}}{1 - \varsigma e^{i \phi(\omega)}} a_{J}^{V_{wg,in}}(0^-, t; \omega)
    \end{split}
\end{equation}

When adding back the loss channels, solving for the asymptotic-in state for the waveguide channel corresponds to setting the input in each loss channel input port to the vacuum. Passing through a coupling point with a phantom channel coupled to the ring would then attenuate the amplitude of the ring field by a factor $\sigma_i^{ph}$, where $\sigma_i^{ph}$ corresponds to the self-coupling factor associated with the $i^{th}$ phantom channel. Consequently, the transmission for the lossy single ring is identical to equation (\ref{eq:LosslessSingleRingTransmission}) upon taking $e^{i \phi_J(\omega)} \rightarrow \Gamma e^{i \phi_J(\omega)}$ with $\Gamma = \prod_i \sigma_i^{ph}$.

Returning now to the dual ring structure, we can note that the region of the system containing the auxiliary ring and the interval around the primary/auxiliary coupling point in the primary ring resembles the single ring case (see the inlay in Fig. \ref{fig:simplifiedAppendixRings}(b)). Thus, when solving for the transmission for an input in the waveguide, we can treat the dual ring system as the reduced structure shown in Fig. \ref{fig:simplifiedAppendixRings}(c) in which the effect of the auxiliary resonator is captured by the transmission function $T_J^{ar}(\omega)$ given in equation (\ref{eq:AuxiliaryRingTransmission}). The remaining reduced system is then itself an effective single ring, with the auxiliary contribution resulting in $\Gamma_J^{pr} e^{i \phi^{pr}_J(\omega)} \rightarrow \Gamma_J^{pr} T_J^{ar}(\omega) e^{i \phi^{pr}_J(\omega)}$ in the lossy single ring transmission. This would then lead to the waveguide asymptotic-in field contribution in the waveguide output port given by equation (\ref{eq:SimplifiedTransmissionField}).

\section{Non-classical Equation of Motion Terms} \label{sec:EOMappendix}
In this section, we list the $A_{J, \nu, \nu'}^{\chi}(k, t)$ terms found in the equations of motion for the non-classical resonances, $J= LI$, $S$, and $RI$ (see eq. (\ref{eq:NonClassicalEOM})). From eq. (\ref{eq:generalEOM}), we find
\begin{widetext}
    \begin{equation}
        \begin{split}
            A^{XPM_i}_{J,\nu, \nu'}(k,t) &= 4 \sum_{\vec{\nu}} \delta_{\nu'}^{\nu_3} \left[ \mathcal{C}_J(k) \right]_{\nu, \nu_1} \bar{\Lambda}_{(J, P_i, J, P_i), \vec{\nu}}^{loc} \int dk_2 dk_4 \alpha_{P_i, \nu_2}^{loc *}(k_2, t) \alpha_{P_i, \nu_4}^{loc}(k_4, t) e^{i \Delta_{(J, P_i, J, P_i)} t} \\
            A^{DP}_{S,\nu, \nu'}(k,t) &= 2 \sum_{\vec{\nu}} \delta_{\nu'}^{\nu_2} \left[ \mathcal{C}_S(k) \right]_{\nu, \nu_1} \bar{\Lambda}_{(S, S, P_1, P_2), \vec{\nu}}^{loc} \int dk_3 dk_4 \alpha_{P_1, \nu_3}^{loc}(k_3, t) \alpha_{P_2, \nu_4}^{loc}(k_4, t) e^{i \Delta_{(S, S, P_1, P_2)} t} \\
            A^{SP_1}_{LI,\nu, \nu'}(k,t) &= 2 \sum_{\vec{\nu}} \delta_{\nu'}^{\nu_2} \left[ \mathcal{C}_{LI}(k) \right]_{\nu, \nu_1} \bar{\Lambda}_{(LI, S, P_1, P_1), \vec{\nu}}^{loc} \int dk_3 dk_4 \alpha_{P_1, \nu_3}^{loc}(k_3, t) \alpha_{P_1, \nu_4}^{loc}(k_4, t) e^{i \Delta_{(LI, S, P_1, P_1)} t} \\
            A^{SP_1}_{S,\nu, \nu'}(k,t) &= 2 \sum_{\vec{\nu}} \delta_{\nu'}^{\nu_2} \left[ \mathcal{C}_{S}(k) \right]_{\nu, \nu_1} \bar{\Lambda}_{(S, LI, P_1, P_1), \vec{\nu}}^{loc} \int dk_3 dk_4 \alpha_{P_1, \nu_3}^{loc}(k_3, t) \alpha_{P_1, \nu_4}^{loc}(k_4, t) e^{i \Delta_{(S, LI, P_1, P_1)} t} \\
            A^{SP_2}_{S,\nu, \nu'}(k,t) &= 2 \sum_{\vec{\nu}} \delta_{\nu'}^{\nu_2} \left[ \mathcal{C}_{S}(k) \right]_{\nu, \nu_1} \bar{\Lambda}_{(S, RI, P_2, P_2), \vec{\nu}}^{loc} \int dk_3 dk_4 \alpha_{P_2, \nu_3}^{loc}(k_3, t) \alpha_{P_2, \nu_4}^{loc}(k_4, t) e^{i \Delta_{(S, RI, P_2, P_2)} t} \\
            A^{SP_2}_{RI,\nu, \nu'}(k,t) &= 2 \sum_{\vec{\nu}} \delta_{\nu'}^{\nu_2} \left[ \mathcal{C}_{RI}(k) \right]_{\nu, \nu_1} \bar{\Lambda}_{(RI, S, P_2, P_2), \vec{\nu}}^{loc} \int dk_3 dk_4 \alpha_{P_2, \nu_3}^{loc}(k_3, t) \alpha_{P_2, \nu_4}^{loc}(k_4, t) e^{i \Delta_{(RI, S, P_2, P_2)} t} \\
            A^{HP}_{LI,\nu, \nu'}(k,t) &= 4 \sum_{\vec{\nu}} \delta_{\nu'}^{\nu_2} \left[ \mathcal{C}_{LI}(k) \right]_{\nu, \nu_1} \bar{\Lambda}_{(LI, RI, P_1, P_2), \vec{\nu}}^{loc} \int dk_3 dk_4 \alpha_{P_1, \nu_3}^{loc}(k_3, t) \alpha_{P_2, \nu_4}^{loc}(k_4, t) e^{i \Delta_{(LI, RI, P_1, P_2)} t} \\
            A^{HP}_{RI,\nu, \nu'}(k,t) &= 4 \sum_{\vec{\nu}} \delta_{\nu'}^{\nu_2} \left[ \mathcal{C}_{RI}(k) \right]_{\nu, \nu_1} \bar{\Lambda}_{(RI, LI, P_1, P_2), \vec{\nu}}^{loc} \int dk_3 dk_4 \alpha_{P_1, \nu_3}^{loc}(k_3, t) \alpha_{P_2, \nu_4}^{loc}(k_4, t) e^{i \Delta_{(RI, LI, P_1, P_2)} t} \\
            A^{BS_1}_{LI,\nu, \nu'}(k,t) &= 4 \sum_{\vec{\nu}} \delta_{\nu'}^{\nu_3} \left[ \mathcal{C}_{LI}(k) \right]_{\nu, \nu_1} \bar{\Lambda}_{(LI, P_2, S, P_1), \vec{\nu}}^{loc} \int dk_2 dk_4 \alpha_{P_2, \nu_2}^{loc *}(k_2, t) \alpha_{P_1, \nu_4}^{loc}(k_4, t) e^{i \Delta_{(LI, P_2, S, P_1)} t} \\
            A^{BS_1}_{S,\nu, \nu'}(k,t) &= 4 \sum_{\vec{\nu}} \delta_{\nu'}^{\nu_3} \left[ \mathcal{C}_{S}(k) \right]_{\nu, \nu_1} \bar{\Lambda}_{(S, P_1, LI, P_2), \vec{\nu}}^{loc} \int dk_2 dk_4 \alpha_{P_1, \nu_2}^{loc *}(k_2, t) \alpha_{P_2, \nu_4}^{loc}(k_4, t) e^{i \Delta_{(S, P_1, LI, P_2)} t} \\
            A^{BS_2}_{S,\nu, \nu'}(k,t) &= 4 \sum_{\vec{\nu}} \delta_{\nu'}^{\nu_3} \left[ \mathcal{C}_{S}(k) \right]_{\nu, \nu_1} \bar{\Lambda}_{(S, P_2, RI, P_1), \vec{\nu}}^{loc} \int dk_2 dk_4 \alpha_{P_2, \nu_2}^{loc *}(k_2, t) \alpha_{P_1, \nu_4}^{loc}(k_4, t) e^{i \Delta_{(S, P_2, RI, P_1)} t} \\
            A^{BS_2}_{RI,\nu, \nu'}(k,t) &= 4 \sum_{\vec{\nu}} \delta_{\nu'}^{\nu_3} \left[ \mathcal{C}_{RI}(k) \right]_{\nu, \nu_1} \bar{\Lambda}_{(RI, P_1, S, P_2), \vec{\nu}}^{loc} \int dk_2 dk_4 \alpha_{P_1, \nu_2}^{loc *}(k_2, t) \alpha_{P_2, \nu_4}^{loc}(k_4, t) e^{i \Delta_{(RI, P_1, S, P_2)} t} \\
        \end{split}
    \end{equation}
\end{widetext}
\section{Characteristic Functions} \label{sec:appendix1}
In this section, we derive the Wigner characteristic functions for the full output state of the signal and idler fields (including the contributions within the phantom channels), as well as the Wigner characteristic function for the reduced state in the subspace $\mathcal{H}_A \subseteq \mathcal{H}$ where $\mathcal{H} = \mathcal{H}_A \otimes \mathcal{H}_B$ for $\mathcal{H}_A$ and $\mathcal{H}_B$ representing arbitrary orthogonal spaces. The derivation will follow the method presented in \cite{Quesada:22}.

Let us consider the case in which, at some initial time $t_i$, the signal and idler fields are in the vacuum. Denoting by $\rho(t)$ the density operator representing the signal and idler fields in the asymptotic-out basis, we write 
\begin{equation}
    \rho(t_i) = |vac\rangle \langle vac |.
\end{equation}
For a unitary operator, $\mathcal{U}$, which describes the evolution of the fields from the initial time $t_i$ to some final time $t_f$ such that $\rho (t_f) = \mathcal{U} \rho (t_i) \mathcal{U}^{\dagger} = \mathcal{U}  |vac\rangle \langle vac |\mathcal{U}^{\dagger}$, the action on the Schrodinger picture bosonic creation and annihilation operators in the asymptotic-out basis are, from equation (\ref{eq:ExpandedEvolution}),
\begin{equation}
\begin{split}
    \mathcal{U}^{\dagger}a^{out}_{J,\nu}(k) \mathcal{U} = \sum_{J'}' \sum_{\nu', k'} & \left[ V_{\nu, \nu'}^{J, J'}(k, k') a^{out}_{J', \nu'}(k') \right. \\
    & \quad \left.  + W_{\nu, \nu'}^{J, J'}(k, k') a^{out \dagger}_{J', \nu'}(k') \right] \\
    \mathcal{U}^{\dagger} a^{out \dagger}_{J,\nu, k} \mathcal{U} = \sum_{J'}' \sum_{\nu', k'} & \left[ V_{\nu, \nu'}^{J, J' *}(k, k') a^{out \dagger}_{J', \nu'}(k') \right. \\
    & \quad \left.  + W_{\nu, \nu'}^{J, J' *}(k, k') a^{out}_{J', \nu'}(k') \right]
\end{split}
\end{equation}
From the definition of the Wigner characteristic function of the state $\rho(t_f)$, we then have
\begin{equation} \label{eq:CharacteristicFunctionFullState}
\begin{split}
    \chi_W(\{ \alpha_{J,\nu}^k \}) &= \Tr [\rho (t_f) \mathcal{D}(\{ \alpha_{J,\nu}^k \})] \\
    &= \Tr [\mathcal{U}|vac \rangle \langle vac | \mathcal{U}^{\dagger} \mathcal{D}(\{ \alpha_{J,\nu}^k \})] \\
    &= \langle vac | \mathcal{U}^{\dagger} \mathcal{D}(\{ \alpha_{J,\nu}^k \}) \mathcal{U} | vac \rangle
\end{split}
\end{equation}
where $\mathcal{D}(\{ \alpha_{J,\nu}^k \})$ is the displacement operator on the signal and idler modes, given by
\begin{equation}
    \mathcal{D}(\{ \alpha_{J,\nu}^k \}) = \exp \left[ \sum_{J}' \sum_{\nu, k} \left( \alpha_{J,\nu}^k a^{out \dagger}_{J, \nu}(k) - \alpha_{J,\nu}^{k*} a^{out}_{J, \nu}(k) \right) \right].
\end{equation}
the action of $\mathcal{U}$ on $\mathcal{D}(\{ \alpha_{J,\nu}^k \})$ can then be computed as
\begin{equation}
\begin{split}
    \mathcal{U}^{\dagger} \mathcal{D}(\{ \alpha_{J,\nu}^k \}) \mathcal{U} &= \exp \left[ \sum_{J}' \sum_{\nu, k} \alpha_{J,\nu}^k \mathcal{U}^{\dagger} a^{out \dagger}_{J, \nu}(k) \mathcal{U} \right.  \\
    & \qquad \qquad \qquad \left. - \sum_{J}' \sum_{\nu, k} \alpha_{J,\nu}^{k*} \mathcal{U}^{\dagger} a^{out}_{J, \nu}(k) \mathcal{U}  \right] \\
    &= \exp \left[ \sum_{J,}' \sum_{\nu, k} \left( \tilde{\alpha}_{J, \nu}^k a_{J, \nu}^{out \dagger}(k) - \tilde{\alpha}_{J, k}^{k*} a_{J, \nu}^{out}(k) \right) \right] \\
    &= \mathcal{D}(\{ \tilde{\alpha}_{J,\nu}^k \}),
\end{split}
\end{equation}
where,
\begin{equation}
    \tilde{\alpha}_{J,\nu}^k = \sum_{J'}'\sum_{\nu', k'} \left(\alpha_{J',\nu'}^{k'} V_{\nu',\nu}^{J',J *}(k',k) - \alpha_{J',\nu'}^{k'*} W_{\nu',\nu}^{J',J}(k',k) \right),
\end{equation}
From the disentangling theorem, we then immediately get that the characteristic function is
\begin{equation} \label{eq:CharacteristicFunction_fromAppend}
\begin{split}
    &\chi_W(\{ \alpha_{J,\nu}^k \}) = \exp \left\{ -\frac{1}{2} \sum_J' \sum_{\nu, k} \tilde{\alpha}_{J,\nu}^k \tilde{\alpha}_{J, \nu}^{k*} \right\} \\
    & \qquad = \exp \left\{ -\frac{1}{2} \begin{pmatrix}
        \pmb{\alpha}^{\dagger} & \pmb{\alpha}^{T}
    \end{pmatrix} \begin{pmatrix}
        \frac{1}{2} \mathbb{I} + \textbf{N}^* & -\textbf{M} \\ -\textbf{M}^* & \frac{1}{2} \mathbb{I} + \textbf{N}
    \end{pmatrix} \begin{pmatrix}
        \pmb{\alpha} \\ \pmb{\alpha}^*
    \end{pmatrix}  
    \right\}. 
\end{split}
\end{equation}
where we have used the commutation relations of the asymptotic-out operators and expanded out the definition of $\tilde{\alpha}_{J,\nu}^k$. From equation (\ref{eq:CharacteristicFunction_fromAppend}), it is clear that one needs only $N_{\nu,\nu'}^{J,J'}(k,k')$ and $M_{\nu,\nu'}^{J,J'}(k,k')$ to describe the signal and idler fields at $t=t_f$, and that such a state corresponds to a Gaussian state centered at a mean of zero.

Consider now the partition of the Hilbert space as $\mathcal{H} = \mathcal{H}_A \otimes \mathcal{H}_B$, where for the moment we will let $\mathcal{H}_A$ and $\mathcal{H}_B$ be arbitrary orthogonal spaces with $\dim(\mathcal{H}_A) = N_A$ and $\dim(\mathcal{H}_B) = N_B$ such that $N_A + N_B = N = \dim(\mathcal{H})$. Furthermore, let $\{a_i\}_{i=1}^N$ be an arbitrary ordering of the operators $a^{out}_{J,\nu}(k)$ for $J \in \{LI, S, RI\}$ which forms a basis of $\mathcal{H}$, and let $\{b^A_i\}_{i=1}^{N_A}$ and $\{b^B_i\}_{i=1}^{N_B}$ be some mode basis of $\mathcal{H}_A$ and $\mathcal{H}_B$. We then define the transformation $U_{\mathcal{H}}$ such that
\begin{equation} \label{eq:SubspaceTransformation}
    \begin{pmatrix}
        b^A_1 \\ \vdots \\ b^A_{N_A} \\ b^B_1 \\ \vdots \\ b^B_{N_A}
    \end{pmatrix} = U_{\mathcal{H}} \begin{pmatrix}
        a_1 \\ \vdots \\ a_N
    \end{pmatrix}
\end{equation}

Tracing over the operators in $\mathcal{H}_B$, we can write the reduced density operator in $\mathcal{H}_A$ as $\rho_A^R(t) = \Tr_B [\rho(t)]$, from which the reduced characteristic function is given by 
\begin{equation}
\begin{split}
    \chi_W^A \left(\{\beta^A_i \} \right) &= \Tr_A \left[ \rho^R_A(t_f) \mathcal{D}_A(\{\beta^A_i \}) \right] \\
    &= \Tr \left[ \rho(t_f) \left( \mathcal{D}_A(\{\beta^A_i \}) \otimes \mathcal{I}_B \right) \right] \\
    &= \langle vac| \mathcal{U}^{\dagger} \left( \mathcal{D}_A(\{\beta^A_i \}) \otimes \mathcal{I}_B \right) \mathcal{U} | vac \rangle,
\end{split}
\end{equation}
where,
\begin{equation}
    \mathcal{D}(\{\beta^A_i \}) = \exp \left[ \sum_{i=1}^{N_A} \left( \beta^A_i b^{A\dagger}_{i} - \beta^{A*}_i b^{A}_{i} \right) \right].
\end{equation}
Using equation (\ref{eq:SubspaceTransformation}) and following a similar process as for the full state, we can write the reduced characteristic function as
\begin{equation} \label{eq:ReducedCharacteristicFunction_append}
\begin{split}
    &\chi_W^A(\{\beta^A_i \})  \\
    & = \exp \left\{ -\frac{1}{2} \begin{pmatrix}
        \pmb{\beta}^{\dagger} & \pmb{\beta}^{T}
    \end{pmatrix} \begin{pmatrix}
        \frac{1}{2} \mathbb{I} + \tilde{\textbf{N}}_A^* & -\tilde{\textbf{M}}_A \\ -\tilde{\textbf{M}}_A^* & \frac{1}{2} \mathbb{I} + \tilde{\textbf{N}}_A
    \end{pmatrix} \begin{pmatrix}
        \pmb{\beta} \\ \pmb{\beta}^*
    \end{pmatrix}  
    \right\} 
\end{split}
\end{equation}
where for $\mathcal{P}_A$ representing the projector onto the column space of the $N_A$ first columns, we can write
\begin{equation} \label{eq:ReducedMoments}
\begin{split}
    \tilde{\textbf{N}}_A &= \mathcal{P}_A U_{\mathcal{H}}^* \textbf{N} U_{\mathcal{H}}^T \mathcal{P}_A^T \\
    \tilde{\textbf{M}}_A &= \mathcal{P}_A U_{\mathcal{H}} \textbf{N} U_{\mathcal{H}}^T \mathcal{P}_A^T
\end{split}
\end{equation}

We can then note that, in the case when the subspace $\mathcal{H}_A$ corresponds to the operators describing the signal field in the output channel ($a^{out}_{J, \nu}(k)$ for $J=S$ and $\nu$ corresponding to a given output port), then we can chose for $U_{\mathcal{H}}$ to be a permutation matrix that organizes the $N_K$ operators in $\mathcal{H}_A$ as the set $\{b^{A}_i\}_{i=1}^{N_k}$, from which equation (\ref{eq:ReducedCharacteristicFunction_append}) and (\ref{eq:ReducedMoments}) immediately reduce to equation (\ref{eq:ReducedCharacteristicFunction}).
\section{Characteristic Function Decomposition} \label{sec:appendix2}
Let us consider the decomposition of the matrix $\pmb{\Sigma}$ in equation (\ref{eq:QuadratureCharacteristicFunction}). First, we note that $\mathcal{K}^{out}(t_i, t_f)$ describes the unitary evolution of the asymptotic-out operators from some initial time $t_i$ to a final time $t_f$, and as such preserves the bosonic commutation relations of the operator. By reordering the elements of $\mathcal{K}^{out}(t_i, t_f)$ into a matrix $\bar{\mathcal{K}}^{out}(t_i, t_f)$ such that 
\begin{equation}
    \begin{pmatrix}
        \textbf{a}(t_f) \\ \textbf{a}^{\dagger}(t_f)
    \end{pmatrix} = \bar{\mathcal{K}}^{out}(t_i, t_f) \begin{pmatrix}
        \textbf{a}(t_i) \\ \textbf{a}^{\dagger}(t_i)
    \end{pmatrix} 
\end{equation}
where $\textbf{a}(t)$ is some ordering of the operators $a^{out}_{J,\nu}(k,t)$, then $\bar{\mathcal{K}}^{out}(t_i, t_f)$ represents a Bogliubov transformation that can be decomposed as \cite{Quesada:22}
\begin{equation} \label{eq:KblockDecomp}
    \bar{\mathcal{K}}^{out}(t_i, t_f) = \begin{pmatrix}
        U & 0 \\ 0 & U^*
    \end{pmatrix} \begin{pmatrix}
        \mathcal{C} & \mathcal{S} \\ \mathcal{S} & \mathcal{C}
    \end{pmatrix} \begin{pmatrix}
        U' & 0 \\ 0 & U'^*
    \end{pmatrix}
\end{equation}
for $\mathcal{S} = \otimes_i \sinh (r_i)$ and $\mathcal{C} = \otimes_i \cosh (r_i)$, where each $r_i$ is real and positive, with $U$ and $U'$ unitary matrices.

Choosing the ordering of $\pmb{\alpha}$ in equation (\ref{eq:CharacteristicFunction_NMdef}) to be the same as $\textbf{a}(t)$, it then follows that
\begin{equation}
    \begin{split}
        \textbf{N} &= U^* \mathcal{S}^2 U^T, \\
        \textbf{M} &= U \mathcal{S} \mathcal{C} U^T.
    \end{split}
\end{equation}
From this, we can set
\begin{equation}
    \mathcal{O}_{\Sigma} = \begin{pmatrix}
        \Re \{U\} & -\Im \{U\} \\
        \Im \{U\} & \Re \{U\}
    \end{pmatrix},
\end{equation}
which one can easily verify is real, orthogonal, and symplectic, after which we immediately recover equation (\ref{eq:FullSqueezedState}). On the other hand, we can set $\mathcal{O}_S = \mathcal{O}_{\Sigma}$, $\mathcal{O}_S' = \mathbb{I}$, $R_S = \oplus_i e^{r_i}$, and $D_S = \frac{1}{2} \mathbb{I}$ to recover the expansion of $\pmb{\Sigma}$ following a Williamson and Bloch-Messiah decomposition (see equation (\ref{eq:ReducedCharacteristicDecomp})). As a result, we find from equation (\ref{eq:ThermAndSqueezedPhotonDef}) that $n^{th}_{tot} = 0$ for the full state at $t=t_f$ as expected. 

Now we consider the decomposition of the Hilbert space as $\mathcal{H} = \mathcal{H}_A \otimes \mathcal{H}_B$, and consider the reduced state in $\mathcal{H}_A$. Upon performing a Williamson and Bloch-Messiah decomposition on the matrix $\tilde{\pmb{\Sigma}}$ describing the reduced state, we can recover the matrices $\tilde{D}_S$, $\tilde{R}_S$, $\tilde{\mathcal{O}}_S$ and $\tilde{\mathcal{O}}_S'$, but it is not immediately clear how to obtain the values of $\tilde{D}_S$ from the corresponding matrices describing the full state discussed above. However, we can still construct inequalities relating the values of $D_S$ for the full state to $\tilde{D}_S$ of the reduced state. In particular, letting $\mathcal{H}_B$ be a one dimensional subspace of $\mathcal{H}$, we can write $D_S = diag \{ d_1, \dots, d_N \}$ with $d_1 \geq d_2 \geq \dots \geq d_N$ and $\tilde{D}_S = diag \{ \tilde{d}_1, \dots, \tilde{d}_{N-1} \}$ with $\tilde{d}_1 \geq \tilde{d}_2 \geq \dots \geq \tilde{d}_{N-1}$, from which the interlacing theorem for symplectic eigenvalues states that \cite{Interlacing}
\begin{equation}
\begin{split}
    \tilde{d_1} &\geq d_2 \\
    d_i &\geq \tilde{d}_{i+1} \geq d_{i+2} \qquad \textit{for } 1 \leq i \leq N-2
\end{split}
\end{equation}
So noting that in our case $d_i = \frac{1}{2}$ for all $i$, repeatedly tracing out a single mode and applying the interlacing theorem show that $\tilde{D}_S \geq \frac{1}{2} \mathbb{I}$ for $\tilde{D}_S$ corresponding to any  subspace $\mathcal{H}_A \subseteq \mathcal{H}$. 

Furthermore, in the case in which $\mathcal{H}_A$ and $\mathcal{H}_B$ represent disjoint subspaces with no coupling between the operators $b_i^A \in \mathcal{H}_A$ and $b_i^B \in \mathcal{H}_B$, then the matrix $\tilde{\pmb{\Sigma}}$ representing the reduced state in $\mathcal{H}_A$ would have a decomposition resulting in $\tilde{D}_S = \frac{1}{2} \mathbb{I}$ and as such would correspond to a pure multi-mode squeezed state. This can be seen by noting that, under such an assumption, we could write the evolution operator from $t_i$ to $t_f$ as $\mathcal{U} = \mathcal{U}_A \oplus \mathcal{U}_B$ with $\mathcal{U}_A$ acting on the operators in $\mathcal{H}_A$ and $\mathcal{U}_B$ on the operators in $\mathcal{H}_B$. As a result, we could decompose $\bar{\mathcal{K}}^{out}(t_i, t_f) = \bar{\mathcal{K}}^{out}_A(t_i, t_f) \oplus \bar{\mathcal{K}}^{out}_B(t_i, t_f)$ with each block being expressed as in equation (\ref{eq:KblockDecomp}). Tracing out the subspace $\mathcal{H}_B$ would then leave the reduced state in $\mathcal{H}_A$ unchanged, resulting in a pure squeezed state. Of course, including coupling between the subspaces $\mathcal{H}_A$ and $\mathcal{H}_B$ would, in general, result in a squeezed thermal state in the reduced system with some non-zero thermal photon number, $n^{th}_{tot} > 0$.
\section{Numerical Convergence of Squeezing Method} \label{section:appendix_convergence}
In this appendix, we provide an example of the convergence of the asymptotic scattering method, developed in Section \ref{sec:SectionII} and \ref{sec:SectionIII}, in terms of the number of $k$-bins per resonance, time steps, and loss channels per ring for the dual-ring structure considered in Section \ref{sec:SectionV}. In particular, we consider here the scheme from Example 1 in which the auxiliary ring causes an effective splitting of the $J=LI$ and $J=RI$ resonances, and fix the ring cross-coupling value to $\varkappa_{J}^{aux} = 0.015$. The resulting splitting from such a cross-coupling value corresponds to an intermediate regime in which the parasitic SFWM interactions coupling the fields in the $J=LI, RI$ resonances to those in the $J=S$ resonance are partially suppressed, but still result in a significant reduction in the maximum output squeezing. In all cases, the total propagation time, size of the resonances ranges, $R(J)$, and the total round trip attenuation from the loss channels, $\Gamma_J^{pr}$ and $\Gamma_J^{ar}$, are fixed.

\begin{figure*}
    \centering
    \includegraphics[width=0.98\linewidth]{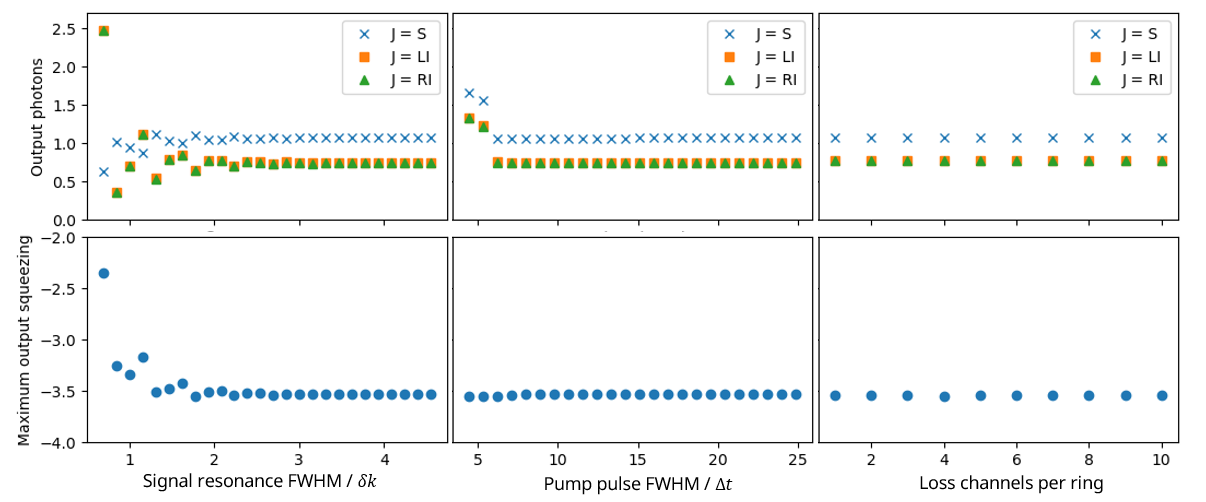}
    \caption{(top) Number of output photons per pulse in the signal and idler resonances ($J = LI, S, RI$)  and (bottom) maximum output signal squeezing while sweeping (left) the bin width, $\delta k$, (center) the length of a single time step, $\Delta T$, and (right) the number of phantom loss channels per ring resonator. In each case, the non-swept values are fixed at $\delta k = (\textit{signal resonance FWHM})/4$, $\Delta t = (\textit{pump pulse FWHM}) / 20$, with 5 loss channels per ring.}
    \label{fig:SampleConvergence}
\end{figure*}

In Fig. \ref{fig:SampleConvergence}, we show the number of photons in the $J = LI, RI,$ and $S$ resonances, as well as the maximum squeezing, within the output waveguide. Here, the $P_1$ and $P_2$ pumps are set to drive the cold cavity resonances (corresponding to the maximum field enhancement when neglecting SPM and XPM) which is slightly off from the detuning which maximizes the output squeezing. As desired, we see that for a sufficiently fine choice of the $k$-space bin width, $\delta k$, and time step, $\Delta t$, relative to the FWHM of the signal resonance and the FWHM of the pump pulse respectively, the output photon number and resulting squeezing converge.

We note also that, for the high finesse and low loss resonators considered here, the output is essentially invariant on the number of coupled phantom channels. Indeed, the low net round trip attenuation of the primary and auxiliary ring ($\Gamma_J^{pr} = 0.9991$ and $\Gamma_J^{ar} = 0.99935$ respectively in this example) due to the presence of the phantom channels result in a field intensity which is nearly uniform within each resonator. Consequentially, the strength of the nonlinear interaction coming from the overlaps of the asymptotic field distributions are negligibly affected by the discretized loss steps. Furthermore, the high finesse of the resonators ensures that the range in $k$-space around each of the resonance centers that can see appreciable field generation are sufficiently small such that the approximation in Eq. (\ref{eq:phaseMatchingAssumption}) is well satisfied even in the case of only a single loss channel coupled to each of the ring structures. Despite this, we still choose to populate each of the rings with 5 loss channels per ring in the examples presented in Section \ref{sec:SectionV}.
\section{Beyond the Five Resonance Approximation} \label{sec:appendix_nineMode}
In this section, we discuss the use of the five resonance approximation as employed in Section \ref{sec:SectionV}. For the purposes of reconstructing the output state of the desired degenerate squeezed signal, we look to include those processes which can have some appreciable impact on the fields in the signal resonance ($J = S$). As such, the inclusion of the $J=LI$ and $J=RI$ resonances in the five resonance approximation are necessary as the SP-SFWM and BS-FWM interactions couple the fields in these resonances to the fields in the signal resonance with a comparable nonlinear strength to the DP-SFWM interaction producing the squeezing. However, for an arbitrary resonator system driven by one or more pumps, SFWM interactions can, in general, result in the generation of photons in all system resonances. If the resonances not included in the five resonance approximation can see a sufficiently large gain, and if there is some phase matched and energy preserving SFWM interaction which can couple the fields in these to the signal resonance, then one would expect additional pollution of the squeezed output state beyond what is seen in the five resonance model. 

For the moment, let us consider a single ring where, for simplicity, we denote the resonances, $J$, by a discrete index rather than by a label as done previously. We choose for $J=0$ to correspond to the signal resonance, with the classical pumps stimulating the $J = \pm p$ resonances for some integer $p$. This would result in the idler resonances (previously indexed by $LI$ and $RI$) being denoted by $J = \pm 2 p$. In the case when group velocity dispersion is negligible, each of the resonances would be equally spaced with SFWM interactions corresponding to Hamiltonian terms of the form $a_{J_1}^{\dagger} a_{J_2}^{\dagger} a_{J_3} a_{J_4}$ preserving energy whenever $J_1 + J_2 - J_3 - J_4 = 0$. 

Initially, when the fields in all resonances are vacuum except for the classical pumps, the most significant SFWM interactions involving the non-pump resonances would be those that correspond to terms in the Hamiltonian which are cubic in the pump creation and annihilation operators. For the pumping scheme considered here, none of these interactions could involve the signal resonance while also preserving energy. The next highest order contributions would be those Hamiltonian terms which are quadratic in the pump operators, all of which including the signal resonance and preserving energy being present in the five resonance approximation (see Fig. \ref{fig:FiveResonanceModel}). Indeed, even when including group velocity dispersion as in Example 2 of Section \ref{sec:SectionV}, the shifts of the resonance positions due to GVD within the range of frequencies considered are on the order of the resonance linewidths. As such, quadratic pump interactions involving the resonances between those included in the five resonance approximation would not share an energy preserving interaction with the signal ($J = 0$) or idler ($J = \pm 2p$) resonances.

However, for sufficiently large pump powers, the SFWM interactions resulting in field generation within these additional resonances (that do not involve the $J = 0, \pm 2p$ resonances) could result in powers large enough that the fields act as a 'secondary pump' generated from the input fields. For the parameters considered in Section \ref{sec:SectionV}, almost all of these higher order pump interactions can safely be neglected. Indeed, for those quadratic pump interactions included in the five resonance approximation (DP-SFWM, SP-SFWM, HP-SFWM, and BS-FWM), the generation within the signal and idler resonances are seen to be on the order of single photons per pulse (see Fig. \ref{fig:splittingPlotsMerged} and Fig. \ref{fig:pushingPlotsMerged}), corresponding to powers which are many orders of magnitude weaker than the input pumps. Additional quadratic pump SFWM interactions involving other system resonances would have comparable generation rates. Within the pertubative regime, the generation rate of photons resulting from these quadratic pump interactions scale with the product of the pump powers \cite{Banic-PhysRevA.106.043707}, from which we can immediately see that such interactions involving these secondary pumps, which themselves are generated from quadratic pump interactions, would be many orders of magnitude weaker than those interactions contained within the five resonance approximation. 

\begin{figure}
    \centering
    \includegraphics[width=0.7\linewidth]{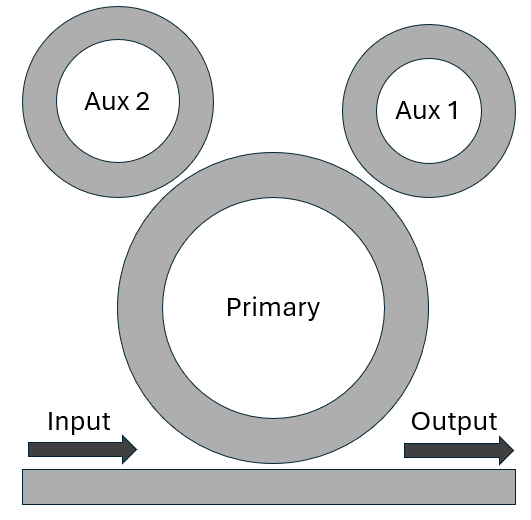}
    \caption{Diagram of a three ring photonic molecule. Here, two auxiliary rings 'Aux1' and 'Aux2' are coupled to the primary ring resonator supporting the squeezing interaction.}
    \label{fig:threeRingDiagram}
\end{figure}

\begin{figure*}
    \centering
    \includegraphics[width=0.75\linewidth]{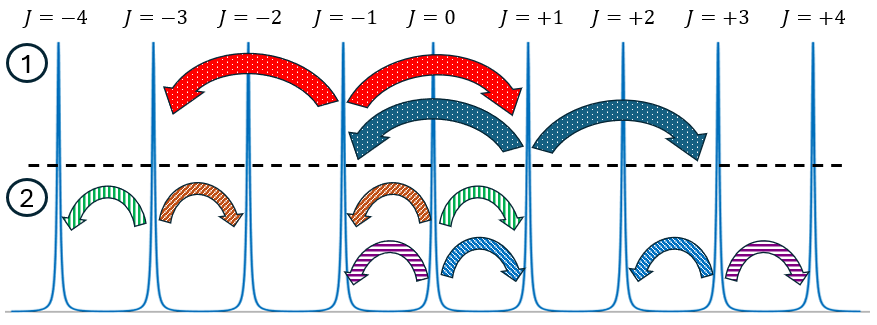}
    \caption{Diagram of the higher order parasitic SFWM interactions in a nine mode approximation. Here the $J = \pm 1$ pumps can generate a strong classical field in the $J = \pm 3$ resonances due to stimulated FWM processes show in region (1) above the dotted line. The generated classical fields in the $J= \pm 3$ resonances can then add thermal noise to the signal resonance through a variety of FWM processes, some of which are shown in region (2) below the dotted line.}
    \label{fig:nineResonanceHigherOrderParasitics}
\end{figure*}

The exception to the above discussion would be those resonances which see gain as the result of a SFWM interactions which are cubic in the pump operators. In particular, the $J = \pm 3p$ resonances can see significant power generation as a result of stimulated single pump interactions of the form $a_{3p}^{\dagger} a_{-p}^\dagger a_{p} a_p$ and $a_{-3p}^{\dagger} a_{p}^\dagger a_{-p} a_{-p}$, and can subsequently couple with the signal resonance through, for example, SFWM interactions of the form $a_0^{\dagger} a_{3p}^{\dagger} a_{-p} a_{4p} + H.c.$ and $a_0^{\dagger} a_{-3p}^{\dagger} a_{p} a_{-4p} + H.c.$, among others. To include all the relevant interactions which coupled to the desired signal field resulting from the input pumps ($J = \pm p$) as well as the secondary pumps ($J = \pm 3p$), one would need to expand the treatment to a nine resonance model, as shown in Fig. \ref{fig:nineResonanceHigherOrderParasitics} for the case of $p=1$ (pumps neighboring the signal resonance).

\begin{figure*}
    \centering
    \includegraphics[width=0.98\linewidth]{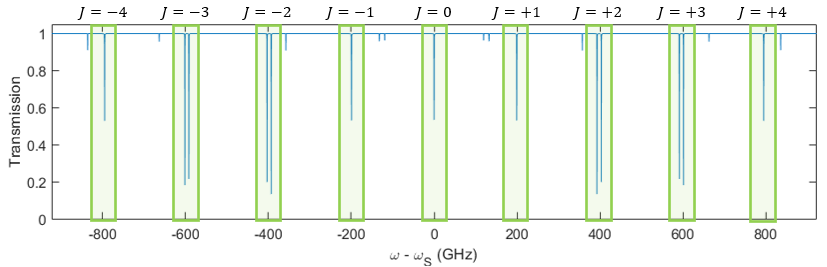}
    \caption{Diagram of the transmission spectrum for a three ring photonic molecule in which the Aux1 ring splits the $J = \pm 2$ resonances and the Aux2 ring splits the $J = \pm 3$ resonances (see Fig. \ref{fig:threeRingDiagram}).}
    \label{fig:threeRing_transmission}
\end{figure*}

\begin{figure}
    \centering
    \includegraphics[width=0.98\linewidth]{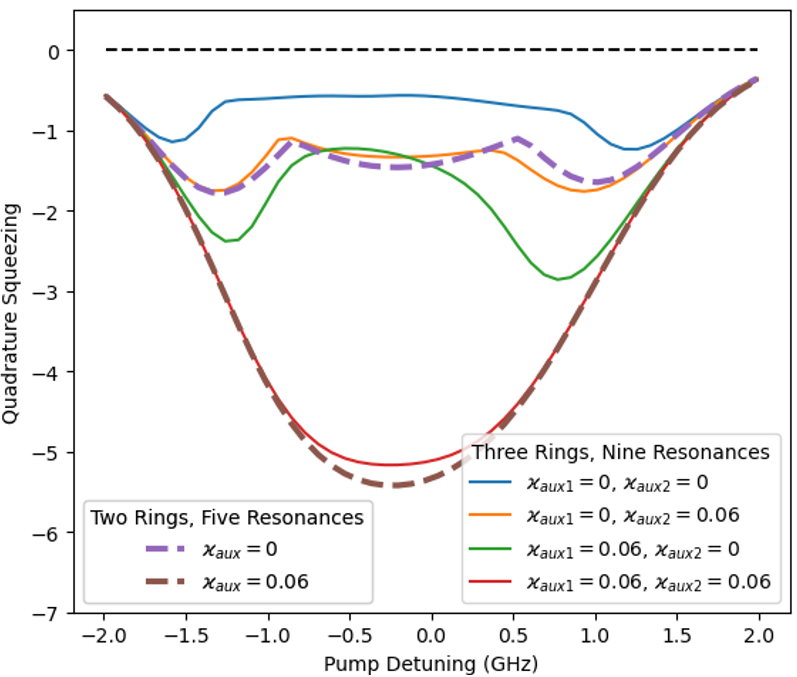}
    \caption{Quadrature squeezing of the maximally squeezed temporal mode of the reduced signal output for the case of a dual-ring photonic molecule (dashed line) and a three ring photonic molecule (solid line) as a function of the symmetric $P_1$ and $P_2$ detuning. Two ring samples use a five resonance approximation with the same parameters from Section \ref{sec:SectionV}. Three ring samples use a nine resonance approximation with the splitting scheme shown schematically in Fig. \ref{fig:threeRing_transmission}.}
    \label{fig:threeRingSqueezing}
\end{figure}

For the setup and parameters used in Example 1 from Section \ref{sec:SectionV}, where we take group velocity dispersion to be negligible, these additional interactions involving the generated secondary pumps are significant (see Fig. \ref{fig:threeRingSqueezing}). But, we can reduce the system down to an effective five resonance model by generalizing the dual ring structure considered in that example to a three ring resonator, as shown in Fig. \ref{fig:threeRingDiagram}. In a similar fashion to the auxiliary ring in the dual-ring case, each of the two auxiliary rings are tuned to hybridize the primary ring resonances and suppress the SFWM which involve them. In particular, we choose for one of the auxiliary ring, referred to here as 'Aux1' with length $L_{aux1} = \frac{3}{4} L_p$, to split the $J = \pm 2p$ while the other auxiliary ring, referred to as 'Aux2' with length $L_{aux2} = \frac{5}{6} L_p$, splits the $J = \pm 3p$ resonances. In such a scheme, the Aux2 ring suppresses the interactions which would generate the strong secondary pump in the $J = \pm 3p$ resonances, where as the Aux1 ring suppresses the interactions which couple the $J = \pm 2p$ resonances to the signal ($J = 0$) as in the dual ring, five resonance case (see Fig. \ref{fig:threeRing_transmission}). We denote by $\varkappa_{aux1}$ and $\varkappa_{aux2}$ the effective cross-coupling coefficient for the Aux1 and Aux2 rings respectively.

In Fig \ref{fig:threeRingSqueezing} we show the maximum squeezing in the signal output as a function of the $P_1$ and $P_2$ pump detunings for the three ring photonic molecule under a nine mode approximation (solid lines) for the cases when each auxiliary ring is coupled to the primary or decoupled from the primary resonator. This is compared with the maximum squeezing in the signal output for the dual ring scheme under a five resonance approximation (dashed lines) as considered in Section \ref{sec:SectionV}. In the three ring, nine resonance calculations, we treat the $J = \pm 3p$ resonance fields classically and identical to the $J = \pm p$ resonances, but with a vacuum input. The classical operators corresponding to the $J = \pm p, \pm 3p$ then seed the field generation in the $J = 0, \pm 2p$ and $\pm 4p$ resonances as with the $J = 0$ and $\pm 2p$ resonances in the five resonance approximation. We include all interactions which satisfy $J_1 + J_2 - J_3 - J_4 = 0$ which, unlike the five resonance case, are too numerous to enumerate here. Indeed, from Fig. \ref{fig:threeRingSqueezing} we note that the output signal squeezing in the three ring, nine resonance model closely resembles the maximum output squeezing in the dual-ring, five resonance model for the case of $\varkappa_{Aux2} = 0.06$, corresponding to a sufficiently large splitting of the $J = \pm 3p$ resonances.

On the other hand, for the parameters considered in Example 2 of Section \ref{sec:SectionV}, the effect of the generated secondary pumps are much less significant as the GVD leads to the relevant cubic pump interaction which generates the secondary pumps being significantly detuned, resulting in a much lower power in the $J = \pm 3p$ resonances. This is shown in Fig. \ref{fig:pushingPlotsMerged} in which the extension of the dual-ring, five resonance model to a dual-ring, nine resonance model manifests as a small decrease to the squeezing of the maximally squeezed mode of the signal output. As in the three ring, nine resonance calculation for the resonance splitting example above, we treat the $J = \pm 3p$ resonances classically, but only use the single auxiliary ring to shift the $P_1$ pump resonance. 

\end{document}